\documentclass[preprint,aps,prd,nofootinbib]{revtex4-2}

\usepackage[T1]{fontenc}
\usepackage[utf8]{inputenc}
\usepackage{amsmath,amssymb,bm}
\usepackage{graphicx}
\graphicspath{{./figures/}}
\usepackage{xcolor}
\usepackage[colorlinks=true,linkcolor=blue,citecolor=blue,urlcolor=blue]{hyperref}
\usepackage{float}
\usepackage{placeins}
\usepackage{subcaption}
\usepackage{slashed}
\usepackage{caption}

\AtBeginDocument{
	\setlength{\abovedisplayskip}{12pt plus 2pt minus 2pt}
	\setlength{\belowdisplayskip}{12pt plus 2pt minus 2pt}
	\setlength{\abovedisplayshortskip}{10pt plus 2pt minus 2pt}
	\setlength{\belowdisplayshortskip}{10pt plus 2pt minus 2pt}
	\setlength{\jot}{7pt}
}

\begin{document}
	
	\title{Electric Dipole Moments in the CP-violating Generational Three-Higgs-Doublet Model}
	
	\author{Hao-Ran Ma$^{1,2,3}$}
	\email{haoranma2024@163.com}
	
	\author{Ti-Bin Hou$^{1,2,3}$}
	\email{tibinhou2312@163.com}
	
	\author{Jin-Lei Yang$^{1,2,3}$}
	\email{jlyang@hbu.edu.cn}

	\author{Tai-Fu Feng$^{1,2,3,4}$}
	\email{fengtf@hbu.edu.cn}
	
	\affiliation{$^1$Department of Physics, Hebei University, Baoding 071002, China}
	\affiliation{$^2$Hebei Key Laboratory of High-precision Computation and Application of Quantum Field Theory, Baoding 071002, China}
	\affiliation{$^3$Hebei Research Center of the Basic Discipline for Computational Physics, Baoding 071002, China}
	\affiliation{$^4$Department of Physics, Chongqing University, Chongqing 401331, China}
	
	\begin{abstract}
		Motivated by the Yukawa interactions in the CP-violating Generational Three-Higgs-Doublet Model (G3HDM), we investigate the electric dipole moments (EDMs) of the $b$ quark, $c$ quark, electron, neutron, and mercury atom, imposing constraints from the observed $125~\mathrm{GeV}$ Higgs signals, flavor physics, and rare $B$-meson decays. We find that the $b$ quark, neutron, and mercury EDMs are predominantly controlled by top-enhanced charged-Higgs amplitudes, whose internal cancellations are generally weak. Meanwhile, pronounced pairwise cancellations between nearly degenerate neutral Higgs states occur in the $c$ quark, electron, neutron, and mercury EDMs. Using degenerate perturbation theory, we show that these cancellations originate from approximate relations among the neutral-Higgs mixing-matrix elements, which enforce the anti-alignment of the corresponding CP-odd coupling invariants.

	\end{abstract}

	\maketitle
	\clearpage   
	
	\section{Introduction}
	\label{sec:intro}
	The discovery of a scalar boson with a mass of approximately $125~\mathrm{GeV}$ at the Large Hadron Collider (LHC) represents an important milestone in particle physics and provides crucial experimental evidence for the electroweak symmetry breaking mechanism in the Standard Model (SM)~\cite{ATLAS_2012_Higgs,CMS_2012_Higgs}. Current experimental measurements show that the properties of this particle are highly consistent with the predictions for the SM Higgs boson~\cite{ATLAS_CMS_2016_Higgs,ATLAS_2022_Higgs,CMS_2022_Higgs}, but still allow for possible deviations in its coupling strengths and decay rates. Such deviations may suggest that the scalar sector possess a richer extended structure. The new scalar degrees of freedom can not only modify the scalar mass spectrum, the scalar mixing structure, and the interaction patterns of the Higgs boson~\cite{Lee_1973,Weinberg_1976,Branco_2012}, but also provide a simple and well-motivated theoretical framework for studying new flavor structures and mechanisms of CP violation (CPV)~\cite{Branco_1999,Branco_Ivanov_2016}.

	CPV is one of the Sakharov conditions and is generally regarded as an indispensable requirement for generating the matter-antimatter asymmetry of the Universe~\cite{Sakharov_1967}. However, in the SM, CPV is mainly induced by the irreducible complex phase in the Cabibbo-Kobayashi-Maskawa (CKM) matrix~\cite{Kobayashi_Maskawa_1973}, and its effect is far too small to account for the currently observed baryon asymmetry~\cite{Gavela_1994_1,Gavela_1994_2}. In contrast, an extended Higgs sector can naturally introduce new CPV phases through the scalar potential and Yukawa interactions, thereby providing an important possible source of CPV beyond the SM. However, new CP interactions are stringently constrained by the experimental upper bounds on electric dipole moment (EDM)~\cite{Pospelov_Ritz_2005,Engel_2013}. To date, no nonzero EDM has been observed in neutron, atomic, or molecular systems. The corresponding experimental results therefore impose severe constraints on CP phases beyond the SM~\cite{Abel_2020,Graner_2016,Roussy_2023}. In the CP-violating Generational Three-Higgs-Doublet Model (G3HDM), new CPV phases in the scalar and Yukawa sectors can be transmitted through loop processes to the EDM operators of quarks and leptons, thereby further affecting EDM observables in hadronic and atomic systems, such as the neutron and mercury EDMs~\cite{Pospelov_Ritz_2005,Engel_2013}. 
	
	In addition, the Yukawa interactions in the CP-violating G3HDM also affect the theoretical predictions on the $125~\mathrm{GeV}$ Higgs boson signals, neutral-meson mixing, the rare $B$-meson decays $\bar{B}\to X_s\gamma$ and $B_s^0\to\mu^+\mu^-$, charged-lepton-flavor-violating processes, and the flavor-changing neutral-current top-quark decays $t\to ch_5$ and $t\to uh_5$. To ensure the phenomenological viability of the analyzed numerical results, we impose the latest experimental constraints from these observables on the parameter space.

	The paper is organized as follows. In Section~\ref{sec:CPVG3HDM and higgs}, we briefly introduce the theoretical framework of the CP-violating G3HDM. In Section~\ref{sec:constraints and edm}, we present the Higgs mediated contributions to the EDMs of the $b$ quark, $c$ quark, electron, neutron and mercury atom, and summarize the relevant phenomenological constraints imposed in the numerical analysis. In Section~\ref{sec:Numerical Analysis}, we present the numerical results and discuss their phenomenological implications. Finally, Section~\ref{sec:Conclusion} summarizes the main conclusions of this work.

	\section{The CP-violating Generational Three-Higgs-Doublet Model}
	\label{sec:CPVG3HDM and higgs}
	
	\subsection{The scalar sector of the CP-violating G3HDM}
	\label{subsec:CPVG3HDM}
	
	\paragraph{Field content and scalar potential.}
	The scalar sector contains three Higgs doublets $\Phi_a$ ($a=1,2,3$), transforming under the SM gauge group $SU(3)_c\times SU(2)_L\times U(1)_Y$ as
	\begin{equation}
		\Phi_a\sim (1,2,\tfrac{1}{2}),
		\qquad a=1,2,3.
		\label{eq:Phi_rep}
	\end{equation}
	We assume that the renormalizable Higgs potential respects, to a good approximation, a softly broken $U(1)^3$ symmetry, with each $U(1)$ factor acting on a single Higgs doublet. The resulting potential can be written as
	\begin{align}
		V_{3\mathrm{HDM}}
		&=
		m_{11}^2(\Phi_1^\dagger \Phi_1)
		+ m_{22}^2(\Phi_2^\dagger \Phi_2)
		+ m_{33}^2(\Phi_3^\dagger \Phi_3)
		\nonumber\\
		&\quad
		-\Big[
		m_{12}^2(\Phi_1^\dagger \Phi_2)
		+ m_{23}^2(\Phi_2^\dagger \Phi_3)
		+ m_{13}^2(\Phi_1^\dagger \Phi_3)
		+ \mathrm{h.c.}
		\Big]
		\nonumber\\
		&\quad
		+ \lambda_1(\Phi_1^\dagger \Phi_1)^2
		+ \lambda_2(\Phi_2^\dagger \Phi_2)^2
		+ \lambda_3(\Phi_3^\dagger \Phi_3)^2
		+ \lambda_4(\Phi_1^\dagger \Phi_1)(\Phi_2^\dagger \Phi_2)
		+ \lambda_5(\Phi_1^\dagger \Phi_1)(\Phi_3^\dagger \Phi_3)
		\nonumber\\
		&\quad
		+ \lambda_6(\Phi_2^\dagger \Phi_2)(\Phi_3^\dagger \Phi_3)
		+ \lambda_7(\Phi_1^\dagger \Phi_2)(\Phi_2^\dagger \Phi_1)
		+ \lambda_8(\Phi_1^\dagger \Phi_3)(\Phi_3^\dagger \Phi_1)
		+ \lambda_9(\Phi_2^\dagger \Phi_3)(\Phi_3^\dagger \Phi_2).
		\label{eq:V3HDM}
	\end{align}
	Here, the diagonal mass parameters $m_{aa}^2$ and all quartic couplings $\lambda_i$ are taken to be real. The off-diagonal mass parameters $m_{ab}^2$ ($a\neq b$) correspond to the soft-breaking terms of the approximate $U(1)^3$ symmetry and can in general be complex.

	\paragraph{Electroweak symmetry breaking and vev parametrization.}
	We assume that electroweak symmetry breaking proceeds as usual,
	$SU(2)_L\times U(1)_Y\to U(1)_{\mathrm{em}}$, with the vacuum aligned such that $U(1)_{\mathrm{em}}$ remains unbroken.
	In the presence of CPV, the neutral components of the Higgs doublets may acquire complex vacuum expectation values (vevs). Without loss of generality, the phase of the first Higgs doublet can be removed by a field redefinition. We therefore parametrize the three Higgs doublets as
	\begin{equation}
		\begin{aligned}[b]
			\Phi_1 &=
			\begin{pmatrix}
				\varphi_1^+ \\
				\dfrac{1}{\sqrt{2}}\left(v_1+\phi_1+i a_1\right)
			\end{pmatrix},
			\qquad
			\Phi_2 = e^{i\theta_2\pi}
			\begin{pmatrix}
				\varphi_2^+ \\
				\dfrac{1}{\sqrt{2}}\left(v_2+\phi_2+i a_2\right)
			\end{pmatrix},
			\\[1.0em]
			\Phi_3 &= e^{i\theta_3\pi}
			\begin{pmatrix}
				\varphi_3^+ \\
				\dfrac{1}{\sqrt{2}}\left(v_3+\phi_3+i a_3\right)
			\end{pmatrix}.
		\end{aligned}
		\label{eq:Phi_decomp}
	\end{equation}
	Here $v_a$ denotes the magnitude of the VEV, which is real and positive, while $\theta_2$ and $\theta_3$ are the relative CPV phases measured in units of $\pi$. The fields are labeled such that $v_1\ll v_2\ll v_3$.
	The electroweak scale is fixed by
	\begin{equation}
		v^2\equiv v_1^2+v_2^2+v_3^2\simeq (246~\mathrm{GeV})^2.
		\label{eq:vsum}
	\end{equation}
	A convenient parametrization trades $(v_1,v_2,v_3)$ for $(v,\beta,\beta')$
	\begin{equation}
		v_1=v\cos\beta',
		\qquad
		v_2=v\sin\beta'\cos\beta,
		\qquad
		v_3=v\sin\beta'\sin\beta,
		\label{eq:vev_angles}
	\end{equation}
	implying
	\begin{equation}
		\tan\beta'=\frac{\sqrt{v_2^2+v_3^2}}{v_1},
		\qquad
		\tan\beta=\frac{v_3}{v_2}.
		\label{eq:tanb_def}
	\end{equation}
	\paragraph{Tadpole conditions.}
	Under rephasings of the Higgs fields, the phases of the vacuum expectation values and complex soft-breaking parameters depend on the chosen field basis, whereas physical observables remain unchanged. Individual phases are therefore not physical by themselves; only rephasing-invariant combinations of the vacuum phases and soft parameters carry physical significance. These combinations enter both the stationarity conditions of the scalar potential and the neutral-Higgs mass matrix, thereby controlling the mixing between the CP-even and CP-odd field components. The stationarity conditions further impose nontrivial relations among their imaginary parts. We therefore introduce the following rephasing-invariant combinations
	\begin{equation}
		\begin{aligned}[b]
			R_{12} &\equiv \mathrm{Re}\left(e^{i\theta_2\pi}m_{12}^2\right),
			&
			I_{12} &\equiv \mathrm{Im}\left(e^{i\theta_2\pi}m_{12}^2\right),\\
			R_{13} &\equiv \mathrm{Re}\left(e^{i\theta_3\pi}m_{13}^2\right),
			&
			I_{13} &\equiv \mathrm{Im}\left(e^{i\theta_3\pi}m_{13}^2\right),\\
			R_{23} &\equiv \mathrm{Re}\left(e^{i(\theta_3-\theta_2)\pi}m_{23}^2\right),
			&
			I_{23} &\equiv \mathrm{Im}\left(e^{i(\theta_3-\theta_2)\pi}m_{23}^2\right).
		\end{aligned}
		\label{eq:eq10}
	\end{equation}
	Then the CP-even tadpole equations can be written as
	\begin{equation}
		\begin{aligned}[b]
			\frac{\partial V}{\partial \phi_1}
			&=
			v_1\left[
			m_{11}^2+\lambda_1 v_1^2
			+\frac{1}{2}\left(\lambda_4+\lambda_7\right)v_2^2
			+\frac{1}{2}\left(\lambda_5+\lambda_8\right)v_3^2
			\right]
			-v_2R_{12}-v_3R_{13}
			=0,\\
			\frac{\partial V}{\partial \phi_2}
			&=
			v_2\left[
			m_{22}^2+\lambda_2 v_2^2
			+\frac{1}{2}\left(\lambda_4+\lambda_7\right)v_1^2
			+\frac{1}{2}\left(\lambda_6+\lambda_9\right)v_3^2
			\right]
			-v_1R_{12}-v_3R_{23}
			=0,\\
			\frac{\partial V}{\partial \phi_3}
			&=
			v_3\left[
			m_{33}^2+\lambda_3 v_3^2
			+\frac{1}{2}\left(\lambda_5+\lambda_8\right)v_1^2
			+\frac{1}{2}\left(\lambda_6+\lambda_9\right)v_2^2
			\right]
			-v_1R_{13}-v_2R_{23}
			=0.
		\end{aligned}
		\label{eq:eq11}
	\end{equation}
	The CP-odd tadpole equations can be written as 
	\begin{equation}
		\begin{aligned}[b]
			\frac{\partial V}{\partial a_1}
			&=
			-\left(v_2I_{12}+v_3I_{13}\right)
			=0,\\
			\frac{\partial V}{\partial a_2}
			&=
			v_1I_{12}-v_3I_{23}
			=0,\\
			\frac{\partial V}{\partial a_3}
			&=
			v_1I_{13}+v_2I_{23}
			=0.
		\end{aligned}
		\label{eq:eq12}
	\end{equation}
	
	\paragraph{Scalar mass matrices and diagonalization.}
	After electroweak symmetry breaking, the CP-even and CP-odd components in the neutral scalar sector generally mix, so the physical neutral Higgs bosons no longer have definite CP quantum numbers. After removing the neutral Goldstone boson $G^0$, five physical neutral Higgs bosons remain in this sector, denoted by $(h_1,h_2,h_3,h_4,h_5)$. In the charged scalar sector, besides the charged Goldstone bosons $G^\pm$, there are two physical charged Higgs fields, $H_1^\pm$ and $H_2^\pm$.
	
	In the basis $(\phi_1,\phi_2,\phi_3,a_1,a_2,a_3)$, the neutral Higgs mass-squared matrix is written as
	\begin{equation}
		M_h^2=
		\begin{pmatrix}
			M_{h,11}^2 & M_{h,12}^2 & M_{h,13}^2 & 0        & I_{12}  & I_{13}  \\
			M_{h,12}^2 & M_{h,22}^2 & M_{h,23}^2 & -I_{12}  & 0       & I_{23}  \\
			M_{h,13}^2 & M_{h,23}^2 & M_{h,33}^2 & -I_{13}  & -I_{23} & 0       \\
			0        & -I_{12}  & -I_{13}  & M_{h,44}^2 & -R_{12} & -R_{13} \\
			I_{12}  & 0        & -I_{23}  & -R_{12} & M_{h,55}^2 & -R_{23} \\
			I_{13}  & I_{23}  & 0        & -R_{13} & -R_{23} & M_{h,66}^2
		\end{pmatrix}.
		\label{eq:MH_matrix}
	\end{equation}
	where
	\begin{equation}
		\begin{aligned}[b]
			M_{h,11}^2 &=
			\frac{1}{v_1}
			\left(
			v_2R_{12}+v_3R_{13}+2\lambda_1v_1^3
			\right),
			\\
			M_{h,22}^2 &=
			\frac{1}{v_2}
			\left(
			v_1R_{12}+v_3R_{23}+2\lambda_2v_2^3
			\right),
			\\
			M_{h,33}^2 &=
			\frac{1}{v_3}
			\left(
			v_1R_{13}+v_2R_{23}+2\lambda_3v_3^3
			\right),
			\\
			M_{h,44}^2 &=
			\frac{1}{v_1}
			\left(
			v_2R_{12}+v_3R_{13}
			\right),
			\\
			M_{h,55}^2 &=
			\frac{1}{v_2}
			\left(
			v_1R_{12}+v_3R_{23}
			\right),
			\\
			M_{h,66}^2 &=
			\frac{1}{v_3}
			\left(
			v_1R_{13}+v_2R_{23}
			\right),
			\\
			M_{h,12}^2 &=
			(\lambda_4+\lambda_7)v_1v_2-R_{12},
			\\
			M_{h,13}^2 &=
			(\lambda_5+\lambda_8)v_1v_3-R_{13},
			\\
			M_{h,23}^2 &=
			(\lambda_6+\lambda_9)v_2v_3-R_{23}.
		\end{aligned}
		\label{eq:MH_elements}
	\end{equation}

	In the basis $(\varphi_1^+,\varphi_2^+,\varphi_3^+)$, the charged Higgs mass-squared matrix is written as
	\begin{equation}
		M_{H^\pm}^2=
		\begin{pmatrix}
			M_{\pm,11}^2 & M_{\pm,12}^2 & M_{\pm,13}^2 \\
			\left(M_{\pm,12}^2\right)^\ast & M_{\pm,22}^2 & M_{\pm,23}^2 \\
			\left(M_{\pm,13}^2\right)^\ast & \left(M_{\pm,23}^2\right)^\ast & M_{\pm,33}^2
		\end{pmatrix}.
		\label{eq:charged_mass_matrix}
	\end{equation}
	where
	\begin{equation}
		\begin{aligned}[b]
			M_{\pm,11}^2 
			&= 
			\frac{1}{v_1} 
			\left( 
			v_2R_{12}+v_3R_{13} 
			\right) 
			-\frac{1}{2} 
			\left( 
			\lambda_7v_2^2+\lambda_8v_3^2 
			\right), 
			\\
			M_{\pm,22}^2 
			&= 
			\frac{1}{v_2} 
			\left( 
			v_1R_{12}+v_3R_{23} 
			\right) 
			-\frac{1}{2} 
			\left( 
			\lambda_7v_1^2+\lambda_9v_3^2 
			\right), 
			\\
			M_{\pm,33}^2 
			&= 
			\frac{1}{v_3} 
			\left( 
			v_1R_{13}+v_2R_{23} 
			\right) 
			-\frac{1}{2} 
			\left( 
			\lambda_8v_1^2+\lambda_9v_2^2 
			\right), 
			\\
			M_{\pm,12}^2 
			&= 
			\frac{1}{2}\lambda_7v_1v_2 
			-R_{12} 
			+iI_{12}, 
			\\
			M_{\pm,13}^2 
			&= 
			\frac{1}{2}\lambda_8v_1v_3 
			-R_{13} 
			+iI_{13}, 
			\\
			M_{\pm,23}^2 
			&= 
			\frac{1}{2}\lambda_9v_2v_3 
			-R_{23} 
			+iI_{23}.
		\end{aligned}
		\label{eq:charged_elements}
	\end{equation}
	
	The neutral Higgs mass-squared matrix is diagonalized as a whole by the unitary matrix $Z^H$, the charged Higgs mass-squared matrix is diagonalized by the unitary matrix $Z^\pm$
	\begin{equation}
		Z^h M_h^2 \left(Z^h\right)^T
		=
		\mathrm{diag}
		\left(
		m_{h_1}^2,
		m_{h_2}^2,
		m_{h_3}^2,
		m_{h_4}^2,
		m_{h_5}^2,
		m_{G^0}^2
		\right),
		\label{eq:neutral_diag}
	\end{equation}
	\begin{equation}
		Z^\pm M_{H^\pm}^2 \left(Z^\pm\right)^\dagger
		=
		\mathrm{diag}
		\left(
		m_{H_2^\pm}^2,
		m_{H_1^\pm}^2,
		m_{G^\pm}^2
		\right).
		\label{eq:charged_diag}
	\end{equation}
	here $h_i$ $(i=1,\cdots,5)$ denote the five physical neutral Higgs bosons, while $G^0$ is the neutral Goldstone boson. In the charged sector, $H_1^\pm$ and $H_2^\pm$ are the two physical charged Higgs bosons, and $G^\pm$ is the charged Goldstone boson.
	
	\paragraph{Perturbative unitarity and vacuum stability.}
	
	Imposing the tree-level perturbative unitarity bounds~\cite{Lee_1977}, we adopt the following parameter ranges
	\begin{equation}
		|\lambda_1|,\,|\lambda_2|,\,|\lambda_3| \le \frac{4\pi}{3}, \qquad
		|\lambda_4 + \lambda_7|,\,|\lambda_5 + \lambda_8|,\,|\lambda_6 + \lambda_9| \le 8\pi.
		\label{eq:111}
	\end{equation}
	
	Furthermore, vacuum stability requires the scalar potential to be bounded from below, leading to the following conditions
	\begin{equation}
		\begin{aligned}[b]
			\text{(a)}\quad & \lambda_1>0,\;\lambda_2>0,\;\lambda_3>0,\\[4pt]
			\text{(b)}\quad & \Lambda_{12}+2\sqrt{\lambda_1\lambda_2}\ge0,\;
			\Lambda_{13}+2\sqrt{\lambda_1\lambda_3}\ge0,\;
			\Lambda_{23}+2\sqrt{\lambda_2\lambda_3}\ge0,\\[6pt]
			\text{(c)}\quad & \sqrt{\lambda_1\lambda_2\lambda_3}
			+\frac{1}{2}\Big(\Lambda_{12}\sqrt{\lambda_3}
			+\Lambda_{13}\sqrt{\lambda_2}
			+\Lambda_{23}\sqrt{\lambda_1}\Big) \\
			& +\frac{1}{2}\sqrt{\Big(\Lambda_{12}+2\sqrt{\lambda_1\lambda_2}\Big)
				\Big(\Lambda_{13}+2\sqrt{\lambda_1\lambda_3}\Big)
				\Big(\Lambda_{23}+2\sqrt{\lambda_2\lambda_3}\Big)} \ge 0.
		\end{aligned}
		\label{eq:222}
	\end{equation}
	where
	\begin{equation}
		\Lambda_{12} = \lambda_4 + \min(0,\lambda_7),\quad
		\Lambda_{13} = \lambda_5 + \min(0,\lambda_8),\quad
		\Lambda_{23} = \lambda_6 + \min(0,\lambda_9).
		\label{eq:333}
	\end{equation}
	\subsection{The fermion sector in the CP-violating G3HDM}
	
	The Yukawa interactions of the three Higgs doublets $\Phi_a$ ($a=1,2,3$) with the SM fermions are described by the most general renormalizable Yukawa Lagrangian
	\begin{align}
		&-\mathcal{L}^{\mathrm{Yuk}}_{3\mathrm{HDM}}=
		\sum_{a=1}^{3}\sum_{i,j=1}^{3}
		\Big(
		\lambda^{ua}_{ij}\,\bar q_{Li}\,\widetilde{\Phi}_a\,u_{Rj}
		+\lambda^{da}_{ij}\,\bar q_{Li}\,\Phi_a\,d_{Rj}
		+\lambda^{\ell a}_{ij}\,\bar \ell_{Li}\,\Phi_a\,e_{Rj}
		\Big)
		+\mathrm{h.c.}.
		\label{eq:Yukawa_Lagrangian}
	\end{align}
	where $\widetilde{\Phi}_a\equiv i\sigma_2\Phi_a^{\ast}$ and $i,j$ are flavor indices.
	Neutrino masses and mixing are neglected in the present setup.
	
	To realize a \emph{generational} structure, we adopt Yukawa textures of the form~\cite{Altmannshofer_2016_FlavorfulHiggs,Altmannshofer_2018_FlavorLocked,Altmannshofer_2018_Twist,Altmannshofer_2025_G3HDM}
	\begin{subequations}\label{eq:Yukawa_textures}
		\begin{align}
			&\lambda_{u1} \sim \frac{\sqrt{2}}{v_1}
			\begin{pmatrix}
				m_u & m_u & m_u\\
				m_u & m_u & m_u\\
				m_u & m_u & m_u
			\end{pmatrix},
			\qquad
			\lambda_{u2} \sim \frac{\sqrt{2}}{v_2 e^{i\theta_2\pi}}
			\begin{pmatrix}
				0 & 0 & 0\\
				0 & m_c & m_c\\
				0 & m_c & m_c
			\end{pmatrix},
			\qquad
			\lambda_{u3} \sim \frac{\sqrt{2}}{v_3 e^{i\theta_3\pi}}
			\begin{pmatrix}
				0 & 0 & 0\\
				0 & 0 & 0\\
				0 & 0 & m_t
			\end{pmatrix},
			\label{eq:Yukawa_textures_u}\\[2mm]
			&\lambda_{d1} \sim \frac{\sqrt{2}}{v_1}
			\begin{pmatrix}
				m_d & m_s\,\lambda & m_b\,\lambda^3\\
				m_d & m_d & m_d\\
				m_d & m_d & m_d
			\end{pmatrix},
			\qquad
			\lambda_{d2} \sim \frac{\sqrt{2}}{v_2 e^{i\theta_2\pi}}
			\begin{pmatrix}
				0 & 0 & 0\\
				0 & m_s & m_b\,\lambda^2\\
				0 & m_s & m_s
			\end{pmatrix},
			\qquad
			\lambda_{d3} \sim \frac{\sqrt{2}}{v_3 e^{i\theta_3\pi}}
			\begin{pmatrix}
				0 & 0 & 0\\
				0 & 0 & 0\\
				0 & 0 & m_b
			\end{pmatrix},
			\label{eq:Yukawa_textures_d}\\[2mm]
			&\lambda_{\ell 1} \sim \frac{\sqrt{2}}{v_1}
			\begin{pmatrix}
				m_e & m_e & m_e\\
				m_e & m_e & m_e\\
				m_e & m_e & m_e
			\end{pmatrix},
			\qquad
			\lambda_{\ell 2} \sim \frac{\sqrt{2}}{v_2 e^{i\theta_2\pi}}
			\begin{pmatrix}
				0 & 0 & 0\\
				0 & m_\mu & m_\mu\\
				0 & m_\mu & m_\mu
			\end{pmatrix},
			\qquad
			\lambda_{\ell 3} \sim \frac{\sqrt{2}}{v_3 e^{i\theta_3\pi}}
			\begin{pmatrix}
				0 & 0 & 0\\
				0 & 0 & 0\\
				0 & 0 & m_\tau
			\end{pmatrix}.
			\label{eq:Yukawa_textures_l}
		\end{align}
	\end{subequations}
	Here, “$\sim$” indicates the parametric order of the matrix elements, and $\lambda\simeq |V_{us}|$ is the Wolfenstein parameter. Each Yukawa matrix is assumed to have rank one, implying that every Higgs doublet couples to only one linear combination of the three fermion generations. The nonzero elements of a given matrix may nevertheless differ by complex $\mathcal{O}(1)$ coefficients. By an appropriate phase convention, the phases of the vacuum expectation values can be absorbed into the complex Yukawa parameters. These phases will therefore be treated implicitly as part of the Yukawa matrix elements and will not be displayed separately. 
	
	Assuming that the CKM matrix originates from the diagonalization of the down quark mass matrix, we define in the fermion mass-eigenstate basis the following mass parameters
	\begin{align}
		&m^{f1}_{ff'} = \frac{v_1}{\sqrt{2}}\,\langle f_L|\lambda_{f1}|f'_R\rangle,
		\qquad
		m^{f2}_{ff'} = \frac{v_2}{\sqrt{2}}\,\langle f_L|\lambda_{f2}|f'_R\rangle,
		\qquad
		m^{f3}_{ff'} = \frac{v_3}{\sqrt{2}}\,\langle f_L|\lambda_{f3}|f'_R\rangle,
		\label{eq:mf_def}
	\end{align}
	which satisfy
	\begin{equation}
		m^{f3}_{ff'}+m^{f2}_{ff'}+m^{f1}_{ff'}=m_f\,\delta_{ff'}.
		\label{eq:mf_sumrule}
	\end{equation}
	with $m_f$ denoting the physical fermion masses.
	
	Expanding to leading order in the ratios of first-to-second and second-to-third generation masses, one obtains~\cite{Altmannshofer_2025_G3HDM}
	\begin{align}
		&\frac{m^{u1}_{qq'}}{m_u} \simeq
		\begin{pmatrix}
			1 & O^{u}_{uc} & O^{u}_{ut}\\
			O^{u}_{cu} & O^{u}_{cu}O^{u}_{uc} & O^{u}_{cu}O^{u}_{ut}\\
			O^{u}_{tu} & O^{u}_{tu}O^{u}_{uc} & O^{u}_{tu}O^{u}_{ut}
		\end{pmatrix},
		\qquad
		\frac{m^{u2}_{qq'}}{m_c} \simeq
		\begin{pmatrix}
			\dfrac{m_u^2}{m_c^2}O^{u}_{uc}O^{u}_{cu} & -\dfrac{m_u}{m_c}O^{u}_{uc} & -\dfrac{m_u}{m_c}O^{u}_{uc}O^{u}_{ct}\\
			-\dfrac{m_u}{m_c}O^{u}_{cu} & 1 & O^{u}_{ct}\\
			-\dfrac{m_u}{m_c}O^{u}_{tc}O^{u}_{cu} & O^{u}_{tc} & O^{u}_{tc}O^{u}_{ct}
		\end{pmatrix},
		\label{eq:mu12}\\[1mm]
		&\frac{m^{u3}_{qq'}}{m_t} \simeq
		\begin{pmatrix}
			\dfrac{m_u^2}{m_t^2}(O^{u}_{ut}-O^{u}_{uc}O^{u}_{ct})(O^{u}_{tu}-O^{u}_{tc}O^{u}_{cu}) & \dfrac{m_um_c}{m_t^2}(O^{u}_{ut}-O^{u}_{uc}O^{u}_{ct})O^{u}_{tc} & -\dfrac{m_u}{m_t}(O^{u}_{ut}-O^{u}_{uc}O^{u}_{ct})\\
			\dfrac{m_um_c}{m_t^2}O^{u}_{ct}(O^{u}_{tu}-O^{u}_{tc}O^{u}_{cu}) & \dfrac{m_c^2}{m_t^2}O^{u}_{ct}O^{u}_{tc} & -\dfrac{m_c}{m_t}O^{u}_{ct}\\
			-\dfrac{m_u}{m_t}(O^{u}_{tu}-O^{u}_{tc}O^{u}_{cu}) & -\dfrac{m_c}{m_t}O^{u}_{tc} & 1
		\end{pmatrix},
		\label{eq:mu3}\\[1mm]
		&\frac{m^{d1}_{qq'}}{m_d} \simeq
		\begin{pmatrix}
			1 & \dfrac{m_s}{m_d}V_{ud}^\ast V_{us} & \dfrac{m_b}{m_d}V_{ud}^\ast V_{ub}\\
			O^{d}_{sd} & O^{d}_{sd}\dfrac{m_s}{m_d}V_{ud}^\ast V_{us} & O^{d}_{sd}\dfrac{m_b}{m_d}V_{ud}^\ast V_{ub}\\
			O^{d}_{bd} & O^{d}_{bd}\dfrac{m_s}{m_d}V_{ud}^\ast V_{us} & O^{d}_{bd}\dfrac{m_b}{m_d}V_{ud}^\ast V_{ub}
		\end{pmatrix},
		\qquad
		\frac{m^{d2}_{qq'}}{m_s} \simeq
		\begin{pmatrix}
			-\dfrac{m_d}{m_s}V_{cd}^\ast V_{cs}\,O^{d}_{sd} & V_{cd}^\ast V_{cs} & \dfrac{m_b}{m_s}V_{cd}^\ast V_{cb}\\
			-\dfrac{m_d}{m_s}O^{d}_{sd} & 1 & \dfrac{m_b}{m_s}V_{cs}^\ast V_{cb}\\
			-\dfrac{m_d}{m_s}O^{d}_{bs}O^{d}_{sd} & O^{d}_{bs} & O^{d}_{bs}\dfrac{m_b}{m_s}V_{cs}^\ast V_{cb}
		\end{pmatrix},
		\label{eq:md12}\\[1mm]
		&\frac{m^{d3}_{qq'}}{m_b} \simeq
		\begin{pmatrix}
			-\dfrac{m_d}{m_b}V_{td}^\ast V_{tb}(O^{d}_{bd}-O^{d}_{bs}O^{d}_{sd}) & -\dfrac{m_s}{m_b}V_{td}^\ast V_{tb}O^{d}_{bs} & V_{td}^\ast V_{tb}\\
			-\dfrac{m_d}{m_b}V_{ts}^\ast V_{tb}(O^{d}_{bd}-O^{d}_{bs}O^{d}_{sd}) & -\dfrac{m_s}{m_b}V_{ts}^\ast V_{tb}O^{d}_{bs} & V_{ts}^\ast V_{tb}\\
			-\dfrac{m_d}{m_b}(O^{d}_{bd}-O^{d}_{bs}O^{d}_{sd}) & -\dfrac{m_s}{m_b}O^{d}_{bs} & 1
		\end{pmatrix}.
		\label{eq:md3}
	\end{align}
	The lepton mass parameters are completely analogous to those in the up quark sector. In the above expressions, $O^{q}_{ij}$ are free, in general complex, $\mathcal{O}(1)$ parameters, encoding additional sources of flavor and CPV beyond the SM.
	
	\section{Experimental constraints and electric dipole moments}
	\label{sec:constraints and edm}
	
	In this section, we present the relevant experimental constraints and the Higgs-mediated contributions to the EDMs of the $b$ quark, $c$ quark, electron, neutron, and mercury atom.
	
	\subsection{The $125~\mathrm{GeV}$ Higgs boson decays}
	
	The observed $125~\mathrm{GeV}$ Higgs boson is an important probe of extended scalar sectors. In the CP-violating G3HDM, this particle is identified with the lightest neutral Higgs mass eigenstate, whose couplings to gauge bosons and fermions are generally modified by scalar mixing effects. Therefore, the Higgs signal strengths measured at the LHC can impose important constraints on the scalar mixing structure of the model. The signal strength of the $125~\mathrm{GeV}$ Higgs boson is defined as~\cite{Navas_2024} 
	\begin{align}
		&\mu_{\gamma\gamma}(h_5)
		=
		\frac{
			\sigma(gg \to h_5^{\mathrm{NP}})\,
			\mathrm{BR}(h_5^{\mathrm{NP}} \to \gamma\gamma)
		}{
			\sigma(gg \to h_5^{\mathrm{SM}})\,
			\mathrm{BR}(h_5^{\mathrm{SM}} \to \gamma\gamma)
		}
		= 1.10 \pm 0.06 ,
		\label{eq:mu_h2_gammagamma}
		\\
		&\mu_{WW^*}(h_5)
		=
		\frac{
			\sigma(gg \to h_5^{\mathrm{NP}})\,
			\mathrm{BR}(h_5^{\mathrm{NP}} \to WW^*)
		}{
			\sigma(gg \to h_5^{\mathrm{SM}})\,
			\mathrm{BR}(h_5^{\mathrm{SM}} \to WW^*)
		}
		= 1.00 \pm 0.08 ,
		\label{eq:mu_h2_WW}
		\\
		&\mu_{ZZ^*}(h_5)
		=
		\frac{
			\sigma(gg \to h_5^{\mathrm{NP}})\,
			\mathrm{BR}(h_5^{\mathrm{NP}} \to ZZ^*)
		}{
			\sigma(gg \to h_5^{\mathrm{SM}})\,
			\mathrm{BR}(h_5^{\mathrm{SM}} \to ZZ^*)
		}
		= 1.02 \pm 0.08 ,
		\label{eq:mu_h2_ZZ}
		\\
		&\mu_{b\bar b}(h_5)
		\simeq
		\frac{
			\Gamma(h_5^{\mathrm{NP}} \to VV^*)\,
			\mathrm{BR}(h_5^{\mathrm{NP}} \to b\bar b)
		}{
			\Gamma(h_5^{\mathrm{SM}} \to VV^*)\,
			\mathrm{BR}(h_5^{\mathrm{SM}} \to b\bar b)
		}
		= 0.99 \pm 0.12 ,
		\label{eq:mu_h2_bb}
		\\
		&\mu_{\tau^+\tau^-}(h_5)
		\simeq
		\frac{
			\Gamma(h_5^{\mathrm{NP}} \to VV^*)\,
			\mathrm{BR}(h_5^{\mathrm{NP}} \to \tau^+\tau^-)
		}{
			\Gamma(h_5^{\mathrm{SM}} \to VV^*)\,
			\mathrm{BR}(h_5^{\mathrm{SM}} \to \tau^+\tau^-)
		}
		= 0.91 \pm 0.09 .
		\label{eq:mu_h2_tautau}
	\end{align}
	
	For reference, the corresponding SM predictions are~\cite{Navas_2024} 
	\begin{align} 
		&\Gamma_{\mathrm{tot},125}^{\mathrm{SM}} 
		\simeq 0.0041\,\mathrm{GeV},  
		\label{eq:Gamma_tot_SM_125} 
		\\ 
		&\mathrm{BR}(h_5^{\mathrm{SM}} \to \gamma\gamma) 
		\simeq 0.00227,  
		\label{eq:BR_SM_h2_gammagamma} 
		\\ 
		&\mathrm{BR}(h_5^{\mathrm{SM}} \to WW^*) 
		\simeq 0.214, 
		\label{eq:BR_SM_h2_WW} 
		\\ 
		&\mathrm{BR}(h_5^{\mathrm{SM}} \to ZZ^*) 
		\simeq 0.0262, 
		\label{eq:BR_SM_h2_ZZ} 
		\\
		&\mathrm{BR}(h_5^{\mathrm{SM}} \to b\bar b) 
		\simeq 0.582,  
		\label{eq:BR_SM_h2_bb} 
		\\ 
		&\mathrm{BR}(h_5^{\mathrm{SM}} \to \tau^+\tau^-) 
		\simeq 0.0627. 
		\label{eq:BR_SM_h2_tautau} 
	\end{align}
	
	In the present NP framework, the total decay width is approximated by 
	\begin{align}
		&\Gamma_{\mathrm{tot}}^{\mathrm{NP}}(h_i)
		\simeq
		\Gamma(h_i \to b\bar b)
		+ \Gamma(h_i \to c\bar c)
		+ \Gamma(h_i \to \tau^+\tau^-)
		\nonumber
		\\
		&\quad
		+ \sum_{V=W,Z}\Gamma(h_i \to VV^*)
		+ \Gamma(h_i \to gg).
		\qquad
		\label{eq:Gamma_tot_NP_hi}
	\end{align}
	where subleading channels such as $\gamma\gamma$, $Z\gamma$, and $\mu^+\mu^-$, as well as possible exotic decay modes, are neglected.
	
	The partial decay widths of the Higgs bosons into fermions and gauge bosons in the CP-violating G3HDM can be expressed as~\cite{Djouadi_2008,Ellis_1976,Shifman_1979,Bergstrom_1985}
	\begin{align}
		&\Gamma^{\mathrm{NP}}(h_i \to gg)
		=
		\frac{G_F \alpha_s^2 m_{h_i}^3}{64\sqrt{2}\pi^3}
		\left[
		\left|
		\sum_{q}
		g^{S}_{h_i q\bar q}
		A_{1/2}(x_q)
		\right|^2
		+
		\left|
		\sum_{q}
		g^{A}_{h_i q\bar q}
		A_{2}(x_q)
		\right|^2
		\right],
		\label{eq:Gamma_hi_gg}
		\\[2mm]
		&\Gamma^{\mathrm{NP}}(h_i \to \gamma\gamma)
		=
		\frac{G_F \alpha^2 m_{h_i}^3}{128\sqrt{2}\pi^3}
		\left[
		\left|
		\sum_{f=u,d,\ell}
		N_c^f e_f^2
		g^{S}_{h_i f\bar f}
		A_{1/2}(x_f)
		+
		g^{S}_{h_i WW}
		A_1(x_W)
		\right.\right.
		\nonumber
		\\
		&\left.\left.\hspace{3.0cm}
		+
		\sum_{a=1}^{2}
		g^{S}_{h_i H_a^+H_a^-}
		\frac{m_Z^2}{m_{H_a^\pm}^2}
		A_0(x_{H_a^\pm})
		\right|^2
		+
		\left|
		\sum_{f=u,d,\ell}
		N_c^f e_f^2
		g^{A}_{h_i f\bar f}
		A_2(x_f)
		\right|^2
		\right],
		\label{eq:Gamma_hi_gammagamma}
		\\[2mm]
		&\Gamma^{\mathrm{NP}}(h_i \to VV^*)
		=
		\left|
		g^{S}_{h_iVV}
		\right|^2
		\Gamma^{\mathrm{SM}}(h_i \to VV^*),
		\qquad V=W,Z ,
		\label{eq:Gamma_hi_VV}
		\\[2mm]
		&\Gamma^{\mathrm{NP}}(h_i \to f\bar f)
		=
		N_c^f
		\frac{G_F m_f^2 m_{h_i}}{4\sqrt{2}\pi}
		\left[
		\left|
		g^{S}_{h_i f\bar f}
		\right|^2
		\left(
		1-\frac{4m_f^2}{m_{h_i}^2}
		\right)^{3/2}
		+
		\left|
		g^{A}_{h_i f\bar f}
		\right|^2
		\left(
		1-\frac{4m_f^2}{m_{h_i}^2}
		\right)^{3/2}
		\right].
		\label{eq:Gamma_hi_ff}
	\end{align}
	Here, $G_F$ is the Fermi constant, $g^S$ denotes the CP-even scalar Higgs coupling, and $g^A$ denotes the CP-odd scalar Higgs coupling. Moreover, $N_c$ is the color factor, with $N_c=3$ for quarks, $N_c=1$ for leptons, and $e_q$ denotes the electric charge of the quark $q$ in units of the elementary charge $e$. The loop functions in the calculations are 
	\begin{equation}
		\label{eq:higgs-loop-functions}
		\begin{aligned}[b]
			&A_{1/2}(x)
			=
			\frac{2}{x^2}
			\left[
			x+(x-1)g(x)
			\right],
			\;A_0(x)
			=
			-\frac{x-g(x)}{x^2},
			\\[2mm]
			&A_1(x)
			=
			-\frac{1}{x^2}
			\left[
			2x^2+3x+3(2x-1)g(x)
			\right],\;A_2(x)
			=
			\frac{2g(x)}{x},
			\\[2mm]
			&g(x)
			=
			\begin{cases}
				\arcsin^2\sqrt{x},
				& x\leq 1,
				\\[2mm]
				-\dfrac{1}{4}
				\left[
				\ln
				\left(
				\dfrac{1+\sqrt{1-1/x}}
				{1-\sqrt{1-1/x}}
				\right)
				-i\pi
				\right]^2,
				& x>1.
			\end{cases}
		\end{aligned}
	\end{equation}
	
	To quantify the overall consistency between the model predictions and the measured properties of the $125~\mathrm{GeV}$ Higgs boson, we construct a combined $\chi^2$ function, which includes the mass of the SM-like state $h_5$ and the five signal strengths introduced above
	\begin{align}
		\chi^2_{h_5}
		=
		\left(
		\frac{m_{h_5}^{\mathrm{th}}-m_{h_5}^{\mathrm{exp}}}{\delta_{m_h}}
		\right)^{2}
		+
		\sum_{X}
		\left(
		\frac{\mu_{X}^{\mathrm{th}}-\mu_{X}^{\mathrm{exp}}}{\delta_{X}}
		\right)^{2},
		\qquad
		X\in\{\gamma\gamma,\,ZZ^{*},\,WW^{*},\,b\bar b,\,\tau^{+}\tau^{-}\}.
		\label{eq:chi2_higgs}
	\end{align}
	Here, $\mu_{X}^{\mathrm{th}}$ denotes the predicted signal strength obtained from eqs.~(\ref{eq:Gamma_hi_gg})--(\ref{eq:Gamma_hi_ff}), while $\mu_{X}^{\mathrm{exp}}\pm\delta_{X}$ are the corresponding experimental central values and uncertainties listed in eqs.~(\ref{eq:mu_h2_gammagamma})--(\ref{eq:mu_h2_tautau}). A parameter point is regarded as compatible with the Higgs data if it lies within the $2\sigma$, namely $95.45\%$ confidence-level, region of the combined fit.
	
	\subsection{The rare top quark decay processes $t\to ch_5$ and $t\to uh_5$}
	
	In the CP-violating G3HDM, after rotating to the fermion mass basis, the extended Yukawa sector generally gives rise to tree-level Higgs-mediated flavor changing neutral-current (FCNC) couplings, such as $h_i\bar{t}c$ and $h_i\bar{t}u$. These off-diagonal neutral Higgs couplings provide additional contributions to the decay amplitudes of $t\to ch_5$ and $t\to uh_5$. In the numerical analysis, we require the branching ratios of $t\to ch_5$ and $t\to uh_5$ to remain below the corresponding experimental upper bounds. The adopted experimental upper bounds are  
	\begin{figure}[htbp]
		\centering
		
		\includegraphics[width=0.2\textwidth]{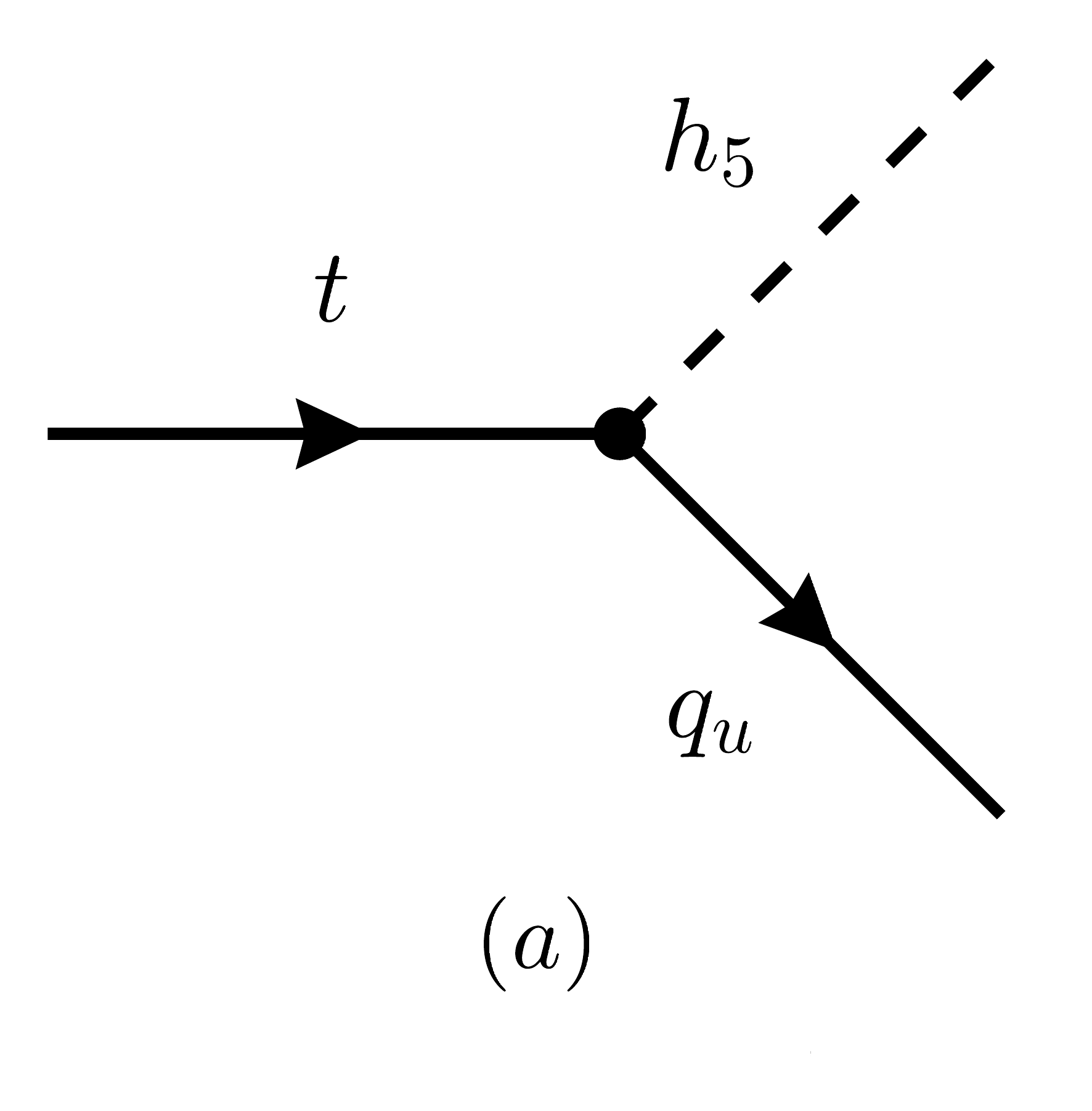}
		
		\vspace{0.15cm}
		
		\includegraphics[width=0.2\textwidth]{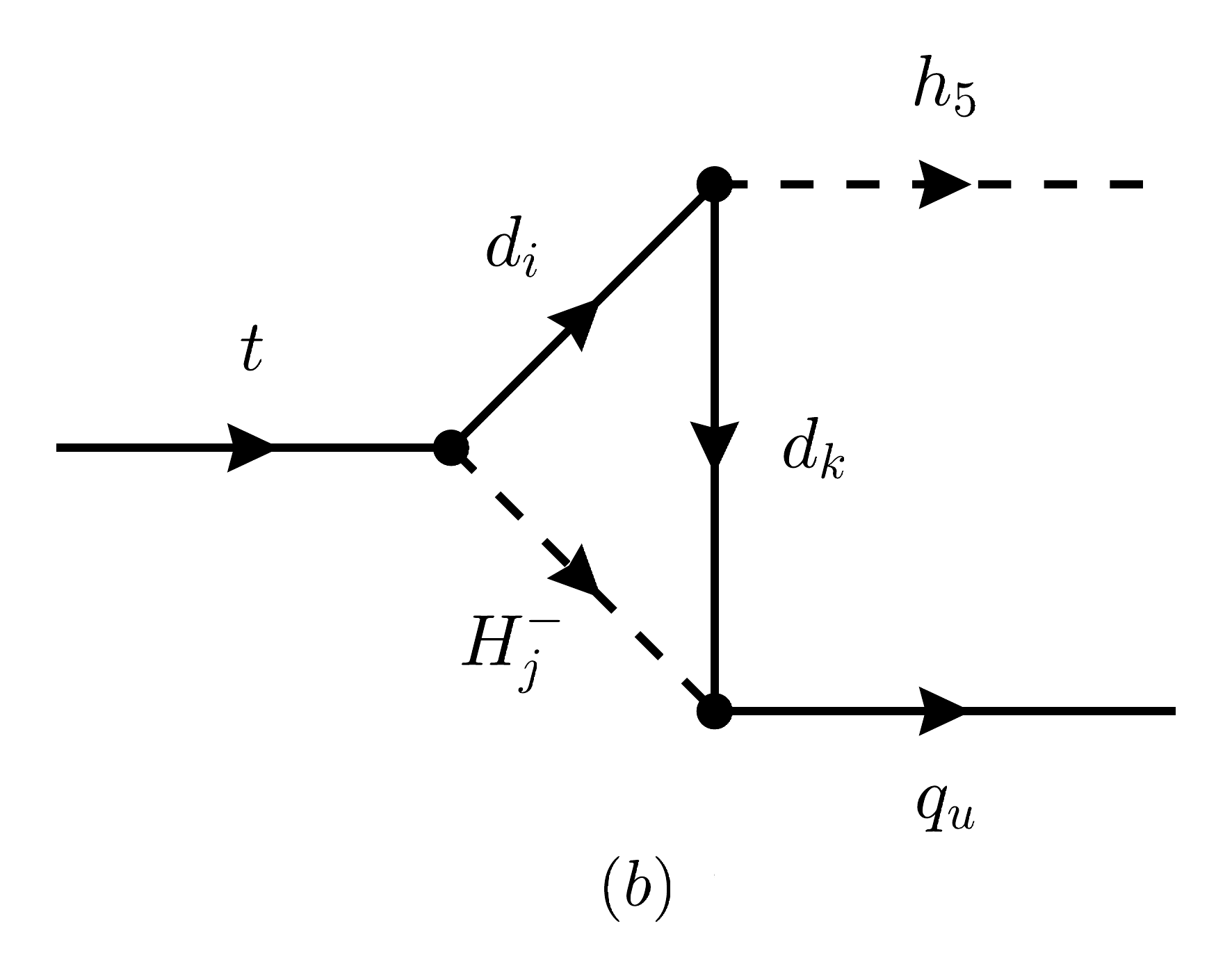}
		\hspace{0.02\textwidth}
		\includegraphics[width=0.2\textwidth]{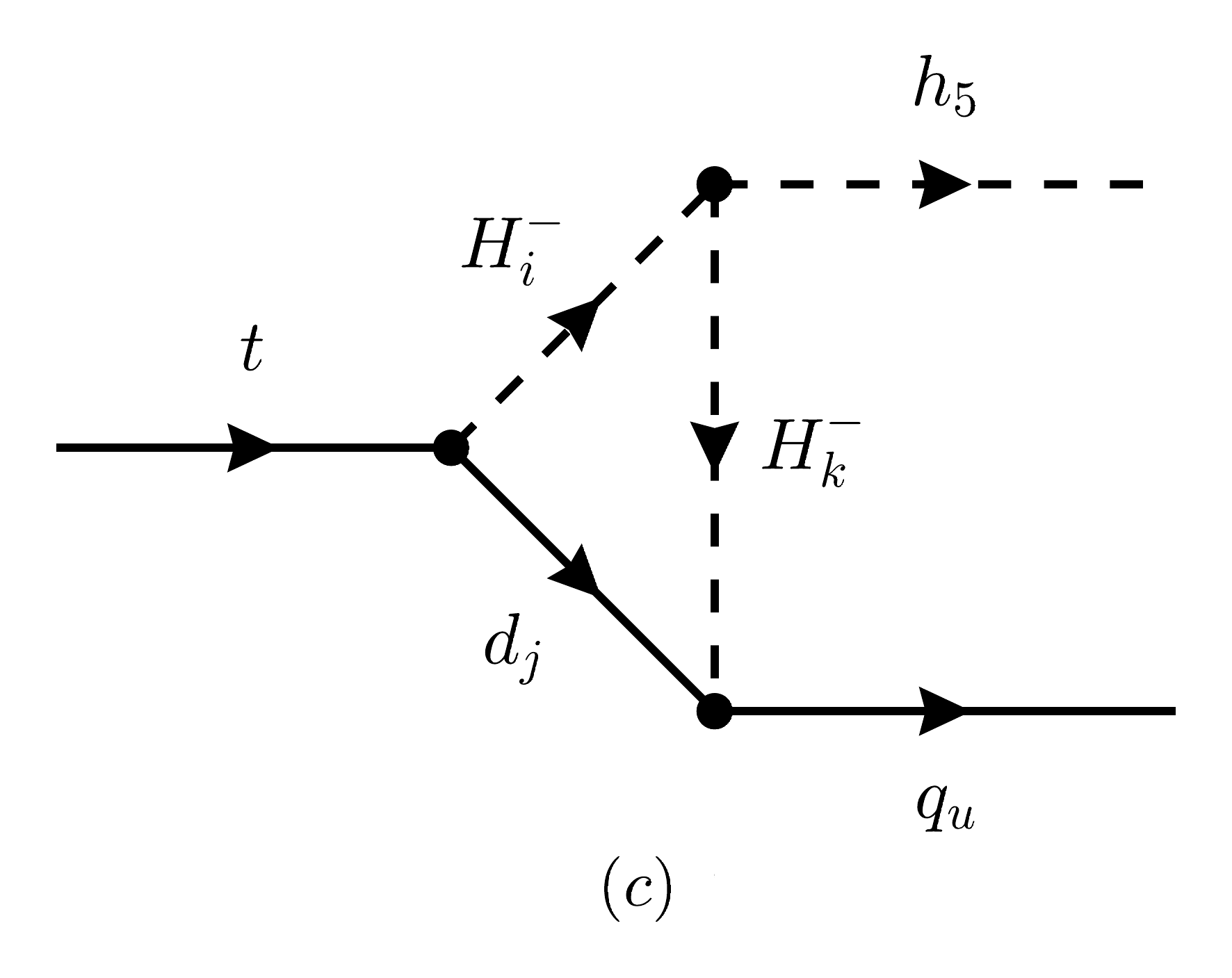}
		\hspace{0.02\textwidth}
		\includegraphics[width=0.2\textwidth]{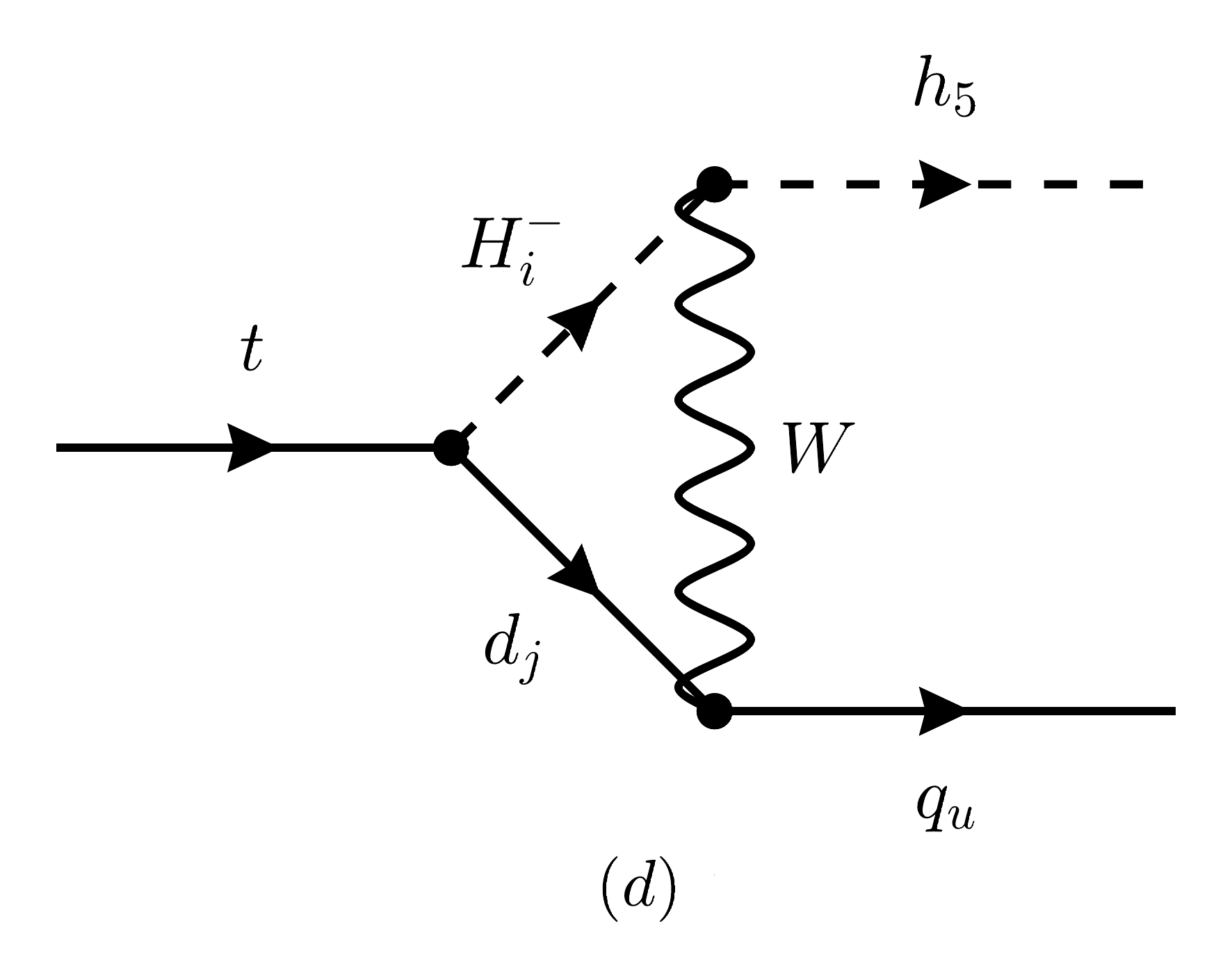}
		
		\vspace{0.15cm}
		
		\includegraphics[width=0.2\textwidth]{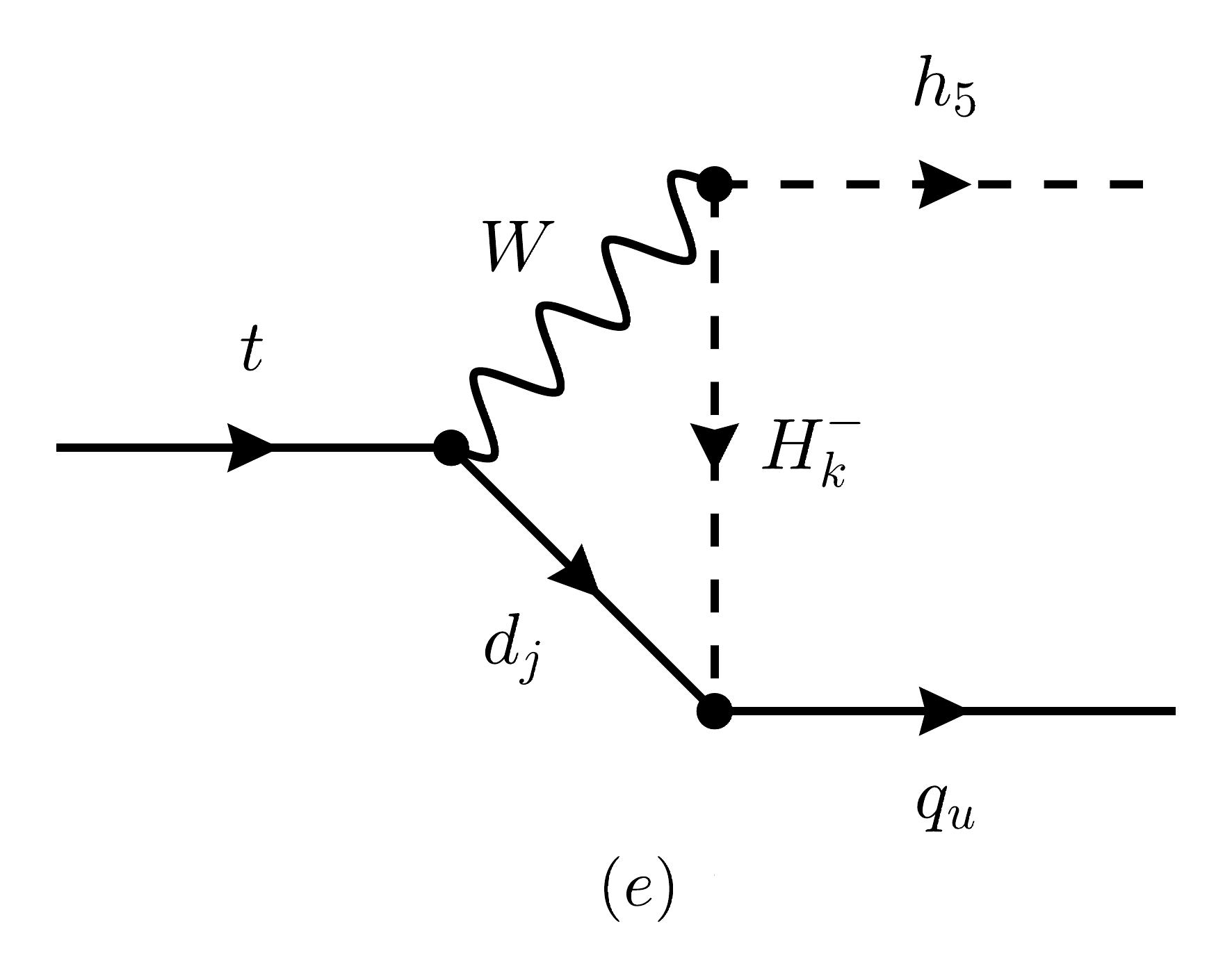}
		\hspace{0.02\textwidth}
		\includegraphics[width=0.2\textwidth]{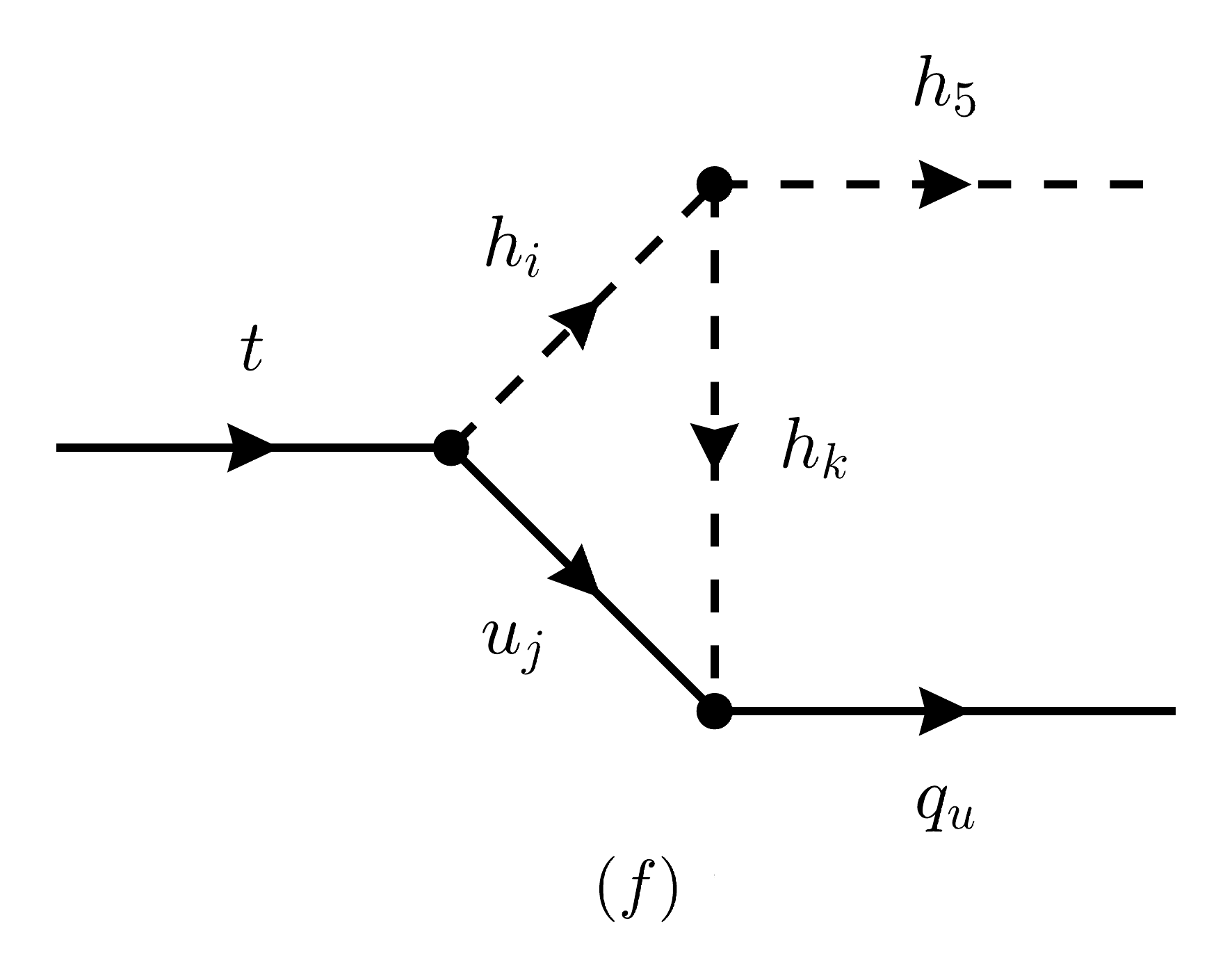}
		\hspace{0.02\textwidth}
		\includegraphics[width=0.2\textwidth]{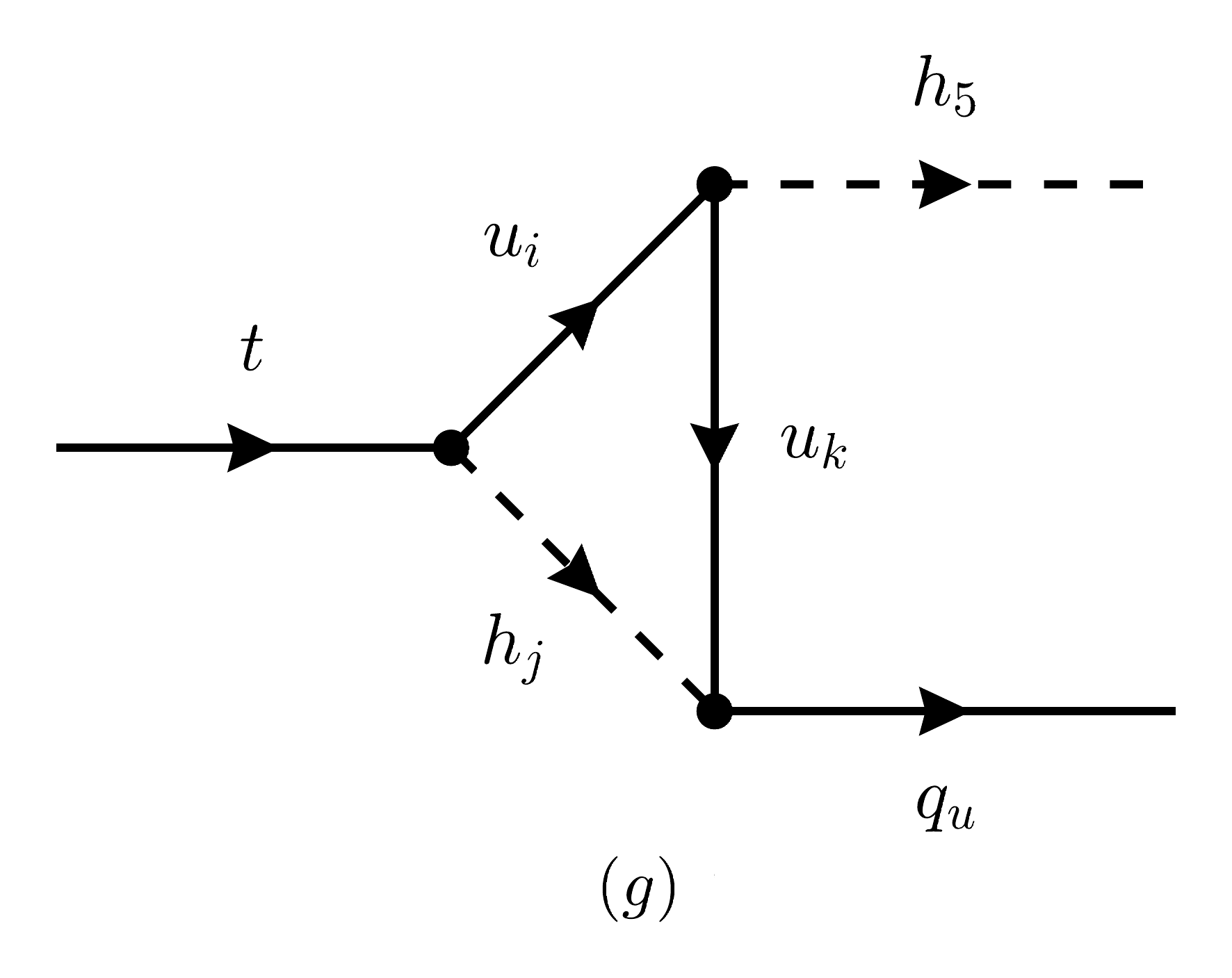}
		
		\caption{\raggedright
			The Feynman diagrams contributing to $t\to q_u h_5$ in the CP-violating G3HDM.}
		\label{fig:four_figures_one_row}
	\end{figure}
	\begin{equation}
		\begin{aligned}[b]
			&\mathrm{Br}(t\to ch_5) < 3.4\times 10^{-4},\\
			&\mathrm{Br}(t\to uh_5) < 1.9\times 10^{-4}.
		\end{aligned}
		\label{eq:top_rare_decay_bounds}
	\end{equation}
	
	The branching ratios of the rare top quark decay processes $t\to ch_5$ and $t\to uh_5$ can be written as
	\begin{equation}
		\mathrm{Br}(t\to q_u h_5)
		=
		\frac{
			\left|\mathcal{M}_{t q_u h_5}\right|^2
			\sqrt{
				\left[(m_t+m_{h_5})^2-m_{q_u}^2\right]
				\left[(m_t-m_{h_5})^2-m_{q_u}^2\right]
			}
		}{
			32\pi m_t^3 \Gamma_{\mathrm{total}}^t
		}.
		\label{eq:Br_t_to_quh}
	\end{equation}
	where $q_u=u,c$ and $\Gamma_{\mathrm{total}}^t=1.42~\mathrm{GeV}$~\cite{Navas_2024} is the total decay width of the top quark.
	
	\subsection{The $B$ meson rare decay processes $\bar{B}\to X_s\gamma$ and $B_s^0\to \mu^+\mu^-$}
	Rare $B$ meson decays can provide important constraints on the flavor structure of the CP-violating G3HDM. In particular, $\bar{B}\to X_s\gamma$ and $B_s^0\to\mu^+\mu^-$ are FCNC processes, which are mainly induced at the loop level in the SM and are therefore strongly suppressed. In the numerical analysis, we calculate the branching ratios of these two processes for each parameter point and require them to satisfy the corresponding experimental constraints. The experimental inputs adopted in this work are
	\begin{equation}
		\begin{aligned}[b]
			&\mathrm{Br}(\bar{B}\to X_s\gamma)
			=
			(3.49\pm0.19)\times10^{-4},
			\\
			&\mathrm{Br}(B_s^0\to\mu^+\mu^-)
			=
			(3.01\pm0.35)\times10^{-9}.
		\end{aligned}
		\label{eq:B-rare-exp}
	\end{equation}
	Parameter points that fall outside the allowed experimental ranges are excluded from the viable parameter space.
	
	To obtain the theoretical predictions for these observables, we employ the effective Hamiltonian formalism. The effective Hamiltonian governing $b\to s$ transitions at the low-energy scale can be written as
	\begin{equation}\label{eq:Heff}
		H_{\mathrm{eff}}
		=
		-\frac{4G_F}{\sqrt{2}}\,V_{ts}^*V_{tb}
		\left[
		C_1 \mathcal{O}_1^{c}
		+
		C_2 \mathcal{O}_2^{c}
		+
		\sum_{i=3}^{6} \mathcal{O}_i
		+
		\sum_{i=7}^{10}\left(C_i \mathcal{O}_i + C_i' \mathcal{O}_i'\right)
		+
		\sum_{i=S,P}\left(C_i \mathcal{O}_i + C_i' \mathcal{O}_i'\right)
		\right].
	\end{equation}
	where $\mathcal{O}_i,(i=1,2,\ldots,10,S,P)$ and $\mathcal{O}_i',(i=7,8,\ldots,10,S,P)$ are defined as~\cite{k1,Buchalla_1996,Altmannshofer_2009,L.LIN,Yang_2010,Goertz_2011}  
	\begin{equation}\label{eq:Oeff}
		\begin{aligned}[b]
			\mathcal{O}_1^u &= (\bar{s}_L \gamma_\mu T^a u_L)(\bar{u}_L \gamma^\mu T^a b_L), 
			&\qquad
			\mathcal{O}_2^u &= (\bar{s}_L \gamma_\mu u_L)(\bar{u}_L \gamma^\mu b_L), \\
			\mathcal{O}_3 &= (\bar{s}_L \gamma_\mu b_L)\sum_q (\bar{q}\gamma^\mu q), 
			&\qquad
			\mathcal{O}_4 &= (\bar{s}_L \gamma_\mu T^a b_L)\sum_q (\bar{q}\gamma^\mu T^a q), \\
			\mathcal{O}_5 &= (\bar{s}_L \gamma_\mu \gamma_\nu \gamma_\rho b_L)\sum_q (\bar{q}\gamma^\mu \gamma^\nu \gamma^\rho q), 
			&\qquad
			\mathcal{O}_6 &= (\bar{s}_L \gamma_\mu \gamma_\nu \gamma_\rho T^a b_L)\sum_q (\bar{q}\gamma^\mu \gamma^\nu \gamma^\rho T^a q), \\
			\mathcal{O}_7 &= \frac{e}{16\pi^2} m_b (\bar{s}_L \sigma_{\mu\nu} b_R)F^{\mu\nu}, 
			&\qquad
			\mathcal{O}_7' &= \frac{e}{16\pi^2} m_b (\bar{s}_R \sigma_{\mu\nu} b_L)F^{\mu\nu}, \\
			\mathcal{O}_8 &= \frac{g_s}{16\pi^2} m_b (\bar{s}_L \sigma_{\mu\nu} T^a b_R)G^{a,\mu\nu}, 
			&\qquad
			\mathcal{O}_8' &= \frac{g_s}{16\pi^2} m_b (\bar{s}_R \sigma_{\mu\nu} T^a b_L)G^{a,\mu\nu}, \\
			\mathcal{O}_9 &= \frac{e^2}{g_s^2} (\bar{s}_L \gamma_\mu b_L)\,\bar{l}\gamma^\mu l, 
			&\qquad
			\mathcal{O}_9' &= \frac{e^2}{g_s^2} (\bar{s}_R \gamma_\mu b_R)\,\bar{l}\gamma^\mu l, \\
			\mathcal{O}_{10} &= \frac{e^2}{g_s^2} (\bar{s}_L \gamma_\mu b_L)\,\bar{l}\gamma^\mu \gamma_5 l, 
			&\qquad
			\mathcal{O}_{10}' &= \frac{e^2}{g_s^2} (\bar{s}_R \gamma_\mu b_R)\,\bar{l}\gamma^\mu \gamma_5 l, \\
			\mathcal{O}_S &= \frac{e^2}{16\pi^2} m_b (\bar{s}_L b_R)\,\bar{l}l, 
			&\qquad
			\mathcal{O}_S' &= \frac{e^2}{16\pi^2} m_b (\bar{s}_R b_L)\,\bar{l}l, \\
			\mathcal{O}_P &= \frac{e^2}{16\pi^2} m_b (\bar{s}_L b_R)\,\bar{l}\gamma_5 l, 
			&\qquad
			\mathcal{O}_P' &= \frac{e^2}{16\pi^2} m_b (\bar{s}_R b_L)\,\bar{l}\gamma_5 l.
		\end{aligned}
	\end{equation}
	where $g_s$ denotes the strong coupling, $F_{\mu\nu}$ are the electromagnetic field strength tensor, $G_{\mu\nu}$ are the gluon field strength tensors, and $T^a$ $(a=1,\ldots,8)$ are the SU(3) generators.
	
	\begin{figure}[htbp]
		\centering
		
		\begin{subfigure}{0.4\textwidth}
			\centering
			\includegraphics[width=\linewidth]{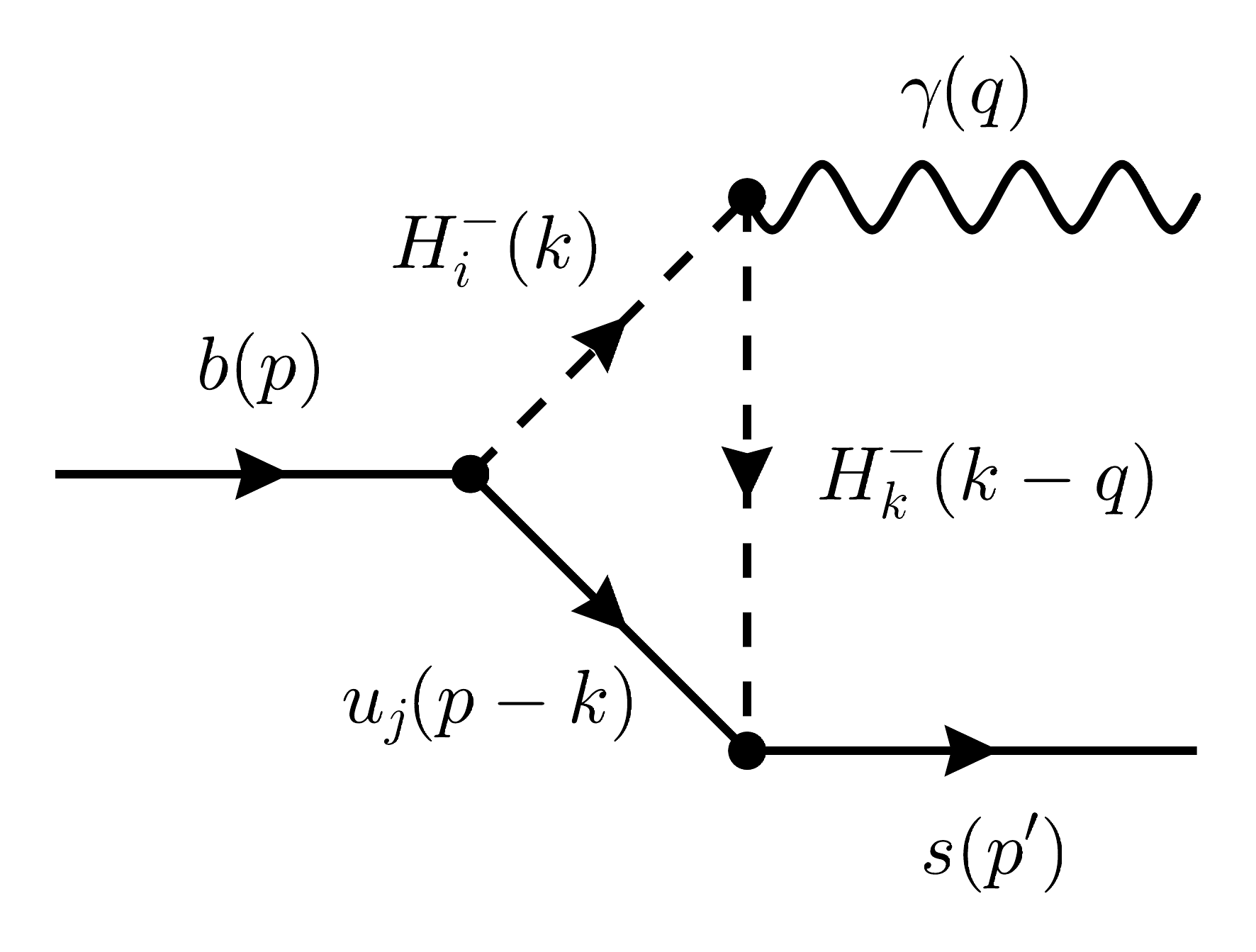}
			\caption*{(1)}
			\label{fig:sub11}
		\end{subfigure}
		\hspace{0.08\textwidth}
		\begin{subfigure}{0.4\textwidth}
			\centering
			\includegraphics[width=\linewidth]{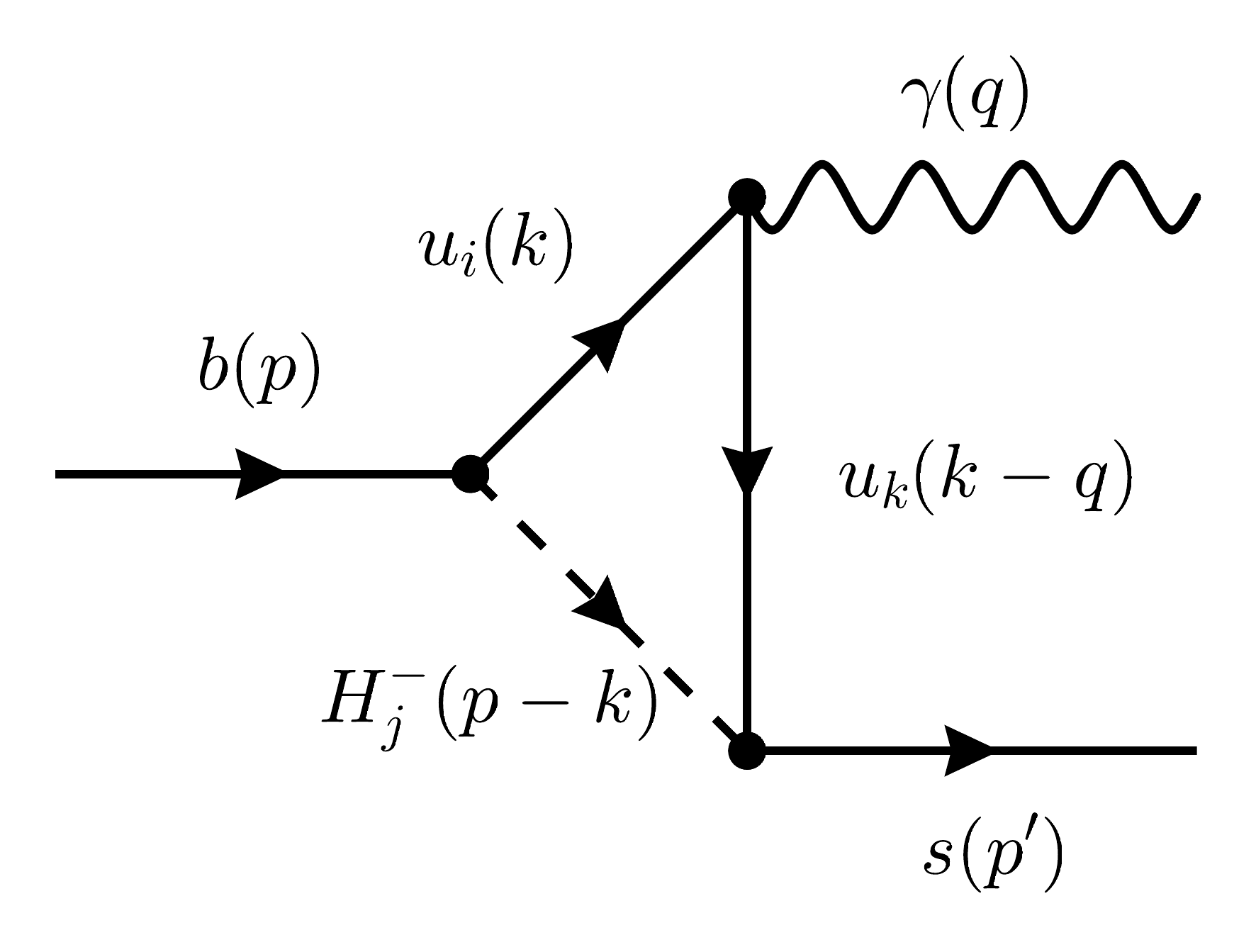}
			\caption*{(2)}
			\label{fig:sub22}
		\end{subfigure}
		
		\caption{\raggedright The one-loop Feynman diagrams contributing to $\bar{B}\to X_s\gamma$ from charged Higgs bosons in the CP-violating G3HDM}
		\label{fig:two-figures}
	\end{figure}
	
	\paragraph{Rare decay $\bar{B}\to X_s\gamma$.}
	
	The dominant contributions to $b\to s\gamma$ come from the charged Higgs
	in the CP-violating G3HDM, and the leading-order Feynman diagrams are
	shown in Fig.~\ref{fig:two-figures}. Then the branching ratio of $\bar{B}\to X_s\gamma$ can
	be written as
	\begin{equation}
		\label{eq:BXsgamma}
		\mathrm{Br}\left(\bar{B}\to X_s\gamma\right)
		=
		R\left[
		\left|C_{7\gamma}(\mu_b)\right|^2
		+
		N(E_\gamma)
		\right] .
	\end{equation}
	where the overall factor $R=2.47\times 10^{-3}$, and the nonperturbative contribution $N(E_\gamma)=(3.6\pm 0.6)\times 10^{-3}$~\cite{Buras_2011}. The Wilson coefficient $C_{7\gamma}(\mu_b)$ can be written as
	\begin{equation}
		\label{eq:C7gamma}
		C_{7\gamma}(\mu_b)
		=
		C_{7\gamma,\mathrm{SM}}(\mu_b)
		+
		C_{7,\mathrm{NP}}(\mu_b) .
	\end{equation}
	where the hadron scale is $\mu_b=2.5~\mathrm{GeV}$, and the SM contribution at NNLO level is given by $C_{7\gamma,\mathrm{SM}}(\mu_b)=-0.3689$~\cite{Gambino_2001,Czakon_2007}. In new physics models, the corresponding Wilson coefficients at the bottom quark scale are~\cite{Buras_1994,T.-J. Gao}
	\begin{equation}
		C_{7,\mathrm{NP}}(\mu_b)
		\simeq
		0.5696\,C_{7,\mathrm{NP}}(\mu_{\mathrm{EW}})
		+
		0.1107\,C_{8,\mathrm{NP}}(\mu_{\mathrm{EW}}) .
	\end{equation}
	where
	\begin{equation}
		\label{eq:C7C8NP-muEW}
		\begin{aligned}[b]
			C_{7,\mathrm{NP}}(\mu_{\mathrm{EW}})
			&=
			C_{7,\mathrm{NP}}^{(1)}(\mu_{\mathrm{EW}})
			+
			C_{7,\mathrm{NP}}^{(2)}(\mu_{\mathrm{EW}})
			+
			C_{7,\mathrm{NP}}^{\prime(1)}(\mu_{\mathrm{EW}})
			+
			C_{7,\mathrm{NP}}^{\prime(2)}(\mu_{\mathrm{EW}}),
			\\
			C_{8,\mathrm{NP}}(\mu_{\mathrm{EW}})
			&=
			C_{8,\mathrm{NP}}(\mu_{\mathrm{EW}})
			+
			C_{8,\mathrm{NP}}^{\prime}(\mu_{\mathrm{EW}}).
		\end{aligned}
	\end{equation}
	The coefficients $C_{7,\mathrm{NP}}^{(1,2)}(\mu_{\mathrm{EW}})$ are Wilson coefficients of the process $b\to s\gamma$ and can be calculated from the diagrams in Fig.~\ref{fig:two-figures}(1) and Fig.~\ref{fig:two-figures}(2), respectively. The results read
	\begin{align}
		C_{7,\mathrm{NP}}^{(1)}(\mu_{\mathrm{EW}})
		&=
		\sum_{H_i^-,u_j}
		\frac{s_W^2}{2e^2 V_{ts}^*V_{tb}}
		\Bigg\{
		\frac{1}{2}
		C_{H_i^-\bar{s}u_j}^{R}
		C_{H_i^-\bar{b}u_j}^{L}
		\left[
		-I_3(x_{u_j},x_{H_i^-})
		+I_4(x_{u_j},x_{H_i^-})
		\right]
		\notag
		\\
		&\qquad
		+
		\frac{m_{u_j}}{m_b}
		C_{H_i^-\bar{s}u_j}^{L}
		C_{H_i^-\bar{b}u_j}^{L}
		\left[
		-I_1(x_{u_j},x_{H_i^-})
		+I_3(x_{u_j},x_{H_i^-})
		\right]
		\Bigg\},
		\notag
		\\[1ex]
		C_{7,\mathrm{NP}}^{(2)}(\mu_{\mathrm{EW}})
		&=
		\sum_{H_j^-,u_i}
		\frac{s_W^2}{3e^2 V_{ts}^*V_{tb}}
		\Bigg\{
		\frac{1}{2}
		C_{H_j^-\bar{s}u_i}^{R}
		C_{H_j^-\bar{b}u_i}^{L}
		\left[
		-I_1(x_{u_i},x_{H_j^-})
		+2I_3(x_{u_i},x_{H_j^-})
		-I_4(x_{u_i},x_{H_j^-})
		\right]
		\notag
		\\
		&\qquad
		+
		\frac{m_{u_i}}{m_b}
		C_{H_j^-\bar{s}u_i}^{L}
		C_{H_j^-\bar{b}u_i}^{L}
		\left[
		I_1(x_{u_i},x_{H_j^-})
		-I_2(x_{u_i},x_{H_j^-})
		-I_3(x_{u_i},x_{H_j^-})
		\right]
		\Bigg\},
		\notag
		\\[1ex]
		C_{7,\mathrm{NP}}^{\prime(a)}(\mu_{\mathrm{EW}})
		&=
		C_{7,\mathrm{NP}}^{(a)}(\mu_{\mathrm{EW}})
		(L\leftrightarrow R),
		\qquad
		(a=1,2).
		\label{eq:C7NP-charged-Higgs}
	\end{align}
	where $x_i=m_i^2/m_W^2$, and $C_{abc}^{L,R}$ denotes the scalar part of the interaction vertex involving $a$, $b$, and $c$, with $a$, $b$, and $c$ denoting the interacting particles. The loop integral functions $I_{1,\ldots,4}$ can be found in Ref.~\cite{Yang_2018}. In addition, $C_{8g,\mathrm{NP}}(\mu_{\mathrm{EW}})$ and $C_{8g,\mathrm{NP}}^{\prime}(\mu_{\mathrm{EW}})$ at the electroweak scale are
	\begin{equation}
		\begin{aligned}[b]
			C_{8g,\mathrm{NP}}(\mu_{\mathrm{EW}})
			&=
			\frac{
				C_{7,\mathrm{NP}}^{(2)}(\mu_{\mathrm{EW}})
				+
				C_{7,\mathrm{NP}}^{(3)}(\mu_{\mathrm{EW}})
			}{Q_u},
			\\
			C_{8g,\mathrm{NP}}^{\prime}(\mu_{\mathrm{EW}})
			&=
			C_{8g,\mathrm{NP}}(\mu_{\mathrm{EW}})
			\left(L\leftrightarrow R\right).
		\end{aligned}
	\end{equation}
	where $Q_u=2/3$.
	
	\paragraph{Rare decay $B_s^0 \to \mu^+\mu^-$.}
	
\begin{figure*}[htbp]
		\centering
		
		\begin{subfigure}{0.23\textwidth}
			\centering
			\includegraphics[width=\linewidth]{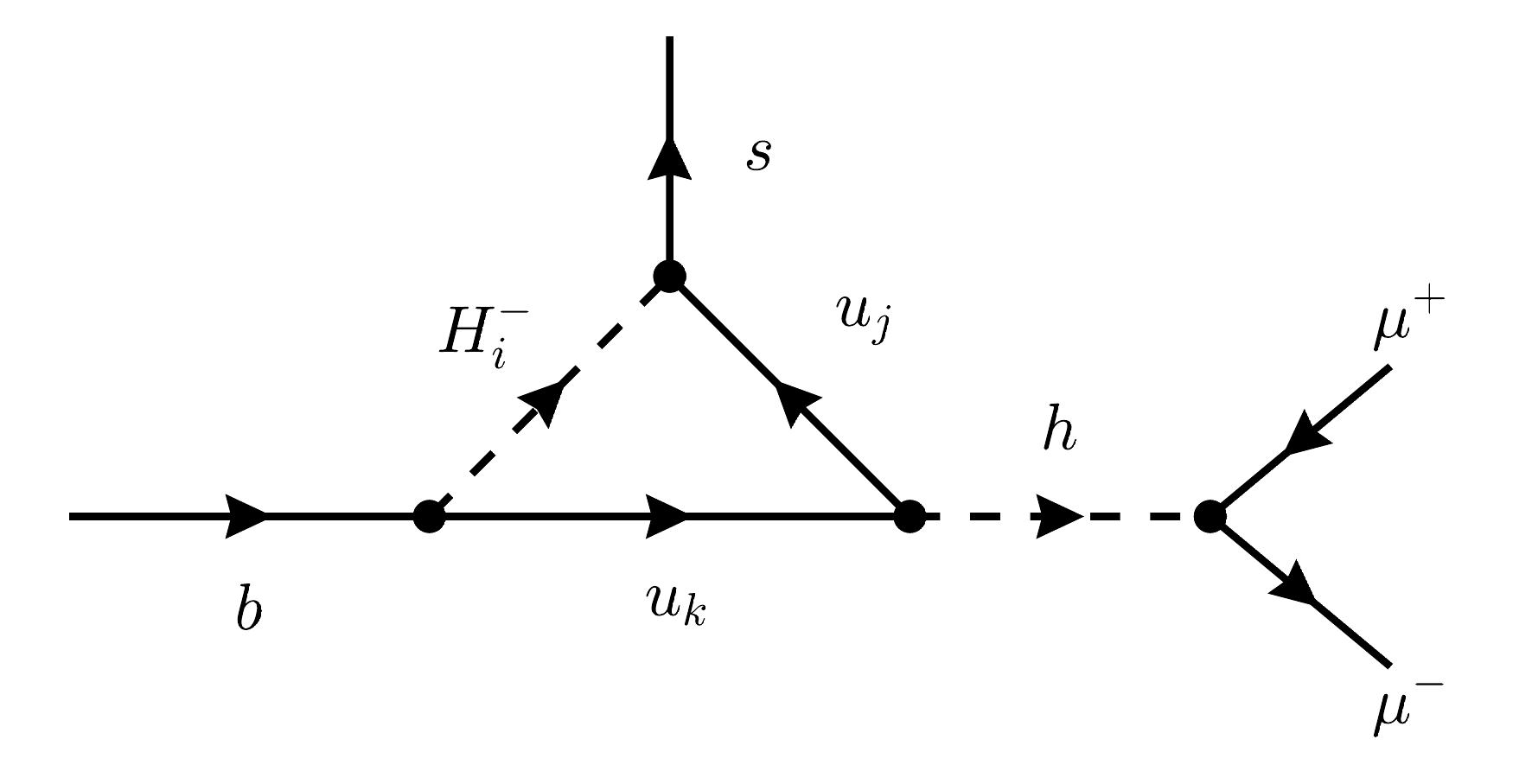}
			\caption*{(1)}
		\end{subfigure}
		\hspace{0.01\textwidth}
		\begin{subfigure}{0.23\textwidth}
			\centering
			\includegraphics[width=\linewidth]{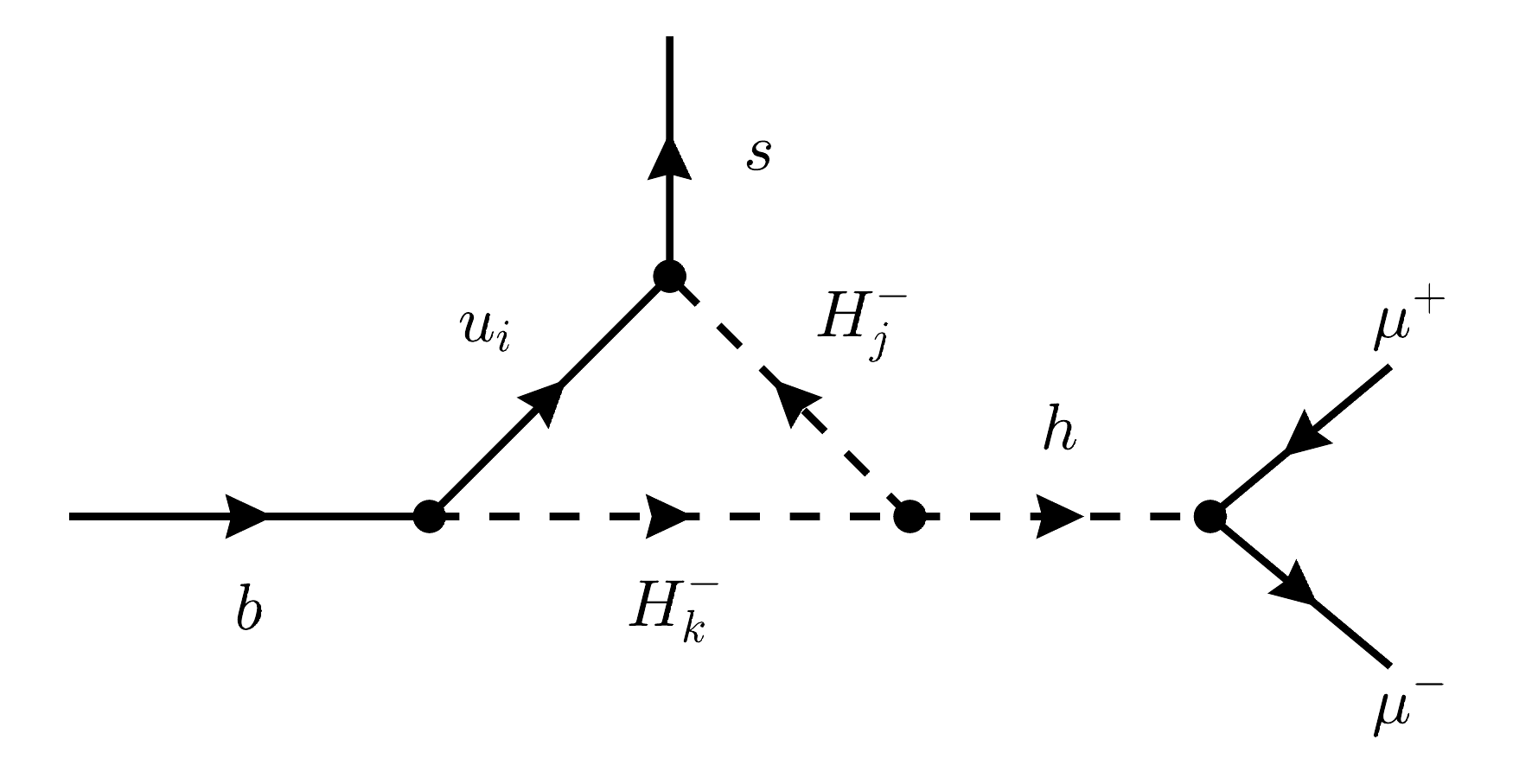}
			\caption*{(2)}
		\end{subfigure}
		\hspace{0.01\textwidth}
		\begin{subfigure}{0.23\textwidth}
			\centering
			\includegraphics[width=\linewidth]{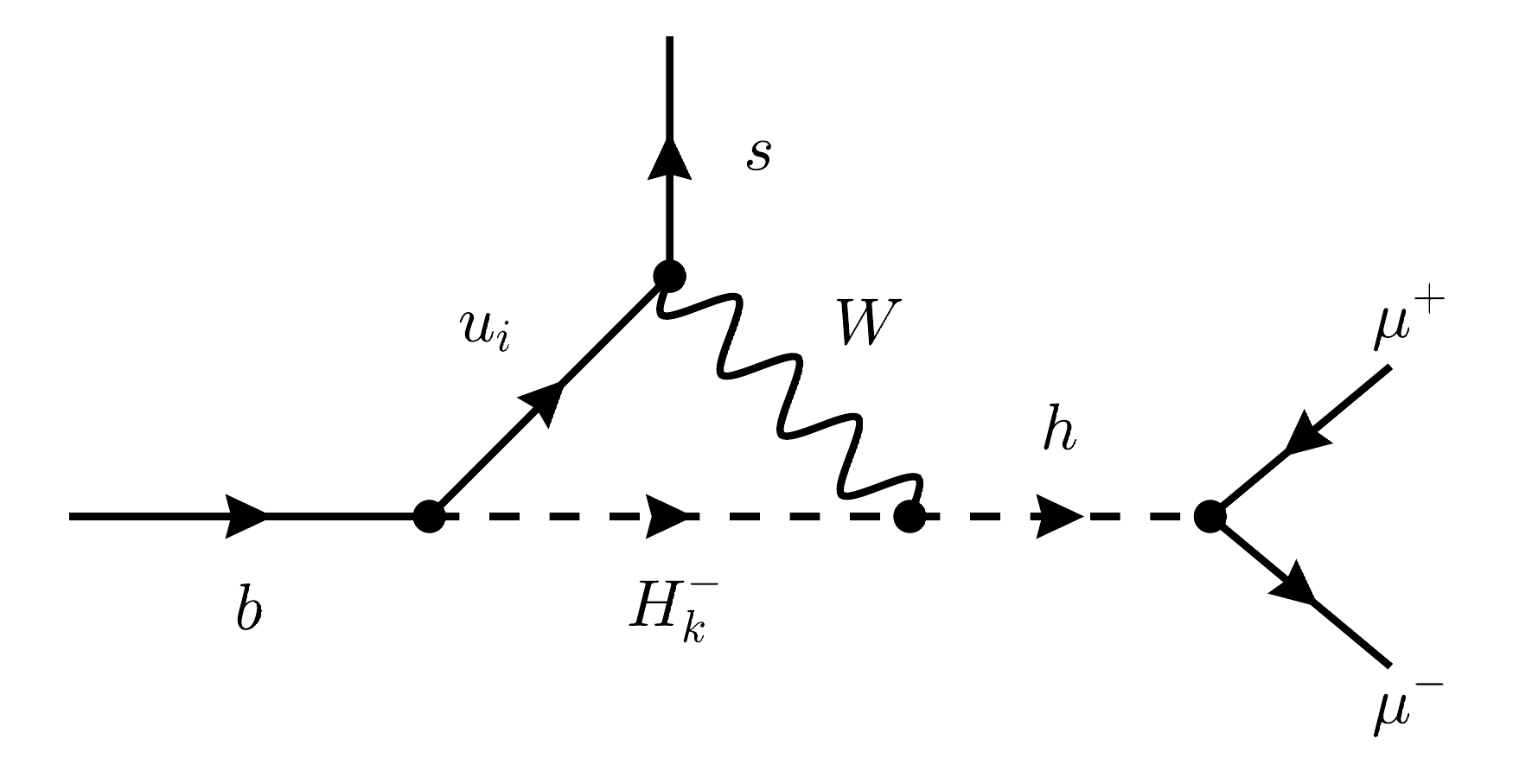}
			\caption*{(3)}
		\end{subfigure}
		\hspace{0.01\textwidth}
		\begin{subfigure}{0.23\textwidth}
			\centering
			\includegraphics[width=\linewidth]{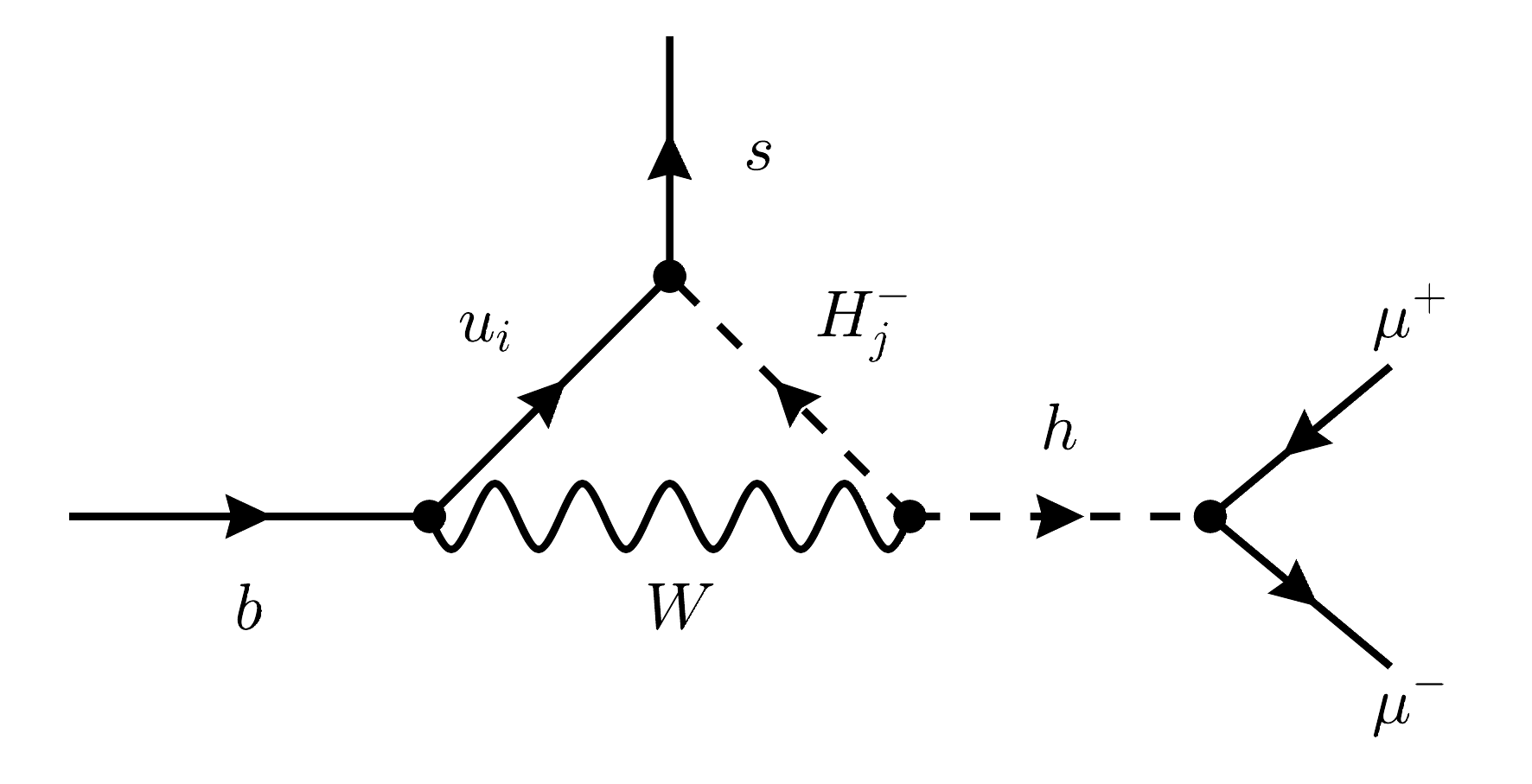}
			\caption*{(4)}
		\end{subfigure}
		
		\vspace{0.3cm}
		
		\begin{subfigure}{0.23\textwidth}
			\centering
			\includegraphics[width=\linewidth]{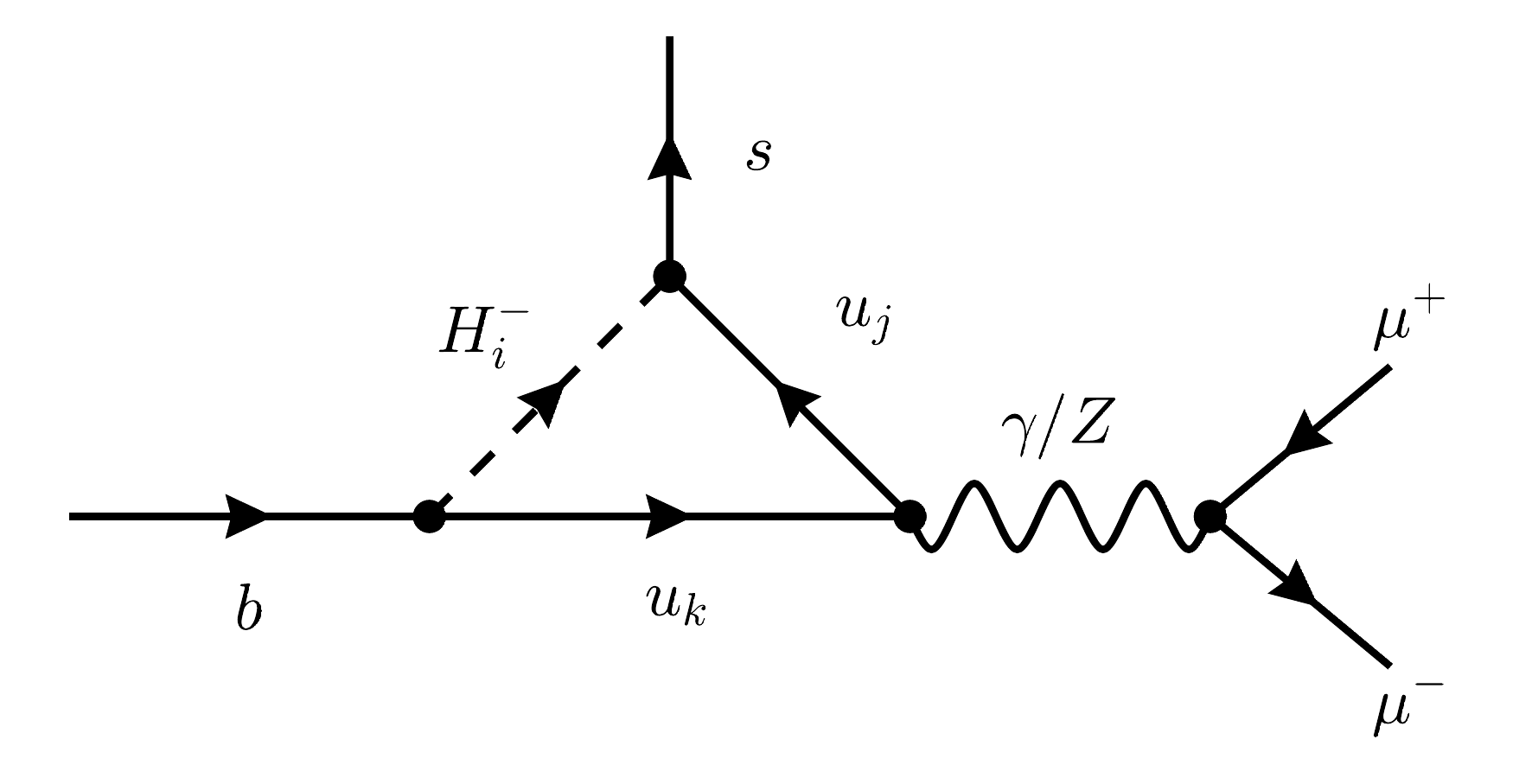}
			\caption*{(5)}
		\end{subfigure}
		\hspace{0.01\textwidth}
		\begin{subfigure}{0.23\textwidth}
			\centering
			\includegraphics[width=\linewidth]{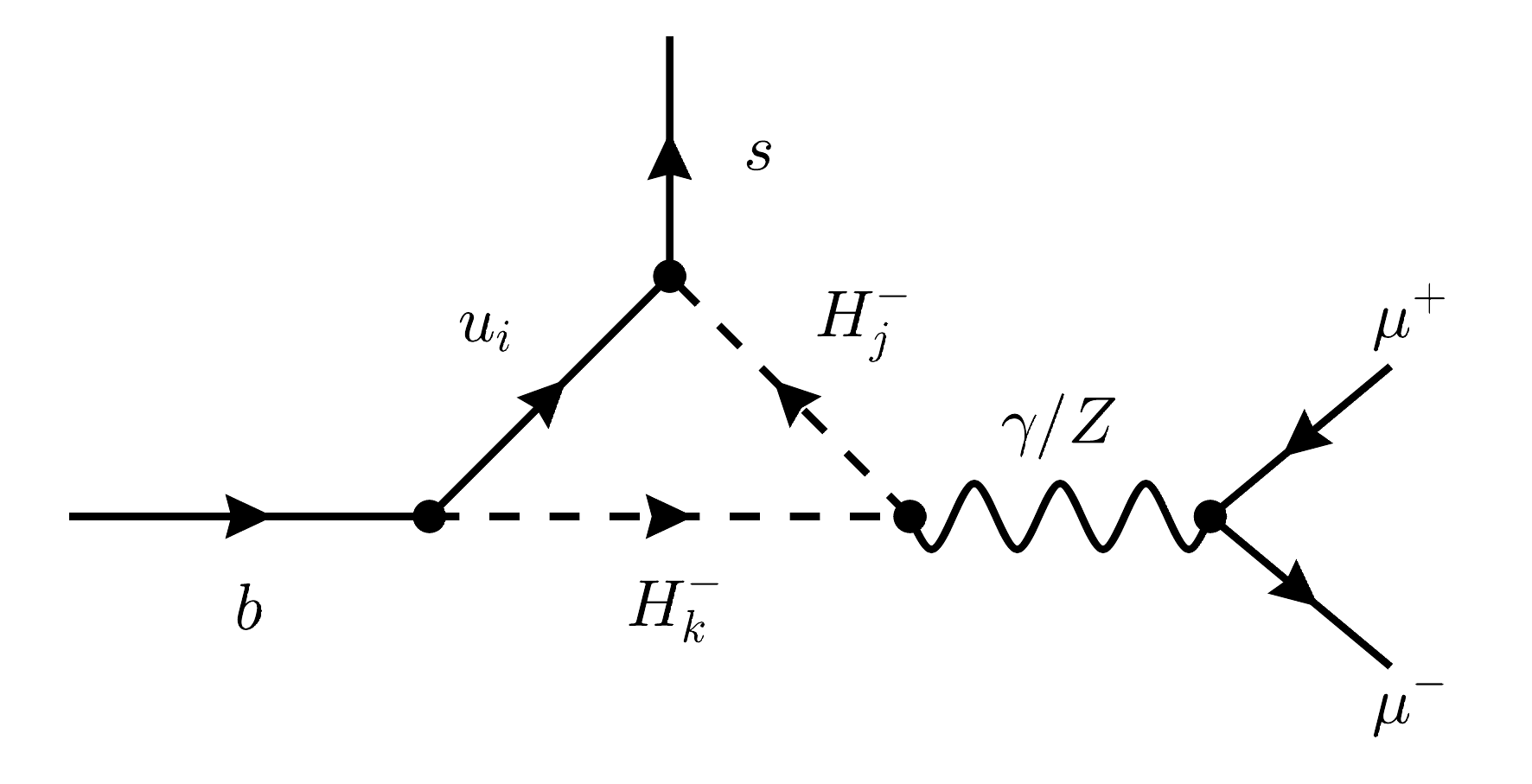}
			\caption*{(6)}
		\end{subfigure}
		\hspace{0.01\textwidth}
		\begin{subfigure}{0.23\textwidth}
			\centering
			\includegraphics[width=\linewidth]{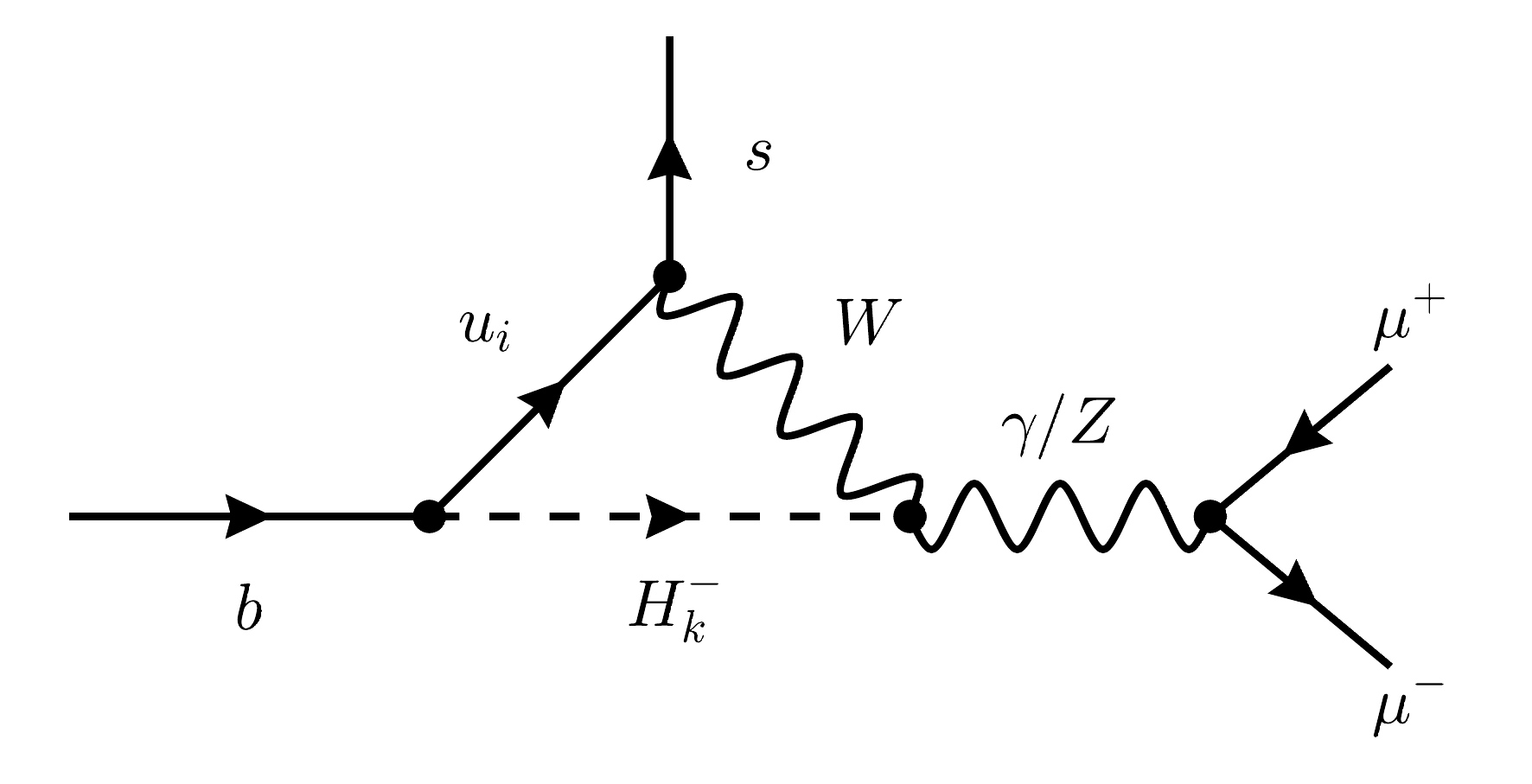}
			\caption*{(7)}
		\end{subfigure}
		\hspace{0.01\textwidth}
		\begin{subfigure}{0.23\textwidth}
			\centering
			\includegraphics[width=\linewidth]{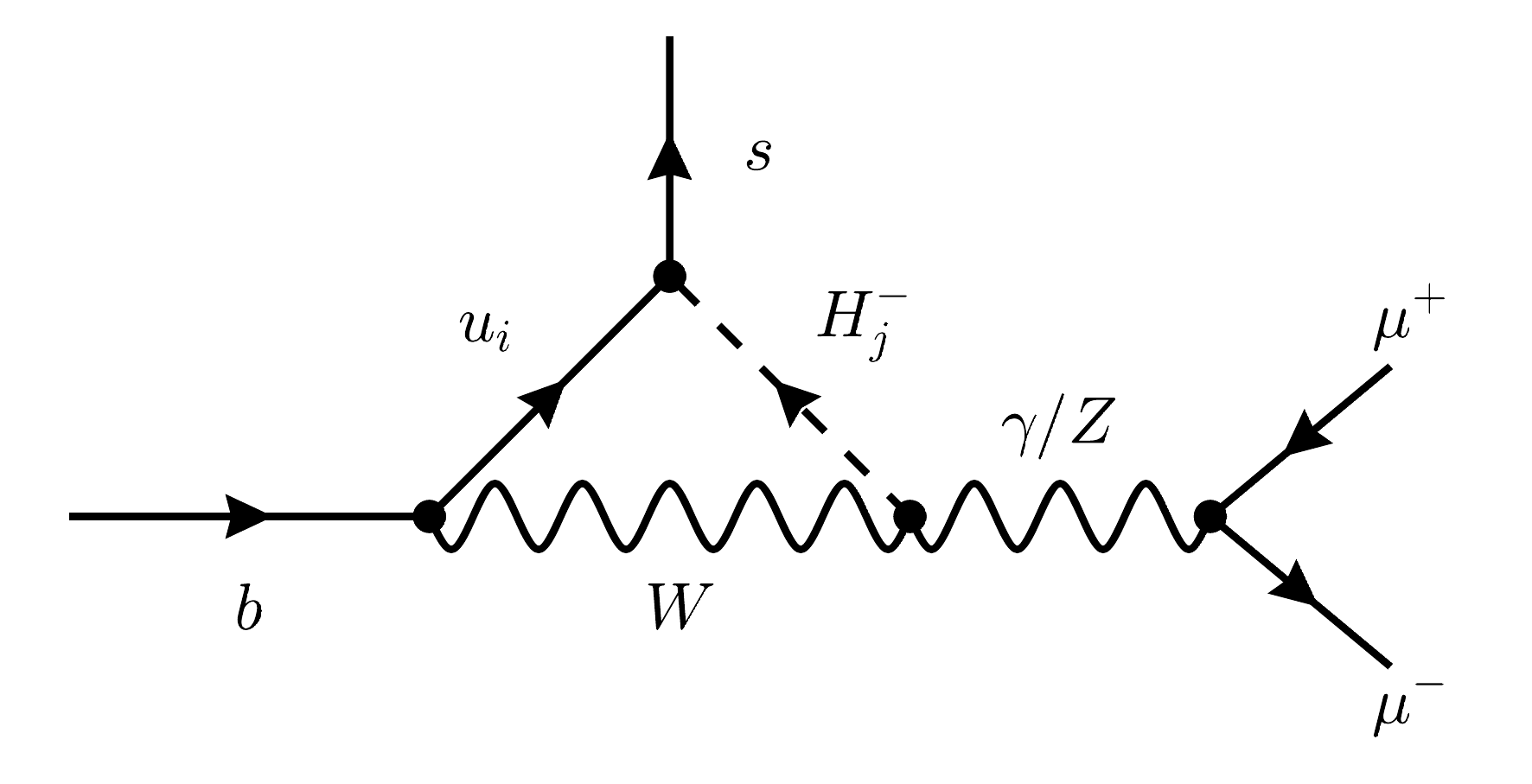}
			\caption*{(8)}
		\end{subfigure}
		
		\caption{\raggedright The main Feynman diagrams contributing to $B_s^0\to \mu^+\mu^-$ in the CP-violating G3HDM.}
		\label{fig:Bs-mumu-diagrams}
	\end{figure*}

	The main Feynman diagrams contributing to $B_s^0\to \mu^+\mu^-$ are shown in Fig.~\ref{fig:Bs-mumu-diagrams}. At the electroweak energy scale $\mu_{\mathrm{EW}}$, the corresponding Wilson coefficients can be written as
	\begin{align}
		C_{S,\mathrm{NP}}(\mu_{\mathrm{EW}})
		&=
		\frac{\sqrt{2}s_W c_W}{4m_b e^3 V_{ts}^{*}V_{tb}}
		\big[
		C_{S,\mathrm{NP}}^{(1)}(\mu_{\mathrm{EW}})
		+
		C_{S,\mathrm{NP}}^{(2)}(\mu_{\mathrm{EW}})
		+
		C_{S,\mathrm{NP}}^{(3)}(\mu_{\mathrm{EW}})
		+
		C_{S,\mathrm{NP}}^{(4)}(\mu_{\mathrm{EW}})
		+
		C_{S,\mathrm{NP}}^{(6)}(\mu_{\mathrm{EW}})
		\big],
		\notag
		\\
		C_{S,\mathrm{NP}}^{\prime}(\mu_{\mathrm{EW}})
		&=
		C_{S,\mathrm{NP}}(\mu_{\mathrm{EW}})
		(L\leftrightarrow R),
		\notag
		\\[1ex]
		C_{P,\mathrm{NP}}(\mu_{\mathrm{EW}})
		&=
		\frac{\sqrt{2}s_W c_W}{4m_b e^3 V_{ts}^{*}V_{tb}}
		\big[
		C_{P,\mathrm{NP}}^{(1)}(\mu_{\mathrm{EW}})
		+
		C_{P,\mathrm{NP}}^{(2)}(\mu_{\mathrm{EW}})
		+
		C_{P,\mathrm{NP}}^{(3)}(\mu_{\mathrm{EW}})
		+
		C_{P,\mathrm{NP}}^{(4)}(\mu_{\mathrm{EW}})
		+
		C_{P,\mathrm{NP}}^{(6)}(\mu_{\mathrm{EW}})
		\big],
		\notag
		\\
		C_{P,\mathrm{NP}}^{\prime}(\mu_{\mathrm{EW}})
		&=
		-
		C_{P,\mathrm{NP}}(\mu_{\mathrm{EW}})
		(L\leftrightarrow R),
		\notag
		\\[1ex]
		C_{9,\mathrm{NP}}(\mu_{\mathrm{EW}})
		&=
		\frac{\sqrt{2}s_W c_W g_s^2}{64\pi^2 e^3 V_{ts}^{*}V_{tb}}
		\big[
		C_{9,\mathrm{NP}}^{(5)}(\mu_{\mathrm{EW}})
		+
		C_{9,\mathrm{NP}}^{(6)}(\mu_{\mathrm{EW}})
		+
		C_{9,\mathrm{NP}}^{(7)}(\mu_{\mathrm{EW}})
		+
		C_{9,\mathrm{NP}}^{(8)}(\mu_{\mathrm{EW}})
		\big],
		\notag
		\\
		C_{9,\mathrm{NP}}^{\prime}(\mu_{\mathrm{EW}})
		&=
		C_{9,\mathrm{NP}}(\mu_{\mathrm{EW}})
		(L\leftrightarrow R),
		\notag
		\\[1ex]
		C_{10,\mathrm{NP}}(\mu_{\mathrm{EW}})
		&=
		\frac{\sqrt{2}s_W c_W g_s^2}{64\pi^2 e^3 V_{ts}^{*}V_{tb}}
		\big[
		C_{10,\mathrm{NP}}^{(5)}(\mu_{\mathrm{EW}})
		+
		C_{10,\mathrm{NP}}^{(6)}(\mu_{\mathrm{EW}})
		+
		C_{10,\mathrm{NP}}^{(7)}(\mu_{\mathrm{EW}})
		+
		C_{10,\mathrm{NP}}^{(8)}(\mu_{\mathrm{EW}})
		\big],
		\notag
		\\
		C_{10,\mathrm{NP}}^{\prime}(\mu_{\mathrm{EW}})
		&=
		-
		C_{10,\mathrm{NP}}(\mu_{\mathrm{EW}})
		(L\leftrightarrow R).
		\label{eq:Bs-mumu-Wilson-EW}
	\end{align}
	
	\begin{table}[htbp]
		\centering
		\caption{\raggedright At hadronic scale $\mu = m_b$, SM Wilson coefficients to next-to-next-to-logarithmic accuracy}
		\label{tab:SMWilson}
		\begin{tabular*}{\textwidth}{@{\extracolsep{\fill}}cccc@{}}
			\hline\hline
			$C_7^{eff,SM}$ & $C_8^{eff,SM}$ & $C_9^{eff,SM}$ & $C_{10}^{eff,SM}$ \\
			\hline
			$-0.304$ & $-0.167$ & $4.211$ & $-4.103$ \\
			\hline\hline
		\end{tabular*}
	\end{table}
	The superscripts $(1,\ldots,8)$ correspond to the contributions from Fig.~\ref{fig:Bs-mumu-diagrams}(1,\ldots,8), respectively and the specific expressions for these Wilson coefficients are detailed in Ref.~\cite{Yang_2018}. The Wilson coefficients at hadronic energy scale from the SM to next-to-next-to-logarithmic accuracy are shown in Table~\ref{tab:SMWilson}. In addition, the Wilson coefficients in Eq.~(\ref{eq:Bs-mumu-Wilson-EW}) should be evolved down to hadronic scale $\mu \sim m_b$ by the renormalization group equations (Table~\ref{tab:SMWilson})
	\begin{align}\label{eq:C1}
		\vec{C}_{NP}(\mu) &= \hat{U}(\mu,\mu_0)\,\vec{C}_{NP}(\mu_0), \nonumber\\
		\vec{C}_{NP}^{\prime}(\mu) &= U^{\prime}(\mu,\mu_0)\,\vec{C}_{NP}^{\prime}(\mu_0).
	\end{align}
	with
	\begin{align}\label{eq:C2}
		\vec{C}_{NP}^{\,T}
		&=
		\left(
		C_{1,NP},\ \cdots,\ C_{6,NP},\ C_{7,NP}^{eff},\ C_{8,NP}^{NP},
		\right.
		\nonumber\\
		&\quad \left.
		C_{9,NP}^{eff}-Y(q^2),\ C_{10,NP}^{eff}
		\right),
		\nonumber\\
		\vec{C}_{\mathrm{NP}}^{\prime\,T}
		&=
		\left(
		C_{7,NP}^{\prime,eff},\ C_{8,NP}^{\prime,eff},\ C_{9,NP}^{\prime,eff},\ C_{10,NP}^{\prime,eff}
		\right).
	\end{align}
	Correspondingly, the evolving matrices are approached as
	\begin{align}
		\hat{U}(\mu,\mu_0) &\simeq 1-\left[\frac{1}{2\beta_0}\ln\frac{\alpha_s(\mu)}{\alpha_s(\mu_0)}\right]\hat{\gamma}^{(0)T}, \nonumber\\
		\hat{U}^{\prime}(\mu,\mu_0) &\simeq 1-\left[\frac{1}{2\beta_0}\ln\frac{\alpha_s(\mu)}{\alpha_s(\mu_0)}\right]\hat{\gamma}^{\prime(0)T}.
	\end{align}
	where the anomalous dimension matrices can be read from Ref.~\cite{Gambino_2003} as
	\begin{equation}
		\hat{\gamma}^{\mathrm{eff}(0)}=
		\left(
		\begin{array}{cccccccccc}
			-4 & \dfrac{8}{3} & 0 & -\dfrac{2}{9} & 0 & 0 & -\dfrac{208}{243} & \dfrac{173}{162} & -\dfrac{2272}{729} & 0 \\[1ex]
			12 & 0 & 0 & \dfrac{4}{3} & 0 & 0 & \dfrac{416}{81} & \dfrac{70}{27} & \dfrac{1952}{243} & 0 \\[1ex]
			0 & 0 & 0 & -\dfrac{52}{3} & 0 & 2 & -\dfrac{176}{81} & \dfrac{14}{27} & -\dfrac{6752}{243} & 0 \\[1ex]
			0 & 0 & -\dfrac{40}{9} & -\dfrac{100}{9} & \dfrac{4}{9} & \dfrac{5}{6} & -\dfrac{152}{243} & -\dfrac{587}{162} & -\dfrac{2192}{729} & 0 \\[1ex]
			0 & 0 & 0 & -\dfrac{256}{3} & 0 & 20 & -\dfrac{6272}{81} & \dfrac{6596}{27} & -\dfrac{84032}{243} & 0 \\[1ex]
			0 & 0 & -\dfrac{256}{9} & \dfrac{56}{9} & \dfrac{40}{9} & -\dfrac{2}{3} & \dfrac{4624}{243} & \dfrac{4772}{81} & -\dfrac{37856}{729} & 0 \\[1ex]
			0 & 0 & 0 & 0 & 0 & 0 & \dfrac{32}{3} & 0 & 0 & 0 \\[1ex]
			0 & 0 & 0 & 0 & 0 & 0 & -\dfrac{32}{9} & \dfrac{28}{3} & 0 & 0 \\[1ex]
			0 & 0 & 0 & 0 & 0 & 0 & 0 & 0 & 0 & 0 \\[1ex]
			0 & 0 & 0 & 0 & 0 & 0 & 0 & 0 & 0 & 0
		\end{array}
		\right).
	\end{equation}
	
	\begin{equation}
		\hat{\gamma}^{\prime\,\mathrm{eff}(0)}=
		\left(
		\begin{array}{cccc}
			\dfrac{32}{3} & 0 & 0 & 0 \\[1ex]
			-\dfrac{32}{9} & \dfrac{28}{3} & 0 & 0 \\[1ex]
			0 & 0 & 0 & 0 \\[1ex]
			0 & 0 & 0 & 0
		\end{array}
		\right).
	\end{equation}
	
	Then, the squared amplitude can be written as
	\begin{equation}
		|\mathcal{M}_s|^2
		=
		16G_F^2\,|V_{tb}V_{ts}^{*}|^2\,M_{B_s^0}^{2}
		\left[
		|F_S^s|^2 + |F_P^s + 2m_{\mu}F_A^s|^2
		\right],
	\end{equation}
	and
	\begin{align*}
		F_S^s &=
		\frac{\alpha_{EW}(\mu_b)}{8\pi}\,
		\frac{m_b M_{B_s^0}^2}{m_b+m_s}\,
		f_{B_s^0}\,(C_S-C_S') .
	\end{align*}
	\begin{align}
		F_P^s &=
		\frac{\alpha_{EW}(\mu_b)}{8\pi}\,
		\frac{m_b M_{B_s^0}^2}{m_b+m_s}\,
		f_{B_s^0}\,(C_P-C_P'),
		\nonumber\\
		F_A^s &=
		\frac{\alpha_{EW}(\mu_b)}{8\pi}\,
		f_{B_s^0}\,
		\left[
		C_{10}^{\mathrm{eff}}(\mu_b)-C_{10}^{\prime\,\mathrm{eff}}(\mu_b)
		\right].
	\end{align}
	where $f_{B_s^0}=(227\pm 8),\mathrm{MeV}$ denotes the decay constant, $M_{B_s^0}=5.367,\mathrm{GeV}$ denotes the mass of the neutral meson $B_s^0$. The branching ratio of $B_s^0 \to \mu^+\mu^-$ can be written as
	\begin{equation}
		\mathrm{Br}(B_s^0 \to \mu^+\mu^-)
		=
		\frac{\tau_{B_s^0}}{16\pi}\,
		\frac{|\mathcal{M}_s|^2}{M_{B_s^0}}\,
		\sqrt{1-\frac{4m_\mu^2}{M_{B_s^0}^2}} \, .
	\end{equation}
	with $\tau_{B_s^0}=1.466(31) \mathrm{ps}$ denoting the lifetime of the meson.
	
	\subsection{Neutral-meson mixing constraints}
	
	Neutral meson--antimeson mixing is highly sensitive to flavor-changing interactions beyond the SM and therefore provides stringent constraints on new physics. In the CP-violating G3HDM, the extended scalar and Yukawa sectors generate flavor-changing neutral-Higgs interactions in the fermion mass basis. These interactions contribute to $\Delta F=2$ transitions and can modify both the mass splittings and the CP-violating phases of neutral-meson mixing.
	
	Following the UTfit convention, the new-physics effects in the $B_q^0$--$\bar B_q^0$ systems are parametrized as
	\begin{equation}
		C_{B_q}e^{2i\phi_{B_q}}
		=
		\frac{M_{12}^{q,\mathrm{full}}}
		{M_{12}^{q,\mathrm{SM}}},
		\qquad q=d,s.
		\label{eq:B-mixing-definition}
	\end{equation}
	where $M_{12}^{q,\mathrm{full}} =M_{12}^{q,\mathrm{SM}}+M_{12}^{q,\mathrm{NP}}$. Thus, $C_{B_q}$ measures the modification of the mixing-amplitude magnitude, while $\phi_{B_q}$ denotes the new-physics phase relative to the SM. For the neutral-kaon system, we define
	\begin{equation}
		C_{\Delta M_K}
		=
		\frac{\operatorname{Re}M_{12}^{K,\mathrm{full}}}
		{\operatorname{Re}M_{12}^{K,\mathrm{SM}}},
		\qquad
		C_{\epsilon_K}
		=
		\frac{\operatorname{Im}M_{12}^{K,\mathrm{full}}}
		{\operatorname{Im}M_{12}^{K,\mathrm{SM}}}.
		\label{eq:K-mixing-definition}
	\end{equation}
	All these quantities reduce to their SM values, $C_{B_q}=C_{\Delta M_K}=C_{\epsilon_K}=1$ and $\phi_{B_q}=0$, in the absence of new-physics contributions.
	
	In the numerical analysis, we impose the following one-dimensional $2\sigma$ intervals based on the UTfit Summer 2025 new-physics fit
	\begin{equation}
		\begin{array}{ll}
			0.90 \leq C_{B_d}\leq 1.22,
			&
			-6.6^\circ \leq \phi_{B_d}\leq 3.0^\circ,
			\\[2pt]
			0.98 \leq C_{B_s}\leq 1.22,
			&
			-1.3^\circ \leq \phi_{B_s}\leq 0.7^\circ,
			\\[2pt]
			0 < C_{\Delta M_K}<2.0,
			&
			0.85 \leq C_{\epsilon_K}\leq1.25 .
		\end{array}
		\label{eq:meson-mixing-constraints}
	\end{equation}
	The $B_d$, $B_s$, and $\epsilon_K$ intervals are obtained from the corresponding one-dimensional UTfit results and are applied independently. For $C_{\Delta M_K}$, we adopt the wider interval above to account conservatively for the sizable uncertainty associated with long-distance contributions to the kaon mass splitting.

	\subsection{Charged-lepton flavor constraints}
	
	Charged-lepton flavor violation provides important constraints on the leptonic Yukawa structure of the model. In the CP-violating G3HDM, after rotating the charged-lepton fields to the mass basis, the neutral scalar interactions can contain flavor changing couplings. These couplings can induce radiative and three-body charged-lepton flavor-violating decays, as well as lepton-flavor-violating Higgs decays. We therefore include the corresponding observables in the numerical scan. The relevant branching ratios are computed with SPheno~\cite{Porod_2003_SPheno} for each parameter point.
	
	In our analysis, a parameter point is accepted only if the predicted branching ratios satisfy
	\begin{equation}
		\begin{aligned}[b]
			{\rm Br}(\mu \to e\gamma) &< 1.5\times 10^{-13}, \\
			{\rm Br}(\tau \to e\gamma) &< 3.3\times 10^{-8}, \qquad
			{\rm Br}(\tau \to \mu\gamma) < 4.2\times 10^{-8}.
		\end{aligned}
	\end{equation}
	for the radiative charged-lepton decays,
	\begin{equation}
		\begin{aligned}[b]
			{\rm Br}(h_5 \to e\mu) &< 4.4\times 10^{-5}, \\
			{\rm Br}(h_5 \to e\tau) &< 2.0\times 10^{-3}, \qquad
			{\rm Br}(h_5 \to \mu\tau) < 1.5\times 10^{-3}.
		\end{aligned}
	\end{equation}
	for the lepton-flavor violating Higgs decays, and
	\begin{equation}
		\begin{aligned}[b]
			{\rm Br}(\mu \to 3e) &< 1.0\times 10^{-12}, \\
			{\rm Br}(\tau \to 3e) &< 2.7\times 10^{-8}, \qquad
			{\rm Br}(\tau \to 3\mu) < 2.1\times 10^{-8}.
		\end{aligned}
	\end{equation}
	for the three-body charged-lepton decays.

	\subsection{Higgs-mediated contributions to electric dipole moments}
	
	Before discussing the Higgs-mediated EDM contributions, we first summarize the current experimental constraints on the EDM observables considered in this work. So far, no nonzero EDM has been observed for the neutron, electron, mercury atom, or heavy quarks. Therefore, the corresponding experimental and phenomenological upper limits provide important constraints~\cite{Graner_2016,Navas_2024,Baron_2014,Andreev_2018,Blinov_2009} on new CPV interactions
	\begin{equation}
		\label{eq:edm-bounds}
		\begin{aligned}[b]
			&|d_n| < 1.8 \times 10^{-26}\, e\cdot {\rm cm},\\
			&|d_e| < 4.1 \times 10^{-30}\, e\cdot {\rm cm},\\
			&|d_{\rm Hg}| < 7.4 \times 10^{-30}\, e\cdot {\rm cm},\\
			&|d_b| < 2.0 \times 10^{-17}\, e\cdot {\rm cm},\\
			&|d_c| < 4.4 \times 10^{-17}\, e\cdot {\rm cm}.
		\end{aligned}
	\end{equation}
	
	Subsequently, a new upper limit $|d_c^g| < 1.0 \times 10^{-22}~{\rm cm}$ can be obtained by considering the limits on $d_n$. Through the calculation of the rigorous constraints on $d_b^g$ in Refs.~\cite{Gisbert_2020,Chang_1990}, new limits for the EDMs of $b$ and $c$ quarks can be derived
	\begin{equation}
		\label{eq:heavy-quark-edm-bounds}
		\begin{aligned}[b]
			|d_b| &< 1.2 \times 10^{-20}\, e\cdot {\rm cm},\\
			|d_c| &< 1.5 \times 10^{-21}\, e\cdot {\rm cm}.
		\end{aligned}
	\end{equation}
	
	\paragraph{The EDMs of the $b$ and $c$ quarks.}
	The quark EDMs at the low scale $\Lambda_\chi$ can be obtained from $d_q^\gamma(\Lambda_\chi)$ and $d_q^g(\Lambda_\chi)$ by
	\begin{align}
		&d_q
		=
		d_q^\gamma(\Lambda_\chi)
		+
		\frac{e}{4\pi}
		d_q^g(\Lambda_\chi).
		\label{eq:dq_low_scale}
	\end{align}
	where $\Lambda_\chi = m_q$ denotes the chirality-breaking scale, and $m_q$ denotes the corresponding quark mass.
	
	The effective Lagrangian for the EDM $d_q$ of the fermion is defined through the dimension-five operator~\cite{Feng_2005}
	\begin{equation}
		\mathcal{L}_{\rm EDM}
		=
		-\frac{i}{2} d_q^\gamma \,
		\bar{q}\sigma^{\mu\nu}\gamma_5 q \,
		F_{\mu\nu}.
		\label{eq:edm-lagrangian}
	\end{equation}
	with $\sigma^{\mu\nu}=i[\gamma^\mu,\gamma^\nu]/2$, $F_{\mu\nu}$ representing the electromagnetic field strength, and $q$ denoting a fermion field. Adopting the effective Lagrangian approach, the quark EDMs can be written as
	\begin{equation}
		d_q^\gamma
		=
		-\frac{2 e Q_q m_q}{(4\pi)^2}
		\left(
		C_2^R + C_2^{L*} + C_6^R
		\right).
		\label{eq:dq-gamma}
	\end{equation}
	where $C_{2}^{L,R}$ and $C_{6}^{R}$ represent the Wilson coefficients of the corresponding operators $O_{2}^{L,R}$ and $O_{6}^{R}$
	\begin{equation}
		\begin{aligned}
			O_{2}^{L,R} &=
			\frac{e Q_q}{(4\pi)^2}
			\left(-iD_\alpha^{*}\right)
			\bar q \gamma^\alpha F_{\mu\nu}\sigma^{\mu\nu} P_{L,R} q ,
			\\
			O_{6}^{L,R} &=
			\frac{e Q_q m_q}{(4\pi)^2}
			\bar q F_{\mu\nu}\sigma^{\mu\nu} P_{L,R} q .
		\end{aligned}
		\label{eq:operators_O2_O6}
	\end{equation}
	
	Besides the operator in Eq.~\eqref{eq:edm-lagrangian}, the chromo electric dipole moment (CEDM) of quarks can also contribute to $d_q$
	\begin{equation}
		\mathcal{L}_{\rm CEDM}
		=
		-\frac{i}{2} d_q^g \,
		\bar{q}\sigma^{\mu\nu}\gamma_5 T^a q \,
		G^a_{\mu\nu}.
		\label{eq:cedm-lagrangian}
	\end{equation}
	where $G^a_{\mu\nu}$ is the $SU(3)$ gauge field strength, and $T^a$ is the $SU(3)$ generator.  Then the quark CEDMs can be written as
	\begin{equation}
		d_q^g
		=
		-\frac{2 g_3 m_q}{(4\pi)^2}
		\left(
		C_7^R + C_7^{L*} + C_8^R
		\right).
		\label{eq:dq-gluon}
	\end{equation}
	where $C_{7}^{L,R}$ and $C_8^R$ represent the Wilson coefficients of the corresponding operators $O_{7}^{L,R}$ and $O_8^R$
	\begin{equation}
		\begin{aligned}[b]
			O_7^{L,R}
			&=
			\frac{g_3}{(4\pi)^2}
			\left(-iD_\alpha^*\right)
			\bar{q}\gamma^\alpha
			G^a\!\cdot\!\sigma \,
			T^a P_{L,R} q,
			\\
			O_8^{L,R}
			&=
			\frac{g_3 m_q}{(4\pi)^2}
			\bar{q}
			G^a\!\cdot\!\sigma \,
			T^a P_{L,R} q.
		\end{aligned}
		\label{eq:O7O8LR}
	\end{equation}
	
\begin{figure}[htbp]
		\centering
		
		\begin{minipage}{0.48\textwidth}
			\centering
			\includegraphics[width=\textwidth]{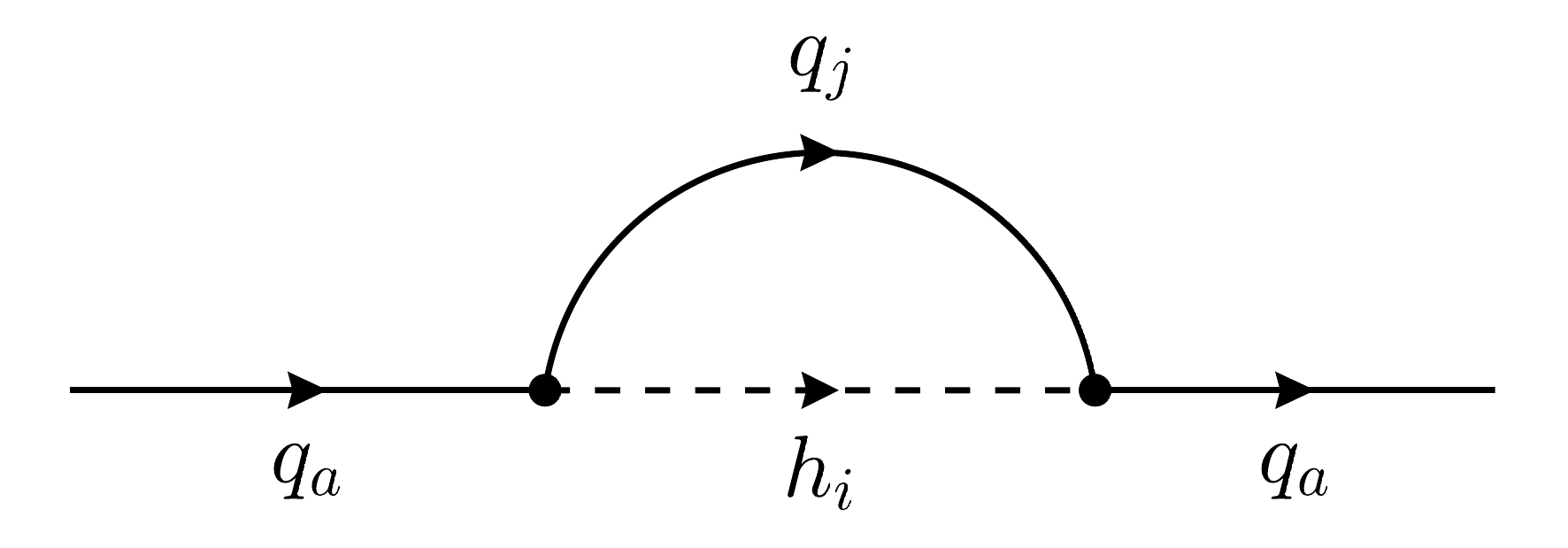}
			\caption*{(a)}
		\end{minipage}
		\hfill
		\begin{minipage}{0.48\textwidth}
			\centering
			\includegraphics[width=\textwidth]{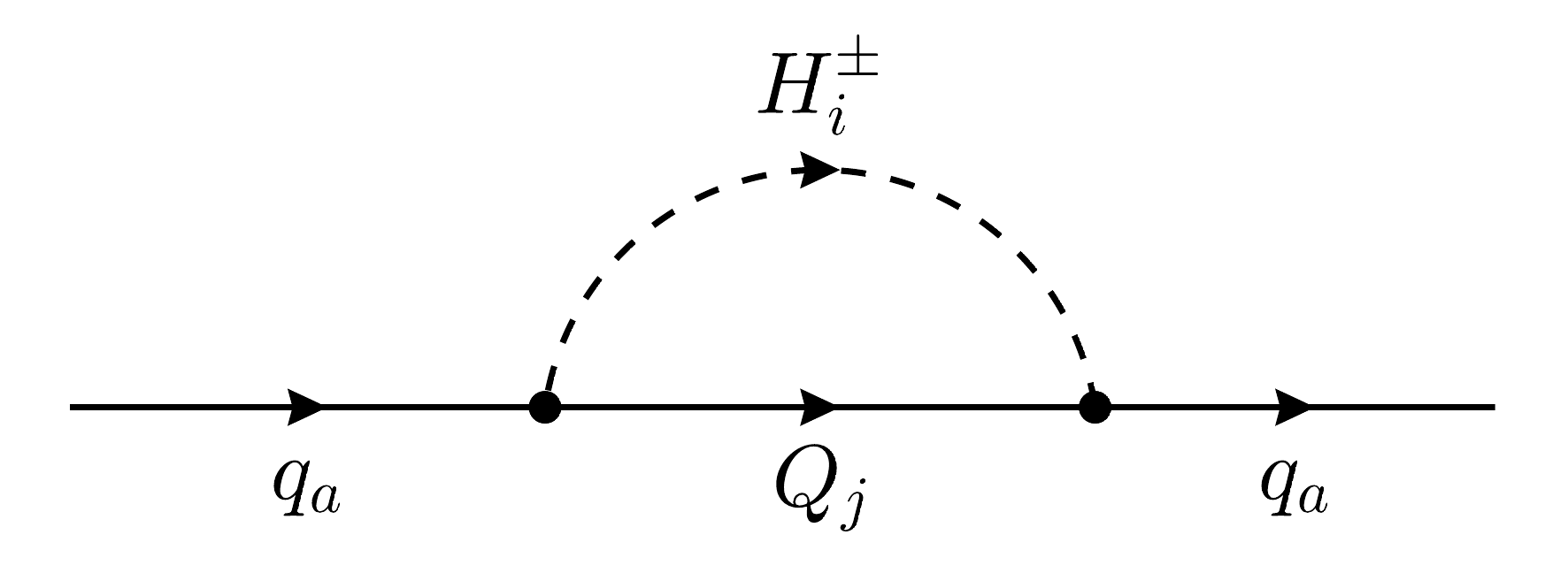}
			\caption*{(b)}
		\end{minipage}
		
		\caption{\raggedright In the Higgs-mediated contributions, the relevant one-loop diagrams can be obtained by attaching a photon and a gluon, respectively, to the internal particle lines in all possible ways, corresponding to $d_q^\gamma$ and $d_q^g$.}
		\label{fig:two_diagrams}
	\end{figure}

	The one-loop Feynman diagrams mediated by Higgs bosons and contributing to the above amplitudes are shown in Fig.~\ref{fig:two_diagrams}. 
	
	By evaluating these diagrams, the neutral Higgs contribution is given by
	\begin{align}
		\left(d_{q_a}^{\gamma}\right)_{h}
		&=
		\frac{e}{16\pi^2}
		\frac{m_{q_j}}{m_{h_i}^2}
		e_q\,
		G_2\!\left(\frac{m_{q_j}^2}{m_{h_i}^2}\right)
		\operatorname{\Im}
		\left[
		C_{h_i\bar q_a q_j}^{L}
		C_{h_i\bar q_a q_j}^{R*}
		\right],
		\nonumber\\
		\left(d_{q_a}^{g}\right)_{h}
		&=
		\frac{g_3}{32\pi^2}
		\frac{m_{q_j}}{m_{h_i}^2}
		G_2\!\left(\frac{m_{q_j}^2}{m_{h_i}^2}\right)
		\operatorname{\Im}
		\left[
		C_{h_i\bar q_a q_j}^{L}
		C_{h_i\bar q_a q_j}^{R*}
		\right].
		\label{eq:neutral-higgs-edm-cedm}
	\end{align}
	where $q_a=u_a,d_a$, with $u_a=(u,c,t)$ and $d_a=(d,s,b)$. Here $e_q$ denotes the electric charge of the quark $q$ in units of the elementary charge $e$, namely $e_{u_a}=2/3$ and $e_{d_a}=-1/3$.
	
	For an external down-type quark $d_a$, the charged Higgs contribution involves an internal up-type quark $u_j$ and can be written as
	\begin{align}
		\left(d_{d_a}^{\gamma}\right)_{H^\pm}
		&=
		\frac{e}{16\pi^2}
		\frac{m_{u_j}}{m_{H_i^\pm}^2}
		\left[
		\frac{2}{3}
		G_2\!\left(\frac{m_{u_j}^2}{m_{H_i^\pm}^2}\right)
		-
		G_1\!\left(\frac{m_{u_j}^2}{m_{H_i^\pm}^2}\right)
		\right]
		\operatorname{\Im}
		\left[
		C_{H_i^\pm\bar d_a u_j}^{L}
		C_{H_i^\pm\bar d_a u_j}^{R*}
		\right],
		\nonumber\\
		\left(d_{d_a}^{g}\right)_{H^\pm}
		&=
		\frac{g_3}{32\pi^2}
		\frac{m_{u_j}}{m_{H_i^\pm}^2}
		G_2\!\left(\frac{m_{u_j}^2}{m_{H_i^\pm}^2}\right)
		\operatorname{\Im}
		\left[
		C_{H_i^\pm\bar d_a u_j}^{L}
		C_{H_i^\pm\bar d_a u_j}^{R*}
		\right].
		\label{eq:charged-higgs-down-edm-cedm}
	\end{align}
	
	Similarly, for an external up-type quark $u_a$, the charged Higgs contribution contains an internal down-type quark $d_j$ and is given by
	\begin{align}
		\left(d_{u_a}^{\gamma}\right)_{H^\pm}
		&=
		\frac{e}{16\pi^2}
		\frac{m_{d_j}}{m_{H_i^\pm}^2}
		\left[
		-\frac{1}{3}
		G_2\!\left(\frac{m_{d_j}^2}{m_{H_i^\pm}^2}\right)
		+
		G_1\!\left(\frac{m_{d_j}^2}{m_{H_i^\pm}^2}\right)
		\right]
		\operatorname{\Im}
		\left[
		C_{H_i^\pm\bar u_a d_j}^{L}
		C_{H_i^\pm\bar u_a d_j}^{R*}
		\right],
		\nonumber\\
		\left(d_{u_a}^{g}\right)_{H^\pm}
		&=
		\frac{g_3}{32\pi^2}
		\frac{m_{d_j}}{m_{H_i^\pm}^2}
		G_2\!\left(\frac{m_{d_j}^2}{m_{H_i^\pm}^2}\right)
		\operatorname{\Im}
		\left[
		C_{H_i^\pm\bar u_a d_j}^{L}
		C_{H_i^\pm\bar u_a d_j}^{R*}
		\right].
		\label{eq:charged-higgs-up-edm-cedm}
	\end{align}
	
	Therefore, the total EDM and CEDM of a generic quark $q$ are obtained as
	\begin{equation}
		d_q^\gamma
		=
		\left(d_q^\gamma\right)_{h}
		+
		\left(d_q^\gamma\right)_{H^\pm},
		\qquad
		d_q^g
		=
		\left(d_q^g\right)_{h}
		+
		\left(d_q^g\right)_{H^\pm}.
		\label{eq:total-quark-edm-cedm}
	\end{equation}
	where $C_{abc}^{L,R}$ denotes the left and right handed constant parts of the interaction vertex involving the fields $a$, $b$, and $c$. These couplings are obtained using SARAH~\cite{F. Staub1,F. Staub2,F. Staub3,F. Staub4,F. Staub5}. The loop functions $G_1$ and $G_2$ are 
	\begin{equation}
		\label{eq:edm-loop-functions}
		\begin{aligned}[b]
			G_1(x)
			&=
			\frac{1}{2(1-x)^2}
			\left(
			1+x+\frac{2x}{1-x}\ln x
			\right),
			\\
			G_2(x)
			&=
			\frac{1}{2(1-x)^2}
			\left(
			3-x+\frac{2}{1-x}\ln x
			\right).
		\end{aligned}
	\end{equation}
	
	\paragraph{The EDM of the electron.}
	
\begin{figure}[htbp]
		\centering
		\includegraphics[width=0.48\textwidth]{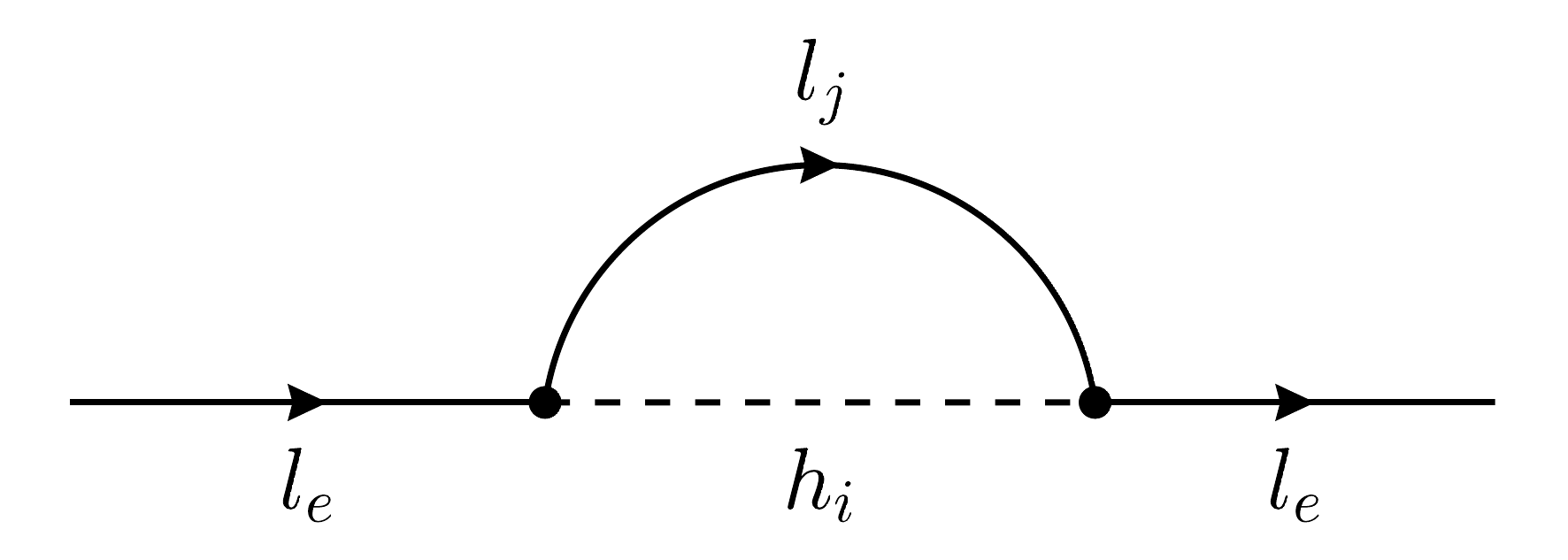}
		\caption{\raggedright The one-loop level diagrams contribute to the electron EDM.}
		\label{fig:figure1}
	\end{figure}

	In the present analysis, we include only the neutral-Higgs-mediated one-loop contributions, while the charged Higgs contributions are neglected because neutrinos are massless in the G3HDM considered here. The effective Lagrangian for the electron EDM can be written as
	\begin{equation}
		\mathcal{L}_{\rm EDM}
		=
		-\frac{i}{2} d_e\,
		\bar{l}_e \sigma^{\mu\nu}\gamma_5 l_e
		F_{\mu\nu}.
		\label{eq:electron-edm-lagrangian}
	\end{equation}
	where $\sigma^{\mu\nu}=i[\gamma^\mu,\gamma^\nu]/2$, and $F_{\mu\nu}$ denotes the electromagnetic field-strength tensor. Adopting the effective Lagrangian approach, we obtain
	\begin{equation}
		d_e
		=
		-\frac{2e Q_f m_e}{(4\pi)^2}
		\left(
		C_2^R + C_2^{L*} + C_6^R
		\right).
		\label{eq:electron-edm-coefficients}
	\end{equation}
	where $Q_f=-1$ is the electric charge of the charged lepton in units of $e$, $m_e$ denotes the electron mass, and $C_2^L$, $C_2^R$, and $C_6^R$ are the Wilson coefficients of the corresponding operators $O_2^L$, $O_2^R$, and $O_6^R$, respectively
	\begin{equation}
		\begin{aligned}
			O_2^{L,R}
			&=
			\frac{e Q_f}{(4\pi)^2}
			(-i\mathcal{D}_\alpha^*)\bar{l}_e\,
			\gamma^\alpha
			F\cdot\sigma\,
			P_{L,R}l_e ,
			\\
			O_6^{L,R}
			&=
			\frac{e Q_f m_e}{(4\pi)^2}
			\bar{l}_e\,
			F\cdot\sigma\,
			P_{L,R}l_e .
		\end{aligned}
		\label{eq:edm-effective-operators}
	\end{equation}
	where $\mathcal{D}_\alpha=\partial_\alpha+iA_\alpha$, $l_e$ denotes the electron field, and $P_{L,R}=(1\mp\gamma_5)/2$. The Feynman diagrams contributing to the above Wilson coefficients are shown in Fig.~\ref{fig:figure1}.
	Calculating the Feynman diagrams, the electron EDM can be written as
	\begin{equation}
		\label{eq:dl_one_loop_chargino}
		d_e
		=
		\frac{-e m_{l_j}}
		{16\pi^2 m_{h_i}^2}
		G_2
		\left(
		\frac{m_{l_j}^2}
		{m_{h_i}^2}
		\right) 
		\operatorname{\Im}
		\left[
		C_{h_i\bar l_e l_j}^{L}
		C_{h_i\bar l_e l_j}^{R*}
		\right].
	\end{equation}
	where $x_i=m_i^2/m_W^2$. 
	\paragraph{The EDMs of the neutron and mercury atom.}
	
	To be consistent with the discussion in Ref.~\cite{Sala_2014}, for the neutron EDM $d_n$, we adopt the values $1.5$ for the coefficients $1\pm0.5$ in Eq.~\eqref{eq:neutron-edm}. In the present analysis, we only consider the one-loop contributions to $d_q^\gamma$ and $d_q^g$. Therefore, $d_q^\gamma$ and $d_q^g$ entering Eq.~\eqref{eq:neutron-edm} are obtained solely from the Higgs-mediated one-loop Feynman diagrams shown in Fig.~\ref{fig:two_diagrams}, whose analytical expressions are given in Eq.~\eqref{eq:total-quark-edm-cedm}.
	\begin{equation}
		d_n
		=
		(1\pm 0.5)
		\left[
		1.4\left(d_d^\gamma-0.25d_u^\gamma\right)
		+
		1.1e\left(d_d^g+0.5d_u^g\right)
		\right].
		\label{eq:neutron-edm}
	\end{equation}
	
	According to Ref.~\cite{Lee_2005}, the mercury atom EDM $d_{\rm Hg}$ can be expressed in terms of the quark CEDMs $d_q^g$ as
	\begin{equation}
		d_{\rm Hg}
		=
		-\left(
		d_d^g
		-
		d_u^g
		-
		0.012\,d_s^g
		\right)
		\times 3.2\times 10^{-2}\, e .
		\label{eq:mercury-edm}
	\end{equation}
	
	\section{Numerical Analysis}
	\label{sec:Numerical Analysis}
	In this section, we present the numerical results for the EDMs of the $b$ quark, $c$ quark, electron, neutron, and mercury atom. For relevant parameters in the SM, we choose 
	\begin{equation}
		\begin{aligned}[b]
			& \alpha_s(m_Z)=0.118,\quad
			\alpha(m_Z)=\frac{1}{128.9},\quad
			m_Z=91.1876\,\mathrm{GeV},\quad
			m_W=80.385\,\mathrm{GeV},\\
			& m_u=0.00216\,\mathrm{GeV},\quad
			m_c=1.2729\,\mathrm{GeV},\quad
			m_t=172.6\,\mathrm{GeV},\quad
			m_d=0.0047\,\mathrm{GeV},\\
			& m_s=0.0929\,\mathrm{GeV},\quad
			m_b=4.186\,\mathrm{GeV},\quad
			m_e=0.000511\,\mathrm{GeV},\quad
			m_\mu=0.105\,\mathrm{GeV},\quad
			m_\tau=1.77\,\mathrm{GeV}.
			\label{eq:SM input}
		\end{aligned}
	\end{equation}
	Here, $\alpha_s(m_Z)$ and $\alpha(m_Z)$ denote the strong coupling constant and electromagnetic fine-structure constant evaluated at the scale $m_Z$, respectively. Once the electroweak scale is fixed, $t_\beta$ and $t_{\beta'}$ determine the three vacuum expectation values, while $\lambda_1,\ldots,\lambda_9$ are taken as independent quartic couplings.
	
	Rather than using the real and imaginary parts of all complex soft-breaking parameters as independent inputs, we adopt $\theta_2$, $\theta_3$, $|m_{12}^2|$, $\phi_{12}$, $R_{13}$, and $R_{23}$, where $\phi_{12}$ is the phase of $m_{12}^2$. As follows from Eq.~(\ref{eq:eq12}), $\theta_3$ alone does not induce CP-violating scalar mixing and hence generates no EDM contribution in the present analysis. We therefore set $\theta_3=0$.
	
	Furthermore, $\theta_2$ and $\phi_{12}$ enter the scalar mass matrix through $I_{12}$ only in the combination $\theta_2+\phi_{12}$. They therefore do not represent independent physical phases. To eliminate this redundancy, we set $\phi_{12}=0$ and retain $\theta_2$ as the independent phase governing CP-violating effects.

	\begin{table}[H]
		\centering
		\small
		\setlength{\tabcolsep}{3.5pt}
		\renewcommand{\arraystretch}{1.1}
		
		\resizebox{0.98\linewidth}{!}{%
			\begin{tabular}{|c|c|c|c|c|c|c|c|}
				\hline
				Parameters
				& $\tan\beta$
				& $\tan\beta'$
				& $\lambda_1$
				& $\lambda_2$
				& $\lambda_3$
				& $\lambda_i\ (i\neq1,2,3)$
				& $\theta_2$
				\\
				\hline
				
				Min
				& 2
				& 15
				& 0
				& 0
				& 0
				& -4
				& -1
				\\
				\hline
				
				Max
				& 10
				& 25
				& 4
				& 4
				& 1
				& 4
				& 1
				\\
				\hline
				
				Parameters
				& $|m_{12}^2|/\mathrm{GeV}^2$
				& $R_{13}/\mathrm{GeV}^2$
				& $R_{23}/\mathrm{GeV}^2$
				& $O^{u}_{ij}\ (i,j=u,c,t)$
				& $O^{d}_{ij}\ (i,j=d,s,b)$
				& $O^{e}_{ij}\ (i,j=e,\mu,\tau)$
				& $\theta_f/\pi$
				\\
				\hline
				
				Min
				& $3\times10^{6}$
				& $3\times10^{6}$
				& $3\times10^{6}$
				& -2
				& -0.1
				& -2
				& -1
				\\
				\hline
				
				Max
				& $5\times10^{6}$
				& $5\times10^{6}$
				& $5\times10^{6}$
				& 2
				& 0.1
				& 2
				& 1
				\\
				\hline
			\end{tabular}%
		}
		
		\caption{Scan ranges of the model parameters.}
		\label{tab:scan-parameters}
	\end{table}
	where $\theta_{f}$ denotes the normalized phase angle corresponding to each Yukawa input parameter.
	
	\begin{figure}[htbp]
		\centering
		\includegraphics[width=1.0\textwidth]{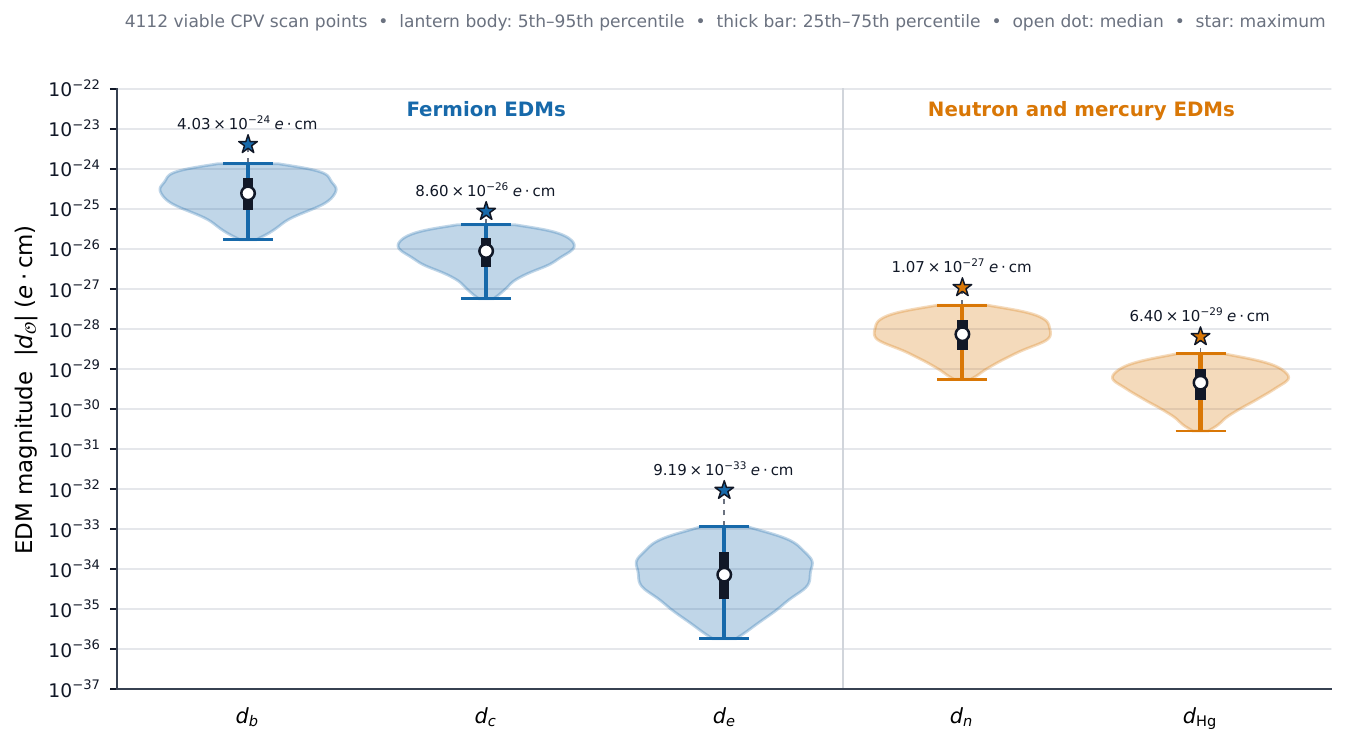}
		\caption{\raggedright Lantern plots of the absolute one-loop EDMs predicted for the 4112 viable parameter points. The horizontal categories represent the $b$ quark, $c$ quark, electron, neutron, and mercury EDMs, while the vertical axis shows $|d_{\mathcal O}|$ in units of $e\cdot{\rm cm}$ on a logarithmic scale. Each lantern profile displays the density distribution between the 5th and 95th percentiles. The thick vertical segment indicates the interquartile range, the open circle marks the median, and the star with its adjacent label denotes the maximum absolute value of each observable.
		}
		\label{fig:edm_lantern}
	\end{figure}
	
	We first present an overview of the one-loop EDM predictions for the $b$ quark, $c$ quark, electron, neutron, and mercury atom. As shown in Fig.~$\ref{fig:edm_lantern}$, the horizontal axis identifies the five observables, while the vertical axis displays their absolute EDMs in units of $e\cdot{\rm cm}$ on a logarithmic scale. The width of each lantern profile represents the density of parameter points at a given EDM value, with wider regions corresponding to higher point densities. The percentile intervals and median of each distribution are also indicated, while the star marks the maximum predicted value.
	
	In particular, the mercury EDM can exceed the current $95\%$ C.L. upper bound, $\left|d_{\rm Hg}\right|<7.4\times10^{-30}e\cdot{\rm cm}$. Among the 4112 parameter points surviving the preceding constraints, 1387 points exceed this limit, accounting for approximately $33.7\%$ of the sample. The largest predicted value is $6.4\times10^{-29}e\cdot{\rm cm}$, approximately $8.6$ times the experimental bound. The mercury EDM therefore provides a particularly stringent additional constraint on the model parameter space.
	
	\begin{figure}[htbp]
		\centering
		
		\begin{subfigure}{0.4\textwidth}
			\centering
			\includegraphics[width=\textwidth]{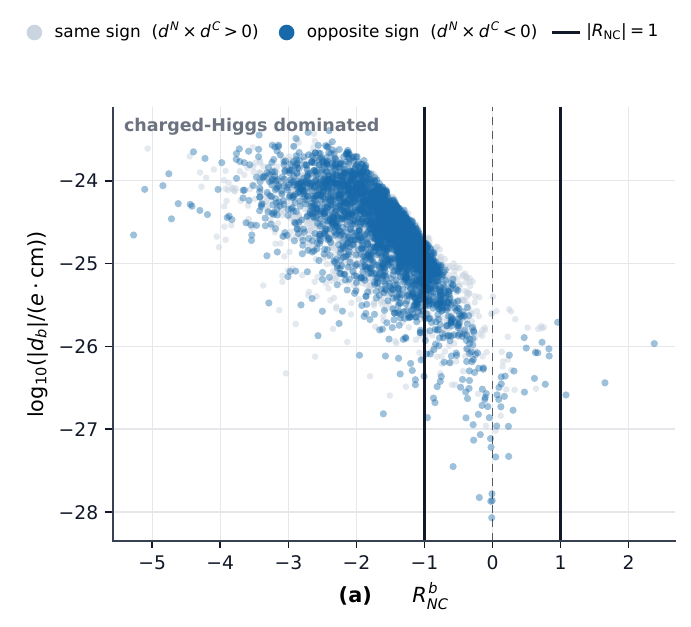}
			\label{fig:subbb1}
		\end{subfigure}
		\hspace{0.5cm}
		\begin{subfigure}{0.4\textwidth}
			\centering
			\includegraphics[width=\textwidth]{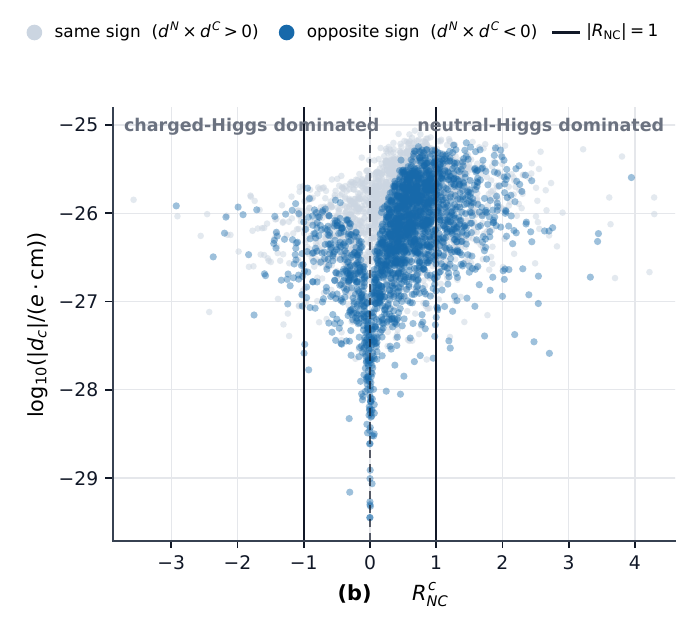}
			\label{fig:subbb2}
		\end{subfigure}
		
		\vspace{0.5cm}
		
		\begin{subfigure}{0.4\textwidth}
			\centering
			\includegraphics[width=\textwidth]{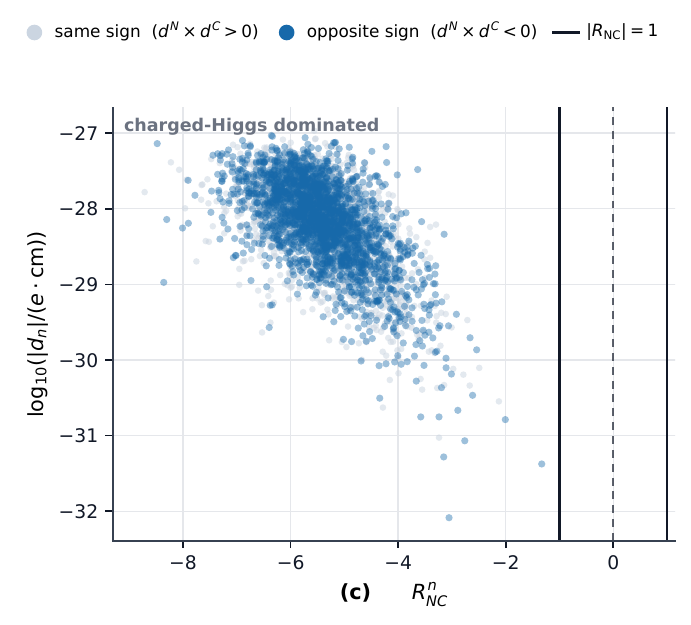}
			\label{fig:subbb3}
		\end{subfigure}
		\hspace{0.5cm}
		\begin{subfigure}{0.4\textwidth}
			\centering
			\includegraphics[width=\textwidth]{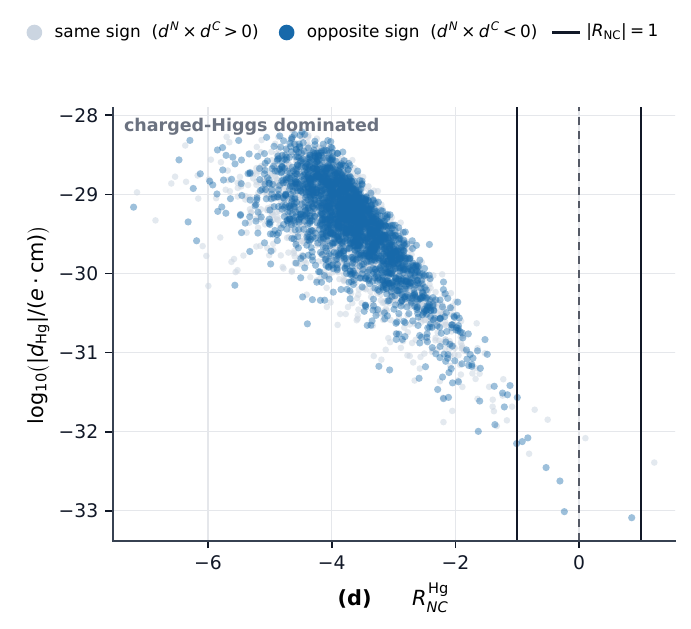}
			\label{fig:subbb4}
		\end{subfigure}
		
		\caption{\raggedright Total EDM magnitude versus $R_{NC}=\log_{10}(|d_{\mathcal O}^{N}|/|d_{\mathcal O}^{C}|)$. Blue and gray points denote opposite-sign and same-sign neutral and charged Higgs contributions, respectively. Positive (negative) $R_{NC}$ indicates neutral Higgs (charged Higgs) dominance, while the vertical lines at $R_{NC}=\pm1$ indicate a one-order-of-magnitude hierarchy between the two contributions. The electron EDM is omitted because $d_e^{C}=0$ at the one-loop level.
		}
		\label{fig:relative_NC}
	\end{figure}

	Since Fig.~$\ref{fig:edm_lantern}$ only shows the distributions of the absolute values of the total EDMs, it is not possible to determine from this figure alone whether a small total EDM originates from the suppression of the individual loop contributions themselves or from destructive interference among different contributions. To distinguish between these two mechanisms, we further decompose each EDM into contributions from the neutral and charged Higgs sectors.
	
	For a fixed loop channel $\gamma=(q,\phi,f)$, we first perform the corresponding EDM/CEDM matching and define
	\begin{equation}
		D_{\mathcal O,\gamma}
		=
		\mathcal C_{\mathcal Oq}^{EDM}\,
		d_{q,f}^{\phi,\rm EDM}
		+
		\mathcal C_{\mathcal Oq}^{CEDM}\,
		\widetilde d_{q,f}^{\phi,\rm CEDM}.
		\label{eq:matched_channel}
	\end{equation}
	here, $\mathcal O=b,c,n,\mathrm{Hg}$. $q$ denotes the quark flavor contributing to the corresponding matching relation, $\phi$ denotes the Higgs mass eigenstate, and $f$ denotes the internal fermion flavor. The coefficients $\mathcal C_{\mathcal Oq}^{\phi,EDM,CEDM}$ contain the corresponding quark, hadronic, and atomic matching factors.
	
	The contribution from each Higgs sector is then defined as
	\begin{equation}
		d_{\mathcal O}^{X}
		=
		\sum_{\gamma\in\Gamma_{\mathcal O}^{X}}
		D_{\mathcal O,\gamma},
		\qquad
		d_{\mathcal O}
		=
		d_{\mathcal O}^{N}+d_{\mathcal O}^{C},
		\qquad
		R_{NC}^{\mathcal O}
		=
		\log_{10}
		\left(
		\frac{|d_{\mathcal O}^{N}|}
		{|d_{\mathcal O}^{C}|}
		\right).
		\label{eq:edm_NC_decomposition}
	\end{equation}
	here $\Gamma_{\mathcal O}^{X}$ denotes the set of all loop diagrams contributing to the observable $\mathcal O$ in the Higgs sector $X=N,C$. $R_{NC}^{\mathcal O}>0$ indicates neutral Higgs dominance, $R_{NC}^{\mathcal O}<0$ indicates charged Higgs dominance, while $R_{NC}^{\mathcal O}\simeq0$ corresponds to contributions of comparable magnitude.

	As shown in Figs.~\ref{fig:relative_NC}, destructive interference requires the two contributions to have not only opposite signs but also comparable magnitudes. Therefore, only the blue points near $R_{NC}=0$ can exhibit substantial cancellation. For opposite sign points with large $|R_{NC}|$, one loop-sector contribution is too small to effectively cancel the other, and the cancellation effect therefore remains weak.
	
	The results show that the EDMs of the $b$ quark, neutron, and mercury atom are predominantly distributed in the region where the charged Higgs contribution dominates. In contrast, a substantial fraction of the parameter points for the $c$ quark EDM are concentrated around $R_{NC}\simeq 0$, indicating that the neutral and charged Higgs contributions are of comparable magnitude. Combined with the fact that these two contributions typically have opposite signs, the $c$ quark EDM consequently exhibits the most pronounced destructive interference between the neutral and charged Higgs contributions.

	\begin{figure}[htbp]
		\centering
		
		\begin{subfigure}{0.4\textwidth}
			\centering
			\includegraphics[width=\textwidth]{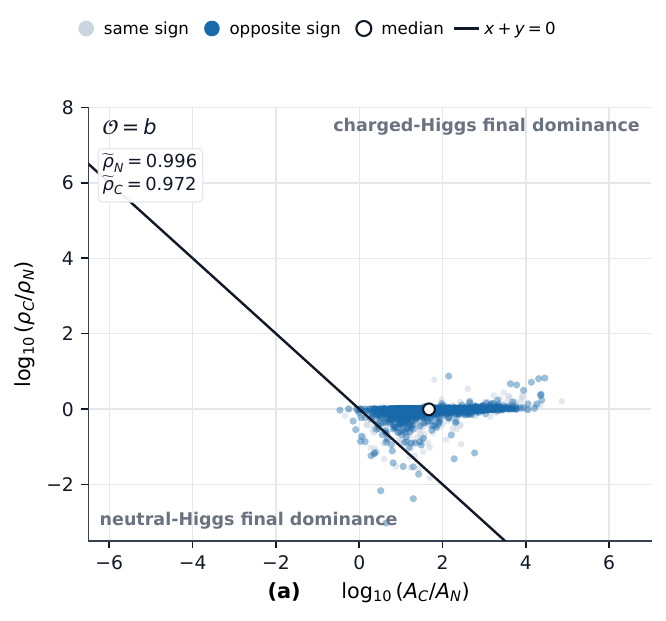}
			\label{fig:subbbb1}
		\end{subfigure}
		\hspace{0.5cm}
		\begin{subfigure}{0.4\textwidth}
			\centering
			\includegraphics[width=\textwidth]{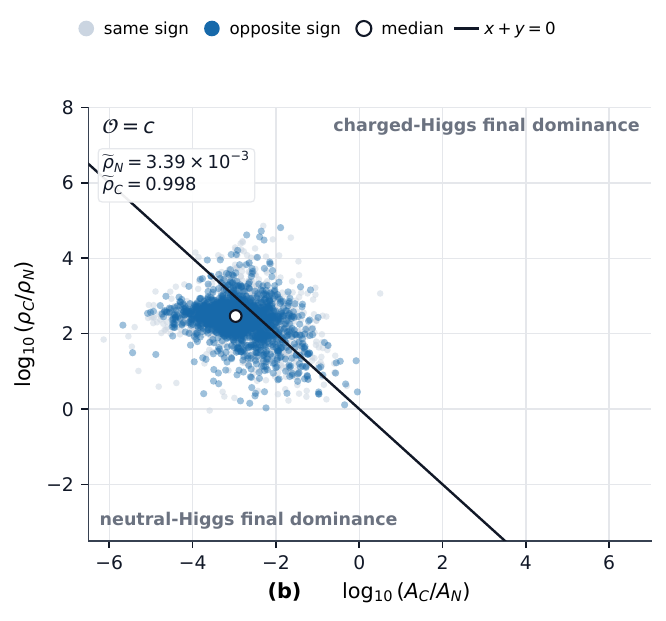}
			\label{fig:subbbb2}
		\end{subfigure}
		
		\vspace{0.5cm}
		
		\begin{subfigure}{0.4\textwidth}
			\centering
			\includegraphics[width=\textwidth]{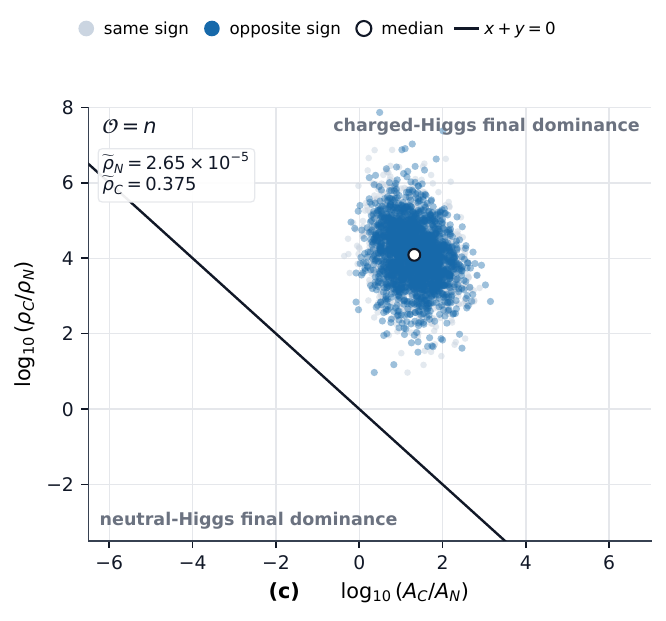}
			\label{fig:subbbb3}
		\end{subfigure}
		\hspace{0.5cm}
		\begin{subfigure}{0.4\textwidth}
			\centering
			\includegraphics[width=\textwidth]{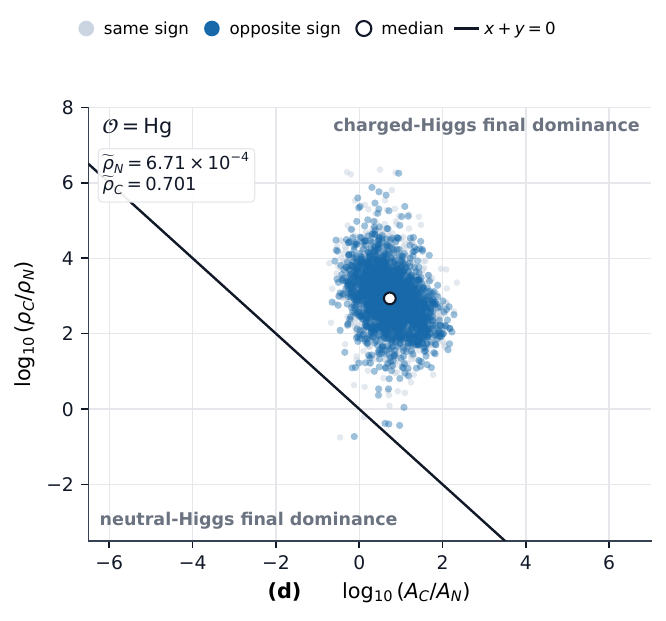}
			\label{fig:subbbb4}
		\end{subfigure}
		
		\caption{\raggedright Decomposition of the hierarchy between the neutral and charged Higgs contributions into the raw-amplitude-scale ratio $x=\log_{10}(A_C/A_N)$ and the coherence correction $y=\log_{10}(\rho_C/\rho_N)$. Points above (below) the line $x+y=0$ indicate charged Higgs (neutral Higgs) dominance. Blue and gray points denote opposite-sign and same-sign neutral and charged Higgs contributions, respectively, while the circular marker denotes the median. The electron EDM is omitted because $d_e^C=0$ at the one-loop level.
		}
		\label{fig:scale_coherence}
	\end{figure}
	
	Although $R_{NC}$ characterizes the relative magnitudes of the neutral and charged Higgs contributions, it cannot distinguish whether this hierarchy originates from a difference in the raw-amplitude scales of the two sectors or from different degrees of internal cancellation within each sector. To further distinguish between these two mechanisms, we define
	\begin{equation}
		A_{X} =
		\sum_{\gamma\in\Gamma_{\mathcal O}^{X}}|D_{\mathcal O,\gamma}|,
		\qquad
		\rho_X =
		\frac{|d_{\mathcal O}^X|}{A_{X}},
		\qquad X=N,C.
		\label{eq:sector_coherence}
	\end{equation}
	here, $A_X$ measures the raw amplitude scale of sector $X$, while $\rho_X$ characterizes the degree of coherence among the contributions from different channels within that sector.
	
	Figs.~\ref{fig:scale_coherence} displays
	\begin{equation}
		x=\log_{10}\frac{A_C}{A_N},
		\qquad
		y=\log_{10}\frac{\rho_C}{\rho_N},
	\end{equation}
	which satisfy
	\begin{equation}
		x+y=\log_{10}\frac{|d_{\mathcal O}^C|}{|d_{\mathcal O}^N|}.
	\end{equation}
	Therefore, parameter points above the line $x+y=0$ indicate charged Higgs dominance, whereas those below the line indicate neutral Higgs dominance.
	
	As shown in Fig.~\ref{fig:scale_coherence} (a), for the $b$ quark, the median raw-amplitude ratio is $A_C/A_N\simeq4.7\times10^1$, while the median coherence factors are $\widetilde{\rho}_N=0.996$ and $\widetilde{\rho}_C=0.972$, respectively. Since both coherence factors are close to unity, internal cancellations are weak in both sectors, and the dominance of the charged Higgs contribution mainly originates from its larger raw amplitude.

	For the $c$ quark EDM, as shown in Fig.~\ref{fig:scale_coherence} (b), we can see that the median raw-amplitude ratio is $A_C/A_N\simeq1.1\times10^{-3}$, while the median coherence factors are $\widetilde{\rho}_N=3.39\times10^{-3}$ and $\widetilde{\rho}_C=0.998$. This indicates that, although the neutral Higgs sector initially possesses a much larger raw amplitude, its strong internal cancellation suppresses the final neutral Higgs contribution to a magnitude comparable to that of the charged Higgs contribution, thereby enhancing the interference between the two sectors.
	
	For the neutron EDM, the corresponding median values are $A_C/A_N\simeq2.1\times10^1$, $\widetilde{\rho}_N=2.65\times10^{-5}$, and $\widetilde{\rho}_C=0.375$. The larger raw amplitude of the charged Higgs contribution, together with the much stronger internal cancellation in the neutral sector, jointly leads to pronounced charged Higgs dominance.
	
	For the mercury EDM, the corresponding median values are $A_C/A_N\simeq5.4$, $\widetilde{\rho}_N=6.71\times10^{-4}$, and $\widetilde{\rho}_C=0.701$. This indicates that the raw-amplitude scales of the charged and neutral Higgs sectors are of the same order of magnitude, while the neutral Higgs sector undergoes stronger internal cancellation. Consequently, the final mercury EDM is dominated by the charged Higgs contribution.
	
	To further elucidate the origin of this exceptionally small coherence factor in the neutral Higgs sector, we further decompose the neutral Higgs contribution into the individual contributions from the five physical neutral Higgs mass eigenstates
	
	\begin{figure}[H]
		\centering
		
		\begin{subfigure}{0.24\textwidth}
			\centering
			\includegraphics[width=\linewidth]{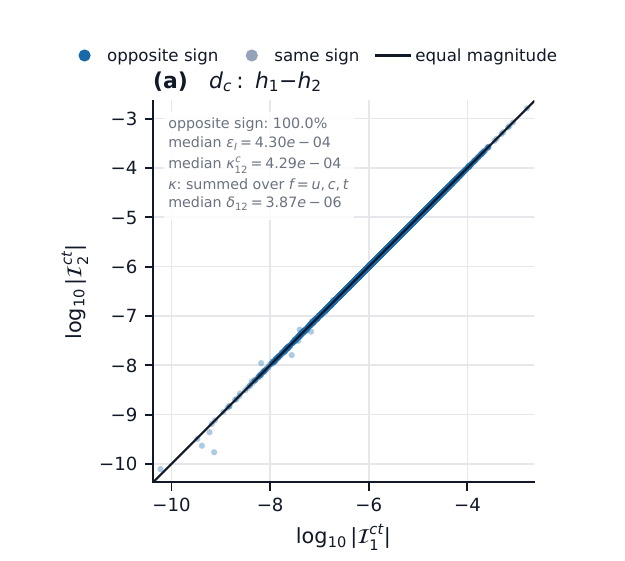}
		\end{subfigure}
		\hfill
		\begin{subfigure}{0.24\textwidth}
			\centering
			\includegraphics[width=\linewidth]{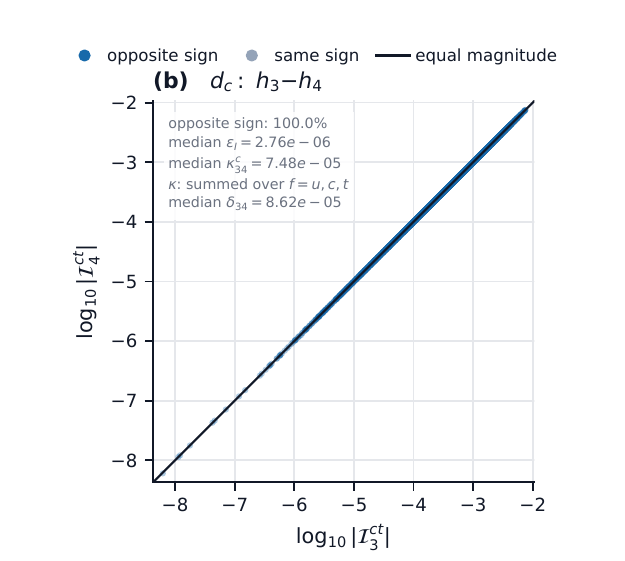}
		\end{subfigure}
		\hfill
		\begin{subfigure}{0.24\textwidth}
			\centering
			\includegraphics[width=\linewidth]{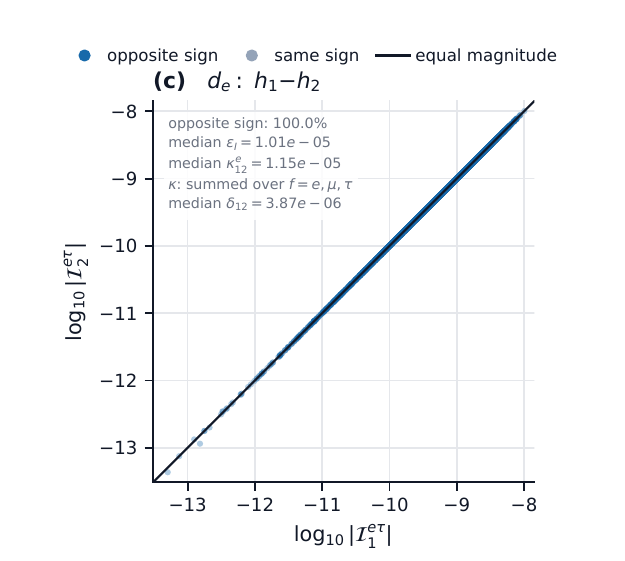}
		\end{subfigure}
		\hfill
		\begin{subfigure}{0.24\textwidth}
			\centering
			\includegraphics[width=\linewidth]{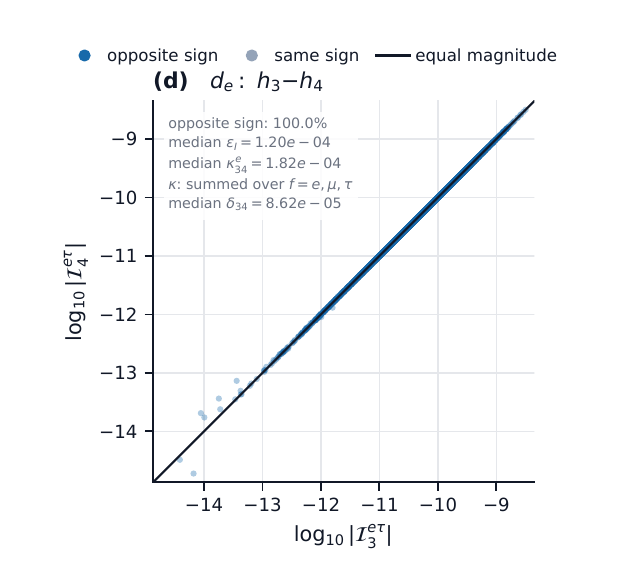}
		\end{subfigure}
		
		\par\vspace{0.3cm}
		
		\begin{subfigure}{0.24\textwidth}
			\centering
			\includegraphics[width=\linewidth]{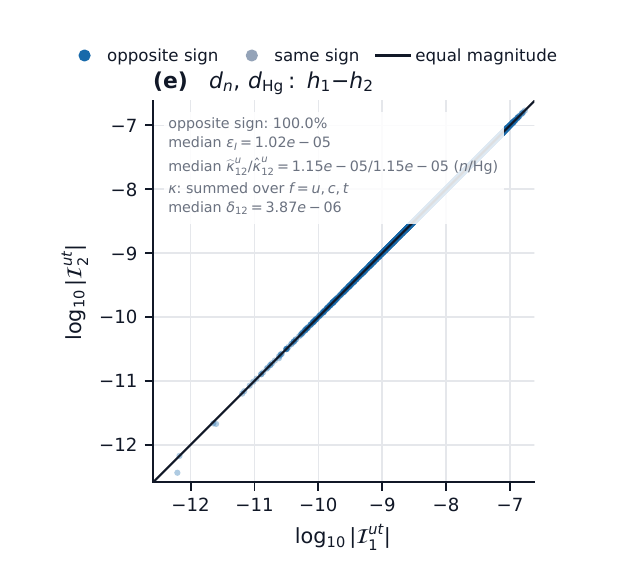}
		\end{subfigure}
		\hfill
		\begin{subfigure}{0.24\textwidth}
			\centering
			\includegraphics[width=\linewidth]{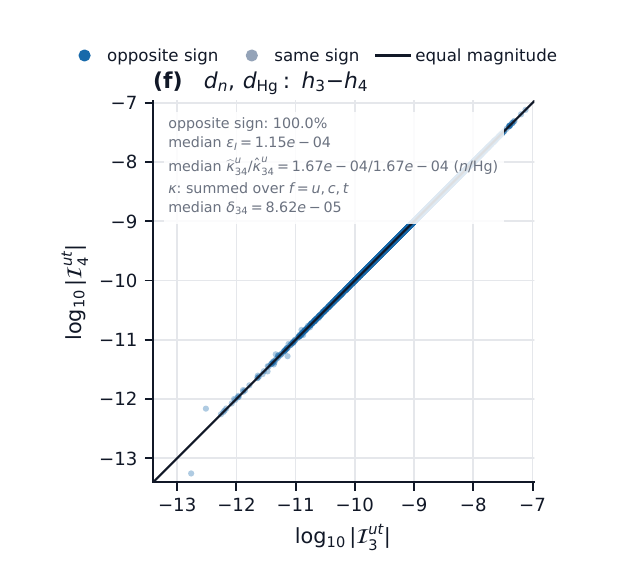}
		\end{subfigure}
		\hfill
		\begin{subfigure}{0.24\textwidth}
			\centering
			\includegraphics[width=\linewidth]{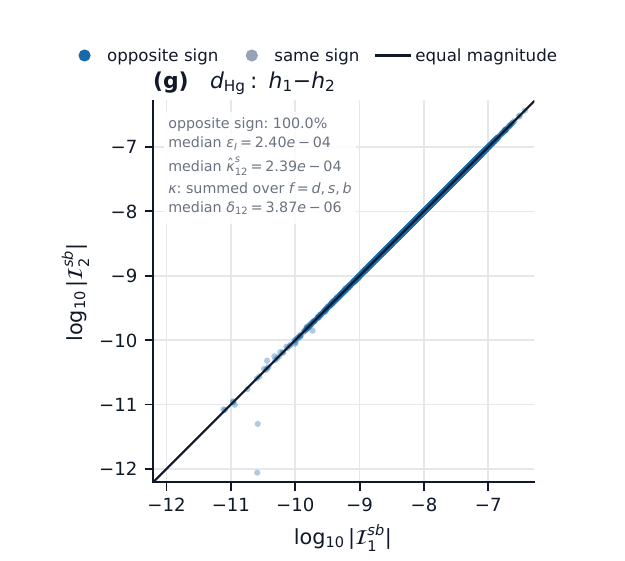}
		\end{subfigure}
		\hfill
		\begin{subfigure}{0.24\textwidth}
			\centering
			\includegraphics[width=\linewidth]{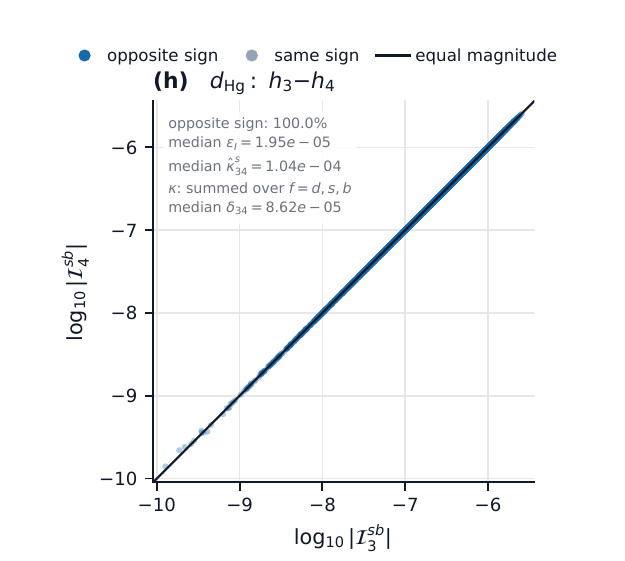}
		\end{subfigure}
		
		\par\vspace{0.3cm}
		
		\begin{subfigure}{0.24\textwidth}
			\centering
			\includegraphics[width=\linewidth]{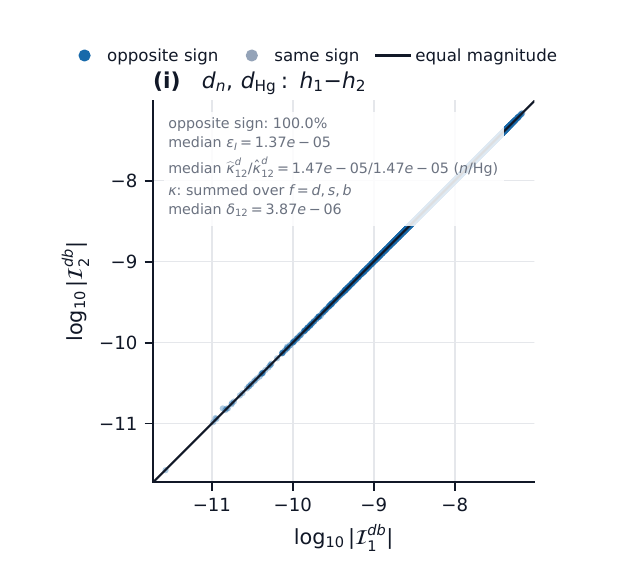}
		\end{subfigure}
		\hspace{0.05\textwidth}
		\begin{subfigure}{0.24\textwidth}
			\centering
			\includegraphics[width=\linewidth]{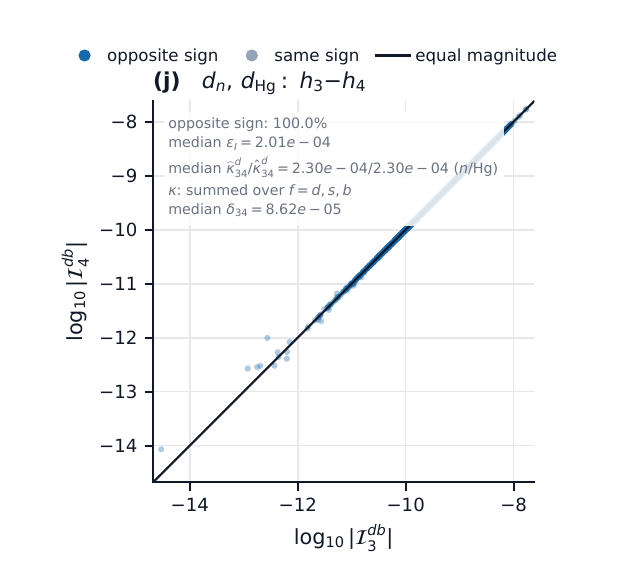}
		\end{subfigure}
		
		\caption{\raggedright
			Correlations between the CP-odd coupling invariants associated with the nearly degenerate Higgs pairs $h_1-h_2$ and $h_3-h_4$. Blue and gray points denote cases in which the two invariants have opposite and identical signs, respectively, while the black diagonal indicates equal absolute magnitudes.
		}
		\label{fig:neutral_invariant}
	\end{figure}
	\begin{equation}
		d_{\mathcal O}^{N} =
		\sum_{i=5}d_{\mathcal O}^{h_i},
		\qquad
		d_{\mathcal O}^{h_i} = 
		\sum_{\gamma\in\Gamma_{\mathcal O}^{h_i}}
		D_{\mathcal O,\gamma}.
		\label{eq:neutral_higgs_decomposition}
	\end{equation}
	
	From Eq.~\eqref{eq:matched_channel}, we further define
	\begin{equation}
		\begin{aligned}[b]
			\widehat d_{u}^{h_i}
			&=
			\sum_{f}
			\left(
			\mathcal C_{n,u}^{EDM}\,
			d_{u,f}^{h_i,\rm EDM}
			+
			\mathcal C_{n,u}^{CEDM}\,
			\widetilde d_{u,f}^{h_i,\rm CEDM}
			\right),
			\\
			\widehat d_{d}^{h_i}
			&=
			\sum_{f}
			\left(
			\mathcal C_{n,d}^{EDM}\,
			d_{d,f}^{h_i,\rm EDM}
			+
			\mathcal C_{n,d}^{CEDM}\,
			\widetilde d_{d,f}^{h_i,\rm CEDM}
			\right),
			\\
			\hat d_{u}^{h_i}
			&=
			\sum_{f}
			\mathcal C_{\mathrm{Hg},u}^{CEDM}\,
			\widetilde d_{u,f}^{h_i,\rm CEDM},
			\\
			\hat d_{d}^{h_i}
			&=
			\sum_{f}
			\mathcal C_{\mathrm{Hg},d}^{CEDM}\,
			\widetilde d_{d,f}^{h_i,\rm CEDM},
			\\
			\hat d_{s}^{h_i}
			&=
			\sum_{f}
			\mathcal C_{\mathrm{Hg},s}^{CEDM}\,
			\widetilde d_{s,f}^{h_i,\rm CEDM}.
		\end{aligned}
		\label{eq:new_define}
	\end{equation}
	then we have
	\begin{equation}
		\begin{aligned}[b]
			d_{n}^{h_i}
			&=
			\widehat d_{u}^{h_i}
			+
			\widehat d_{d}^{h_i}
			,
			\\
			d_{Hg}^{h_i}
			&=
			\hat d_{u}^{h_i}
			+
			\hat d_{d}^{h_i}
			+
			\hat d_{s}^{h_i}.
		\end{aligned}
		\label{eq:n and Hg}
	\end{equation}
	here, $\widehat d_{u,d}^{h_i}$ denote the contributions of the neutral Higgs state $h_i$ to the neutron EDM through the $u$ and $d$ quark EDM and CEDM operators, respectively, while $\hat d_{u,d,s}^{h_i}$ denotes the contribution of the neutral Higgs state $h_i$ to the mercury EDM through the $u$, $d$, and $s$ quark CEDM operators.

	In the present framework, the electron EDM receives only neutral Higgs contributions at the one-loop level, and we therefore include it as a complementary observable. Within the allowed parameter samples, the numerical results indicate that the neutral Higgs mass spectrum contains two pairs of nearly degenerate states, namely $(h_1,h_2)$ and $(h_3,h_4)$. To quantitatively characterize the relevant properties of each Higgs pair, we define
	\begin{equation}
		\begin{aligned}[b]
			\delta_{ij} 
			&= 
			\frac{2\left|m_{h_i}-m_{h_j}\right|} 
			{m_{h_i}+m_{h_j}}, 
			\qquad 
			\epsilon_I 
			= 
			\frac{\left|\mathcal{I}_i^{mn}+\mathcal{I}_j^{mn}\right|} 
			{\left|\mathcal{I}_i^{mn}\right|+\left|\mathcal{I}_j^{mn}\right|}, 
			\qquad 
			\kappa_{ij}^f 
			= 
			\frac{\left|d_{f}^{h_i}+d_{f}^{h_j}\right|} 
			{\left|d_{f}^{h_i}\right|+\left|d_{f}^{h_j}\right|},
			\\[4pt]
			\widehat{\kappa}_{ij}^f
			&=
			\frac{\left|\widehat{d}_{f}^{h_i}+\widehat{d}_{f}^{h_j}\right|}
			{\left|\widehat{d}_{f}^{h_i}\right|+\left|\widehat{d}_{f}^{h_j}\right|},
			\qquad
			\hat{\kappa}_{ij}^f
			=
			\frac{\left|\hat{d}_{f}^{h_i}+\hat{d}_{f}^{h_j}\right|}
			{\left|\hat{d}_{f}^{h_i}\right|+\left|\hat{d}_{f}^{h_j}\right|}.
		\end{aligned}
		\label{eq:delta_kappa_epsilon_neutral_higgs}
	\end{equation}
	here, $\mathcal{I}_i^{mn}=\operatorname{Im}\!\left[ C_{h_i\bar mn}^{L}\left(C_{h_i\bar mn}^{R}\right)^* \right]$. $\delta_{ij}$ measures the relative mass splitting, while $\kappa_{ij}^{f}$, $\widehat{\kappa}_{ij}^f$ and $\hat{\kappa}_{ij}^f$  quantifies the residual contribution after interference. A value $\kappa_{ij}^{f}, \widehat{\kappa}_{ij}^f, \hat{\kappa}_{ij}^f\ll1$ indicates strong pairwise cancellation.
	
	We focus on the loop channels involving the heaviest internal fermions. Since chirality-flipping one-loop amplitudes are generally enhanced by the internal fermion mass, heavier internal fermions tend to yield more significant contributions. It should be emphasized that this choice applies only to the coupling invariants displayed on the two axes of Figs.~$\ref{fig:neutral_invariant}$. In contrast, the cancellation parameters $\kappa_{ij}^f$, $\widehat{\kappa}_{ij}^f$, and $\hat{\kappa}_{ij}^f$, as well as the complete EDM predictions, are obtained after the corresponding EDM and CEDM matching has been performed and the contributions from all relevant internal-fermion flavor channels have been summed with their respective signs.
	
	Figs.~$\ref{fig:neutral_invariant}$ (a) and (b) show the scatter distributions of the CP-odd coupling invariants corresponding to the nearly degenerate $h_1-h_2$ and $h_3-h_4$ Higgs pairs in the $c-t$ loop channel, respectively, while Figs.~$\ref{fig:neutral_invariant}$ (c) and (d) present the corresponding results for the same Higgs pairs in the $e-\tau$ loop channel. The black diagonal line represents the case in which the two coupling invariants have equal absolute values, whereas the blue and gray points correspond to opposite-sign and same-sign configurations, respectively. As can be seen, the vast majority of the parameter points lie close to the diagonal line, and all of them belong to the opposite-sign category. This indicates that, for both nearly degenerate Higgs pairs, the corresponding CP-odd coupling invariants have approximately equal absolute values and opposite signs, thereby exhibiting a pronounced coupling anti-alignment. Moreover, the cancellation parameters $\kappa_{ij}^{c,e}$ defined in all four panels are significantly smaller than unity. Therefore, the mass degeneracy of the nearly degenerate Higgs states, together with the anti-alignment of their CP-odd coupling invariants, gives rise to this pronounced cancellation, thereby substantially suppressing the neutral-Higgs contributions to the $c$ quark and electron EDMs.
	
	The neutron and mercury EDMs are both composed of multiple light-quark EDM and CEDM components, making a more detailed decomposition of their neutral-Higgs contributions necessary. According to the definitions in Eq.~\eqref{eq:new_define} and Eq.~\eqref{eq:n and Hg}, Figs.~$\ref{fig:neutral_invariant}$ (e) and (f) show the scatter distributions of the CP-odd coupling invariants in the $u-t$ channel, which are used to examine the pairwise cancellation features of $\widehat d_{u}^{h_i}$ and $\hat d_{u}^{h_i}$ within the nearly degenerate $h_1-h_2$ and $h_3-h_4$ Higgs pairs. Figs.~$\ref{fig:neutral_invariant}$ (g) and (h) present the corresponding scatter distributions of the CP-odd coupling invariants in the $s-b$ channel, which are used to analyze the pairwise cancellation mechanism of $\hat d_{s}^{h_i}$ within the nearly degenerate $h_1-h_2$ and $h_3-h_4$ Higgs pairs. Similarly, Figs.~$\ref{fig:neutral_invariant}$ (i) and (j) display the corresponding coupling-invariant scatter distributions in the $d-b$ channel, which are used to investigate the pairwise cancellation behavior of $\widehat d_{d}^{h_i}$ and $\hat d_{d}^{h_i}$ within the nearly degenerate $h_1-h_2$ and $h_3-h_4$ Higgs pairs.
	
	The median values of the cancellation parameters in Figs.~$\ref{fig:neutral_invariant}$ (e)--(j) are all of order $10^{-6}$--$10^{-4}$, indicating extremely strong pairwise cancellations between the two nearly degenerate Higgs states. It is worth emphasizing that these strong cancellations are not confined to the heavy-fermion channels; after the remaining internal-flavor channels and the corresponding EDM/CEDM matching contributions are further included, the cancellation effects remain significant.

	In contrast, the neutral Higgs contribution to the $b$ quark EDM exhibits only weak internal cancellation. Although the nearly degenerate $h_1-h_2$ and $h_3-h_4$ Higgs pairs still show pronounced coupling anti-alignment, their contributions are small compared with that of $h_5$, so the corresponding pairwise cancellations are insufficient to significantly modify the total neutral-Higgs contribution. Consequently, the neutral Higgs sector of the $b$ quark EDM remains highly coherent overall, with its coherence factor close to unity.
	
	Although Figs.~\ref{fig:neutral_invariant} clearly demonstrates the anti-alignment of the CP-odd coupling invariants, the numerical results alone do not reveal its underlying origin. This structure can be understood analytically by applying degenerate perturbation theory to the neutral-Higgs mass matrix.
	
	In the weak basis, the neutral-Higgs mass matrix can be decomposed as
	\begin{equation}
		M_h^2
		=
		M_0^2+\Delta M^2,
		\qquad
		M_0^2
		=
		\begin{pmatrix}
			C&B\\
			-B&C
		\end{pmatrix},
		\qquad
		\Delta M^2
		=
		\begin{pmatrix}
			Q&0\\
			0&0
		\end{pmatrix}.
		\label{eq:MH_perturbative_decomposition}
	\end{equation}
	
    According to the Eq.~\eqref{eq:MH_matrix} and Eq.~\eqref{eq:MH_elements}	the three $3\times3$ submatrices are explicitly given by
	\begin{equation}
		Q
		=
		\begin{pmatrix}
			2\lambda_{1}v_{1}^{2}
			&
			(\lambda_{4}+\lambda_{7})v_{1}v_{2}
			&
			(\lambda_{5}+\lambda_{8})v_{1}v_{3}
			\\[1mm]
			(\lambda_{4}+\lambda_{7})v_{1}v_{2}
			&
			2\lambda_{2}v_{2}^{2}
			&
			(\lambda_{6}+\lambda_{9})v_{2}v_{3}
			\\[1mm]
			(\lambda_{5}+\lambda_{8})v_{1}v_{3}
			&
			(\lambda_{6}+\lambda_{9})v_{2}v_{3}
			&
			2\lambda_{3}v_{3}^{2}
		\end{pmatrix},
		\qquad
		Q^T=Q .
		\label{eq:Q_matrix}
	\end{equation}
	
	\begin{equation}
		C
		=
		\begin{pmatrix}
			\dfrac{v_{2}R_{12}+v_{3}R_{13}}{v_{1}}
			&
			-R_{12}
			&
			-R_{13}
			\\[3mm]
			-R_{12}
			&
			\dfrac{v_{1}R_{12}+v_{3}R_{23}}{v_{2}}
			&
			-R_{23}
			\\[3mm]
			-R_{13}
			&
			-R_{23}
			&
			\dfrac{v_{1}R_{13}+v_{2}R_{23}}{v_{3}}
		\end{pmatrix},
		\qquad
		C^T=C .
		\label{eq:C_matrix}
	\end{equation}
	
	\begin{equation}
		B
		=
		\begin{pmatrix}
			0&I_{12}&I_{13}
			\\
			-I_{12}&0&I_{23}
			\\
			-I_{13}&-I_{23}&0
		\end{pmatrix},
		\qquad
		B^T=-B .
		\label{eq:B_matrix}
	\end{equation}
	
	To characterize the special structure of $M_0^2$, we introduce the matrix $J$
	\begin{equation}
		J
		=
		\begin{pmatrix}
			0&\mathbf{1}_{3\times 3}\\
			-\mathbf{1}_{3\times 3}&0
		\end{pmatrix},
		\qquad
		J^T=-J,
		\qquad
		J^2=-\mathbf{1}_{6\times 6}.
		\label{eq:J_matrix}
	\end{equation}

	A direct calculation gives
	\begin{equation}
		[M_0^2,J]=0.
		\label{eq:M0_J_commutator}
	\end{equation}
	Thus, if $r_\alpha$ is a normalized eigenvector of $M_0^2$ with eigenvalue $\mu_\alpha^2$, then $\pm Jr_\alpha$ is an orthogonal eigenvector with the same eigenvalue. In addition, Eq.~\eqref{eq:eq12} and Eq.~\eqref{eq:C_matrix} implies that
	\begin{equation}
		C\boldsymbol v=0,
		\qquad
		B\boldsymbol v=0,
		\qquad
		\boldsymbol v
		=
		\begin{pmatrix}
			v_1\\v_2\\v_3
		\end{pmatrix}.
		\label{eq:CB_stationary_conditions}
	\end{equation}
	
	The unperturbed spectrum therefore takes the form
	\begin{equation}
		\operatorname{spec}(M_0^2)
		=
		\left\{
		\mu_1^2,\mu_1^2,
		\mu_2^2,\mu_2^2,
		0,0
		\right\}.
		\label{eq:MN0_spectrum}
	\end{equation}
	Therefore, the eigenspace of $M_0^2$ decomposes into two doubly degenerate heavy-Higgs subspaces associated with nonzero eigenvalues and a two-dimensional zero-eigenvalue subspace. We denote the eigenvectors corresponding to the zero eigenvalue by
	\begin{equation}
		r_0
		=
		\begin{pmatrix}
			\widehat{\boldsymbol v}\\0
		\end{pmatrix},
		\qquad
		\pm Jr_0
		=
		\begin{pmatrix}
			0\\ \mp\widehat{\boldsymbol v}
		\end{pmatrix},
		\qquad
		\widehat{\boldsymbol v}
		\equiv
		\frac{\boldsymbol v}{v},
		\qquad
		v^2=v_1^2+v_2^2+v_3^2.
		\label{eq:radial_Goldstone_basis}
	\end{equation}
	
	For each doubly degenerate subspace, we introduce the $6\times2$ matrix
	\begin{equation}
		U_\alpha
		\equiv
		\left(
		r_\alpha,\,
		sJr_\alpha
		\right),
		\qquad
		U_\alpha^TU_\alpha=\mathbf{1}_{2\times 2},
		\qquad
		M_0^2U_\alpha
		=
		\mu_\alpha^2U_\alpha,
		\qquad
		 \alpha
		 =
		 0,1,2,
		 \qquad
		 s
		 =
		 \pm 1 .
		\label{eq:degenerate_subspace_basis}
	\end{equation}
	The two columns of $U_\alpha$ form an orthonormal basis of the degenerate subspace associated with $\mu_\alpha^2$. According to degenerate perturbation theory, the perturbation matrix within this degenerate subspace can be written as
	\begin{equation}
		W_\alpha
		\equiv
		U_\alpha^T\Delta M^2U_\alpha,
		\qquad
		W_\alpha\xi_{\alpha a}
		=
		q_{\alpha a}\xi_{\alpha a},
		\qquad
		a=1,2.
		\label{eq:projected_perturbation}
	\end{equation}
	here, $q_{\alpha a}$ and $\xi_{\alpha a}$ are the eigenvalues and normalized eigenvectors of the projected $2\times2$ perturbation matrix $W_\alpha$, respectively.
	
	Within the two-dimensional zero-eigenvalue subspace, we have
	\begin{equation}
		W_0
		=
		U_0^T\Delta M^2U_0
		=
		\begin{pmatrix}
			\widehat{\boldsymbol v}^{\,T}
			Q\widehat{\boldsymbol v}
			&
			0
			\\
			0&0
		\end{pmatrix}.
		\label{eq:zero_subspace_perturbation}
	\end{equation}
	Therefore, the VEV-aligned radial mode acquires the SM-like Higgs mass
	\begin{equation}
		m_{h_{\rm SM}}^2
		\simeq
		\widehat{\boldsymbol v}^{\,T}
		Q\widehat{\boldsymbol v},
		\label{eq:SM_like_mass}
	\end{equation}
	whereas the Goldstone mode remains massless. In the present decomposition, the SM-like Higgs mass originates from the quartic couplings contained in $Q$. In the additional limit $v_3\simeq v$ with negligible off-diagonal quartic contributions, Eq.~\eqref{eq:SM_like_mass} reduces to
	\begin{equation}
		m_{h_{\rm SM}}^2
		\simeq
		2\lambda_3v_3^2.
	\end{equation}
	
	For the two degenerate subspaces associated with nonzero eigenvalues, the corresponding physical masses to first order are given by
	\begin{equation}
		m_{\alpha a}^2
		\simeq
		\mu_\alpha^2+q_{\alpha a},
		\qquad
		a=1,2.
		\label{eq:heavy_pair_masses}
	\end{equation}
	Since the scale of $\mu_{\alpha}$ is mainly controlled by the rephasing invariants and the vev ratios, whereas the scale of $q_{\alpha a}$ is set by the electroweak scale and the quartic couplings, we have
	\begin{equation}
		\frac{\lvert q_{\alpha a}\rvert}{\mu_\alpha^2}
		\ll 1,
		\label{eq:small_intrapair_splitting}
	\end{equation}
	
	The perturbation may also mix different degenerate subspaces. This mixing is negligible provided that
	\begin{equation}
		\frac{
			\left\|
			U_\beta^T\Delta M^2U_\alpha
			\right\|_2
		}{
			\left|\mu_\alpha^2-\mu_\beta^2\right|
		}
		\ll1,
		\qquad
		\beta\neq\alpha,
		\label{eq:interpair_perturbation}
	\end{equation}
	where $\|\cdots\|_2$ denotes the matrix spectral norm, and the denominator represents their unperturbed mass-squared difference.
	
	Under this condition, the two physical mass eigenstates $z_i$ and $z_j$ originating from the same degenerate heavy-Higgs subspace approximately preserve the algebraic relation determined by that subspace
	\begin{equation}
		z_j
		\simeq
		sJz_i.
		\label{eq:J_related_eigenvectors}
	\end{equation}
	here, $z_i=\left(Z^h_{i1},Z^h_{i2},Z^h_{i3},Z^h_{i4},Z^h_{i5},Z^h_{i6}\right)^T$ and $z_j=\left(Z^h_{j1},Z^h_{j2},Z^h_{j3},Z^h_{j4},Z^h_{j5},Z^h_{j6}\right)^T$ respectively represent the mass eigenstates $h_i$ and $h_j$ in the weak basis.
	
	In the CP-violating G3HDM, the neutral-Higgs Yukawa couplings can be expressed as
	\begin{equation}
		\begin{aligned}
			&C_{h_k \bar{f_q} f_{q'}}^{L}
			=
			-
			\Bigg[
			\frac{1}{v_1}
			\left(m_{q'q}^{f1}\right)^{*}
			\left(-iZ^{h}_{k4}+Z^{h}_{k1}\right)
			+
			\frac{1}{v_2}
			\left(m_{q'q}^{f2}\right)^{*}
			\left(-iZ^{h}_{k5}+Z^{h}_{k2}\right)
			+
			\frac{1}{v_3}
			\left(m_{q'q}^{f3}\right)^{*}
			\left(-iZ^{h}_{k6}+Z^{h}_{k3}\right)
			\Bigg],
			\\[2mm]
			&C_{h_k \bar{f_q} f_{q'}}^{R}
			=
			-
			\Bigg[
			\frac{1}{v_1}
			\left(m_{qq'}^{f1}\right)
			\left(iZ^{h}_{k4}+Z^{h}_{k1}\right)
			+
			\frac{1}{v_2}
			\left(m_{qq'}^{f2}\right)
			\left(iZ^{h}_{k5}+Z^{h}_{k2}\right)
			+
			\frac{1}{v_3}
			\left(m_{qq'}^{f3}\right)
			\left(iZ^{h}_{k6}+Z^{h}_{k3}\right)
			\Bigg].
		\end{aligned}
		\label{eq:neutral_higgs_fermion_couplings}
	\end{equation}
	where $Z^{h}$ is defined in Eq.~\eqref{eq:neutral_diag}, while $m_{qq'}^{fi}$ is defined in Eqs.~\eqref{eq:mu12}--\eqref{eq:md3}.
	
	According the Eq.~\eqref{eq:J_related_eigenvectors} and ~\eqref{eq:neutral_higgs_fermion_couplings}, we have 
	\begin{equation}
		C_{h_j \bar{f_q} f_{q'}}^{L}
		\simeq
		-isC_{h_i \bar{f_q} f_{q'}}^{L},
		\qquad
		C_{h_j \bar{f_q} f_{q'}}^{R}
		\simeq
		+isC_{h_i \bar{f_q} f_{q'}}^{R}.
		\label{eq:Yukawa_pair_relation}
	\end{equation}
	then
	\begin{equation}
		\mathcal I_j^{qq'}
		\simeq
		-\mathcal I_i^{qq'},
		\qquad
		\mathcal I_i^{qq'}
		\equiv
		\operatorname{Im}
		\left(
		C_{h_i\bar{f_q} f_{q'}}^{L}
		\left(
		C_{h_i\bar{f_q} f_{q'}}^{R}
		\right)^*
		\right).
		\label{eq:invariant_antialignment}
	\end{equation}
	
	Therefore, the pairwise EDM cancellation observed in Figs.~\ref{fig:neutral_invariant} is not an accidental numerical coincidence. It originates from the approximate $J$-related structure of the neutral-Higgs eigenvectors, which enforces the anti-alignment of the CP-odd coupling invariants, together with the small mass splitting within each physical Higgs pair.

	\section{Conclusion}
	\label{sec:Conclusion}
	
	Within the CP-violating G3HDM, we investigate the one-loop EDMs of the $b$ quark, $c$ quark, electron, neutron, and mercury atom. After imposing theoretical requirements and experimental constraints from the $125~\mathrm{GeV}$ Higgs signals, neutral-meson mixing, rare $B$-meson decays, charged-lepton-flavor-violating processes, and flavor-changing top-quark decays, we obtain 4112 viable parameter points. Among the five observables, the mercury EDM provides the strongest additional constraint: approximately $33.7\%$ of the parameter points exceed the current experimental limit, with the largest prediction reaching about $8.6$ times the bound.
	
	By decomposing the signed EDM amplitudes, we identify distinct cancellation patterns among the five observables. The $b$ quark, neutron, and mercury EDMs are predominantly controlled by top-mass-enhanced charged-Higgs amplitudes. Owing to their large raw scales and relatively weak internal cancellations, the charged-Higgs contributions dominate these observables. For the $c$ quark EDM, pairwise cancellations between nearly degenerate neutral Higgs states strongly suppress the initially large neutral-Higgs contribution, making it comparable to the charged-Higgs contribution and thereby enhancing their destructive interference. The electron EDM, which receives only neutral-Higgs contributions at one loop, is suppressed by the same pairwise cancellation mechanism.
	
	By applying degenerate perturbation theory to the neutral-Higgs mass matrix, we further show that its unperturbed part gives rise to two doubly degenerate heavy-Higgs subspaces. The perturbation splits each degenerate eigenvalue while approximately preserving the pairwise relations among the corresponding mixing-matrix elements. These relations enforce the anti-alignment of the corresponding CP-odd coupling invariants, which, together with the small mass splittings, provides an analytical explanation for the observed pairwise EDM cancellations.

	\section*{Data Availability Statement}
	
	The numerical data generated and analyzed during the present study are available from the corresponding author upon reasonable request. No additional publicly available research data or software were generated beyond those presented in this article.

	\begin{acknowledgments}
		The work has been supported by the National Natural Science Foundation of China
		(NNSFC) with Grants No. 12075074, No. 12235008, No. 11535002, No. 11705045, Natural Science Foundation for Distinguished Young Scholars of Hebei Province with Grant No. A2022201017, Natural Science Foundation of Guangxi Autonomous Region with Grant No. 2022GXNSFDA035068, the youth top-notch talent support program of the Hebei Province, and Midwest Universities Comprehensive Strength Promotion project.
	\end{acknowledgments}

	\bibliography{}

\begin{thebibliography}{99}
        \bibitem{ATLAS_2012_Higgs}
        G. Aad et al. [ATLAS Collaboration],
        ``Observation of a new particle in the search for the Standard Model Higgs boson with the ATLAS detector at the LHC,''
        Phys. Lett. B \textbf{716} (2012), {1--29}.
        doi:10.1016/j.physletb.2012.08.020
        [arXiv:1207.7214 [hep-ex]].
        
        \bibitem{CMS_2012_Higgs}
        S. Chatrchyan et al. [CMS Collaboration],
        ``Observation of a new boson at a mass of 125 GeV with the CMS experiment at the LHC,''
        Phys. Lett. B \textbf{716} (2012), {30--61}.
        doi:10.1016/j.physletb.2012.08.021
        [arXiv:1207.7235 [hep-ex]].
        
        \bibitem{ATLAS_CMS_2016_Higgs}
        G. Aad et al. [ATLAS and CMS Collaborations],
        ``Measurements of the Higgs boson production and decay rates and constraints on its couplings from a combined ATLAS and CMS analysis of the LHC $pp$ collision data at $\sqrt{s}=7$ and $8~\mathrm{TeV}$,''
        JHEP \textbf{08} (2016), {045}.
        doi:10.1007/JHEP08(2016)045
        [arXiv:1606.02266 [hep-ex]].
        
        \bibitem{ATLAS_2022_Higgs}
        G. Aad et al. [ATLAS Collaboration],
        ``A detailed map of Higgs boson interactions by the ATLAS experiment ten years after the discovery,''
        Nature \textbf{607} (2022), {52--59}.
        doi:10.1038/s41586-022-04893-w
        [arXiv:2207.00092 [hep-ex]].
        
        \bibitem{CMS_2022_Higgs}
        A. Tumasyan et al. [CMS Collaboration],
        ``A portrait of the Higgs boson by the CMS experiment ten years after the discovery,''
        Nature \textbf{607} (2022), {60--68}.
        doi:10.1038/s41586-022-04892-x
        [arXiv:2207.00043 [hep-ex]].
        
        \bibitem{Lee_1973}
        T. D. Lee,
        ``A Theory of Spontaneous T Violation,''
        Phys. Rev. D \textbf{8} (1973), {1226--1239}.
        doi:10.1103/PhysRevD.8.1226
        
        \bibitem{Weinberg_1976}
        S. Weinberg,
        ``Gauge Theory of CP Violation,''
        Phys. Rev. Lett. \textbf{37} (1976), {657--661}.
        doi:10.1103/PhysRevLett.37.657
        
        \bibitem{Branco_2012}
        G. C. Branco, P. M. Ferreira, L. Lavoura, M. N. Rebelo, M. Sher, J. P. Silva,
        ``Theory and phenomenology of two-Higgs-doublet models,''
        Phys. Rept. \textbf{516} (2012), {1--102}.
        doi:10.1016/j.physrep.2012.02.002
        [arXiv:1106.0034 [hep-ph]].
        
        \bibitem{Branco_1999}
        G. C. Branco, L. Lavoura, J. P. Silva,
        ``CP Violation,''
        Int. Ser. Monogr. Phys. \textbf{103} (1999), {1--536}.
        Oxford University Press.
        
        \bibitem{Branco_Ivanov_2016}
        G. C. Branco, I. P. Ivanov,
        ``Group-theoretic restrictions on generation of CP-violation in multi-Higgs-doublet models,''
        JHEP \textbf{01} (2016), {116}.
        doi:10.1007/JHEP01(2016)116
        [arXiv:1511.02764 [hep-ph]].
        
        \bibitem{Sakharov_1967}
        A. D. Sakharov,
        ``Violation of CP invariance, C asymmetry, and baryon asymmetry of the universe,''
        JETP Lett. \textbf{5} (1967), {32--35}.
        
        \bibitem{Kobayashi_Maskawa_1973}
        M. Kobayashi, T. Maskawa,
        ``CP-Violation in the Renormalizable Theory of Weak Interaction,''
        Prog. Theor. Phys. \textbf{49} (1973), {652--657}.
        doi:10.1143/PTP.49.652
        
        \bibitem{Gavela_1994_1}
        M. B. Gavela, P. Hernandez, J. Orloff, O. Pene,
        ``Standard Model CP-violation and baryon asymmetry. Part 1: Zero temperature,''
        Nucl. Phys. B \textbf{430} (1994), {345--381}.
        doi:10.1016/0550-3213(94)00409-9
        [arXiv:hep-ph/9406288].
        
        \bibitem{Gavela_1994_2}
        M. B. Gavela, P. Hernandez, J. Orloff, O. Pene, C. Quimbay,
        ``Standard Model CP-violation and baryon asymmetry. Part 2: Finite temperature,''
        Nucl. Phys. B \textbf{430} (1994), {382--426}.
        doi:10.1016/0550-3213(94)00410-2
        [arXiv:hep-ph/9406289].
        
        \bibitem{Pospelov_Ritz_2005}
        M. Pospelov, A. Ritz,
        ``Electric dipole moments as probes of new physics,''
        Annals Phys. \textbf{318} (2005), {119--169}.
        doi:10.1016/j.aop.2005.04.002
        [arXiv:hep-ph/0504231].
        
        \bibitem{Engel_2013}
        J. Engel, M. J. Ramsey-Musolf, U. van Kolck,
        ``Electric Dipole Moments of Nucleons, Nuclei, and Atoms: The Standard Model and Beyond,''
        Prog. Part. Nucl. Phys. \textbf{71} (2013), {21--74}.
        doi:10.1016/j.ppnp.2013.03.003
        [arXiv:1303.2371 [nucl-th]].
        
        \bibitem{Abel_2020}
        C. Abel et al.,
        ``Measurement of the permanent electric dipole moment of the neutron,''
        Phys. Rev. Lett. \textbf{124} (2020), {081803}.
        doi:10.1103/PhysRevLett.124.081803
        [arXiv:2001.11966 [hep-ex]].
        
        \bibitem{Graner_2016}
        B. Graner, Y. Chen, E. G. Lindahl, B. R. Heckel,
        ``Reduced Limit on the Permanent Electric Dipole Moment of $^{199}$Hg,''
        Phys. Rev. Lett. \textbf{116} (2016), {161601}.
        doi:10.1103/PhysRevLett.116.161601
        [arXiv:1601.04339 [physics.atom-ph]].
        
        \bibitem{Roussy_2023}
        T. S. Roussy, L. Caldwell, T. Wright, W. B. Cairncross, Y. Shagam, K. B. Ng, N. Schlossberger, S. Y. Park, A. Wang, J. Ye, E. A. Cornell,
        ``A new bound on the electron's electric dipole moment,''
        Science \textbf{381} (2023), {46--50}.
        doi:10.1126/science.adg4084
        [arXiv:2212.11841 [physics.atom-ph]].
        
      \bibitem{Lee_1977}
      B. W. Lee, C. Quigg and H. B. Thacker,
    ``Weak interactions at very high energies: The role of the Higgs-boson mass,''
     Phys. Rev. D \textbf{16} (1977), {1519--1531}.
      doi:10.1103/PhysRevD.16.1519
        
        \bibitem{Altmannshofer_2016_FlavorfulHiggs}
        W. Altmannshofer, J. Eby, S. Gori, M. Lotito, M. Martone, D. Tuckler,
        ``Collider Signatures of Flavorful Higgs Bosons,''
        Phys. Rev. D \textbf{94} (2016) no.11, {115032}.
        doi:10.1103/PhysRevD.94.115032
        [arXiv:1610.02398 [hep-ph]].
        
        \bibitem{Altmannshofer_2018_FlavorLocked}
        W. Altmannshofer, S. Gori, D. J. Robinson, D. Tuckler,
        ``The Flavor-locked Flavorful Two Higgs Doublet Model,''
        JHEP \textbf{03} (2018), {129}.
        doi:10.1007/JHEP03(2018)129
        [arXiv:1712.01847 [hep-ph]].
        
        \bibitem{Altmannshofer_2018_Twist}
        W. Altmannshofer, B. Maddock,
        ``Flavorful Two Higgs Doublet Models with a Twist,''
        Phys. Rev. D \textbf{98} (2018) no.7, {075005}.
        doi:10.1103/PhysRevD.98.075005
        [arXiv:1805.08659 [hep-ph]].
        
        \bibitem{Altmannshofer_2025_G3HDM}
        W. Altmannshofer, K. Toner,
        ``Flavor Constraints in a Generational Three Higgs Doublet Model,''
        Phys. Rev. D \textbf{111} (2025) no.7, {075009}.
        doi:10.1103/PhysRevD.111.075009
        [arXiv:2502.04579 [hep-ph]].
        
        \bibitem{Navas_2024}
        S. Navas et al. [Particle Data Group],
        ``Review of Particle Physics,''
        Phys. Rev. D \textbf{110} (2024), {030001}.
        doi:10.1103/PhysRevD.110.030001
        
        \bibitem{Djouadi_2008}
        A. Djouadi,
        ``The anatomy of electro-weak symmetry breaking. II: The Higgs bosons in the minimal supersymmetric model,''
        Phys. Rept. \textbf{459} (2008), {1--241}.
        doi:10.1016/j.physrep.2007.10.005
        [arXiv:hep-ph/0503173 [hep-ph]].
        
        \bibitem{Ellis_1976}
        J. R. Ellis, M. K. Gaillard and D. V. Nanopoulos,
        ``A phenomenological profile of the Higgs boson,''
        Nucl. Phys. B \textbf{106} (1976), {292--340}.
        doi:10.1016/0550-3213(76)90382-5
        
        \bibitem{Shifman_1979}
        M. A. Shifman, A. I. Vainshtein, M. B. Voloshin and V. I. Zakharov,
        ``Low-energy theorems for Higgs boson couplings to photons,''
        Sov. J. Nucl. Phys. \textbf{30} (1979), {711--716}
        [Yad. Fiz. \textbf{30} (1979), {1368--1378}].
        
        \bibitem{Bergstrom_1985}
        L. Bergstrom and G. Hulth,
        ``Induced Higgs couplings to neutral bosons in $e^+e^-$ collisions,''
        Nucl. Phys. B \textbf{259} (1985), {137--155}
        [Erratum: Nucl. Phys. B \textbf{276} (1986), {744}].
        doi:10.1016/0550-3213(85)90302-5
        
        
        \bibitem{k1}
        R. Grigjanis, P.J. ODonnell, M. Sutherland, H. Navelet,
        Phys. Rep \textbf{22} (1993), 93.
        
        \bibitem{Buchalla_1996}
        G. Buchalla, A. J. Buras and M. E. Lautenbacher,
        ``Weak decays beyond leading logarithms,''
        Rev. Mod. Phys \textbf{68} (1996), 1125-1244.
        doi:10.1103/revmodphys.68.1125
        [arXiv:hep-ph/9512380].
        
        \bibitem{Altmannshofer_2009}
        W. Altmannshofer, P. Ball, A. Bharucha, A. J. Buras, D. M. Straub and M. Wick,
        ``Symmetries and Asymmetries of $B \to K^{*}\mu^+\mu^-$ Decays in the Standard Model and Beyond,''
        JHEP \textbf{2009} (2009), 019–019.
        doi:10.1088/1126-6708/2009/01/019
        [arXiv:0811.1214 [hep-ph]].
        
        \bibitem{L.LIN}
        L. Lin, T.-F. Feng, F. Sun,
        Mod. Phys. Lett. A \textbf{24} (2009), 2181-2186.
        
        \bibitem{Yang_2010}
        X.-Y. Yang and T.-F. Feng,
        ``Heavy fermions and two-loop electroweak corrections to $b \to s+\gamma$,''
        JHEP \textbf{2010} (2010), 1029-8479.
        doi:10.1007/jhep05(2010)059
        [arXiv:1005.4543 [hep-ph]].
        
        \bibitem{Goertz_2011}
        F. Goertz, T. Pfoh,
        ``Randall-Sundrum Corrections to the Width Difference and CP-Violating Phase in $B_s^0$-Meson Decays,''
        Phys. Rev. D \textbf{84} (2011), 1550-2368.
        doi:10.1103/physrevd.84.095016
        [arXiv:1105.1507 [hep-ph]].
        
        \bibitem{Buras_2011}
        A. J. Buras, L. Merlo and E. Stamou,
        ``The Impact of Flavour Changing Neutral Gauge Bosons on $B \to X_s\gamma$,''
        JHEP \textbf{2011} (2011), 1029-8479.
        doi:10.1007/jhep08(2011)124
        [arXiv:1105.5146 [hep-ph]].
        
        \bibitem{Gambino_2001}
        P. Gambino and M. Misiak,
        ``Quark mass effects in $B \to X_s\gamma$,''
        Nucl. Phys. B \textbf{611} (2001), 338–366.
        doi:10.1016/s0550-3213(01)00347-9
        [arXiv:hep-ph/0104034].
        
        \bibitem{Czakon_2007}
        M. Czakon, U. Haisch and M. Misiak,
        ``Four-loop anomalous dimensions for radiative flavour-changing decays,''
        JHEP \textbf{2007} (2007), {008–008}.
        doi:10.1088/1126-6708/2007/03/008
        [arXiv:hep-ph/0612329].
        
        \bibitem{Buras_1994}
        A.J. Buras, M. Misiak, M. Müunz, S. Pokorski,
        ``Theoretical Uncertainties and Phenomenological Aspects of $B \to X_s\gamma$ Decay,''
        Nucl. Phys. B \textbf{424} (1994), {374–398}.
        doi:10.1016/0550-3213(94)90299-2
        [arXiv:hep-ph/9311345].
        
        \bibitem{T.-J. Gao}
        T.-J. Gao, T.-F. Feng, J.-B. Chen,
        Phys. Lett. A \textbf{27} (2012), {1250011}.
        
        \bibitem{Yang_2018}
        J.-L. Yang, T.-F. Feng, S.-M. Zhao, R.-F. Zhu, X.-Y. Yang and H.-B. Zhang,
        ``Two loop electroweak corrections to $\bar{B}\to X_s\gamma$ and $B_s^0\to \mu^+\mu^-$ in the B-LSSM,''
        EPJC \textbf{78} (2018), {1434-6052}.
        doi:10.1140/epjc/s10052-018-6174-5
        [arXiv:1803.09904 [hep-ph]].
        
        \bibitem{Gambino_2003}
        P. Gambino, M. Gorbahn, U. Haisch,
        ``Anomalous Dimension Matrix for Radiative and Rare Semileptonic B Decays up to Three Loops,''
        Nucl. Phys. B \textbf{673} (2003), {238–262}.
        doi:10.1016/j.nuclphysb.2003.09.024
        [arXiv:hep-ph/0306079].

      \bibitem{Porod_2003_SPheno}
       W. Porod,
      ``SPheno, a program for calculating supersymmetric spectra, SUSY particle decays and SUSY particle production at $e^+e^-$ colliders,''
       Comput. Phys. Commun. \textbf{153} (2003), {275--315}.
        doi:10.1016/S0010-4655(03)00222-4
         [arXiv:hep-ph/0301101 [hep-ph]].
        
        \bibitem{Baron_2014}
        J. Baron et al. [ACME Collaboration],
        ``Order of Magnitude Smaller Limit on the Electric Dipole Moment of the Electron,''
        Science \textbf{343} (2014), 269--272.
        doi:10.1126/science.1248213
        [arXiv:1310.7534 [physics.atom-ph]].
        
        \bibitem{Andreev_2018}
        V. Andreev et al. [ACME Collaboration],
        ``Improved limit on the electric dipole moment of the electron,''
        Nature \textbf{562} (2018), 355--360.
        doi:10.1038/s41586-018-0599-8
        
        \bibitem{Blinov_2009}
        A. E. Blinov and A. S. Rudenko,
        ``Upper Limits on Electric and Weak Dipole Moments of Tau-Lepton and Heavy Quarks from $e^+e^-$ Annihilation,''
        Nucl. Phys. Proc. Suppl. \textbf{189} (2009), 257--259.
        doi:10.1016/j.nuclphysbps.2009.03.043
        [arXiv:0811.2380 [hep-ph]].
        
        \bibitem{Gisbert_2020}
        H. Gisbert and J. Ruiz Vidal,
        ``Improved bounds on heavy quark electric dipole moments,''
        Phys. Rev. D \textbf{101} (2020), {115010}.
        doi:10.1103/PhysRevD.101.115010
        [arXiv:1905.02513 [hep-ph]].
        
        \bibitem{Chang_1990}
        D. Chang, W.-Y. Keung, C. S. Li and T. C. Yuan,
        ``QCD Corrections to CP Violation From Color Electric Dipole Moment of $b$ Quark,''
        Phys. Lett. B \textbf{241} (1990), {589--592}.
        doi:10.1016/0370-2693(90)91875-C
        
        \bibitem{Feng_2005}
        T. F. Feng, X. Q. Li, J. Maalampi and X. M. Zhang,
        ``Two-loop gluino contributions to neutron electric dipole moment in CP-violating MSSM,''
        Phys. Rev. D \textbf{71} (2005), {056005}.
        doi:10.1103/PhysRevD.71.056005
        [arXiv:hep-ph/0412147 [hep-ph]].
        
        \bibitem{F. Staub1}
        F. Staub,
        [arXiv:0806.0538 [hep-ph]].
        
        \bibitem{F. Staub2}
        F. Staub,
        Comput. Phys. Commun \textbf{181} (2010), {1077-1086}.
        doi:10.1016/j.cpc.2010.01.011
        [arXiv:0909.2863 [hep-ph]].
        
        \bibitem{F. Staub3}
        F. Staub,
        Comput. Phys. Commun \textbf{182} (2011), {808-833}.
        doi:10.1016/j.cpc.2010.11.030
        [arXiv:1002.0840 [hep-ph]].
        
        \bibitem{F. Staub4}
        F. Staub,
        Comput. Phys. Commun \textbf{184} (2013), {1792}.
        doi:10.1016/j.cpc.2013.02.019
        [arXiv:1207.0906 [hep-ph]].
        
        \bibitem{F. Staub5}
        F. Staub,
        Comput. Phys. Commun \textbf{185} (2014), {1773}.
        doi:10.1016/j.cpc.2014.02.018
        [arXiv:1309.7223 [hep-ph]].
        
        
        \bibitem{Sala_2014}
        F. Sala,
        ``A bound on the charm chromo-EDM and its implications,''
        JHEP \textbf{03} (2014), {061}.
        doi:10.1007/JHEP03(2014)061
        [arXiv:1312.2589 [hep-ph]].
        
        \bibitem{Lee_2005}
        C. Lee, V. Cirigliano and M. J. Ramsey-Musolf,
        ``Resonant relaxation in electroweak baryogenesis,''
        Phys. Rev. D \textbf{71} (2005), {075010}.
        doi:10.1103/PhysRevD.71.075010
        [arXiv:hep-ph/0412354 [hep-ph]].
        
        
        	
        \end{thebibliography}

\end{document}